\documentclass[11pt]{article}
\usepackage{axodraw2}
\usepackage{epsfig}
\usepackage{amsfonts}
\usepackage{slashed}
\usepackage{amsmath,amssymb}
\usepackage{bbm,bm}
\usepackage{soul}
\usepackage{cite}
\usepackage{url}
\allowdisplaybreaks[1]

\usepackage{tikz}
\usetikzlibrary{shapes,arrows,shadows,automata,positioning}
\newdimen\nodeDist
\usepackage{comment}
\usepackage[normalem]{ulem} 

 \usepackage[
   colorlinks=true,
   linkcolor={red!50!black},
   citecolor={blue!50!black},
   filecolor=black,
   urlcolor=black,
   breaklinks=true
   ]{hyperref}
\usepackage{orcidlink}

\newcommand{\nn}{\nonumber \\}
\renewcommand{\vec}[1]{{\bf #1}}
\newcommand{\rmi}[1]{{\mbox{\scriptsize #1}}}
\newcommand{\rmii}[1]{{\mbox{\tiny\rm{#1}}}}

\newcommand{\bmu}{\bar{\mu}}
\newcommand{\Lamd}{\bmu_{\rmii{3d}}}
\newcommand{\LamD}{\bmu_{\rmii{4d}}}

\newcommand{\gammaE}{{\gamma_\rmii{E}}}
\newcommand{\Veff}{V_{\rmi{eff}}}
\newcommand{\Seff}{{S_{\rmii{eff}}}}
\newcommand{\f}{\varphi}
\newcommand{\fo}{\varphi_0}
\newcommand{\fluc}{\delta\varphi}

\newcommand{\aS}{a_{\rmii{$S$}}}
\newcommand{\aL}{a_{\rmii{$L$}}}
\newcommand{\aT}{a_{\rmii{$T$}}}
\newcommand{\aG}{a_{\rmii{$G$}}}
\newcommand{\aGfv}{a_{\rmii{$G$;F}}}
\newcommand{\aSfv}{a_{\rmii{$S$;F}}}
\newcommand{\aLfv}{a_{\rmii{$L$;F}}}
\newcommand{\TS}{T_{\rmii{$S$}}}
\newcommand{\TL}{T_{\rmii{$L$}}}
\newcommand{\TG}{T_{\rmii{$G$}}}

\newcommand{\yS}{y_{\rmii{$S$}}}
\newcommand{\yL}{y_{\rmii{$L$}}}
\newcommand{\yG}{y_{\rmii{$G$}}}

\newcommand{\Tp}{T_{\rm p}}

\newcommand{\gX}{g_\rmii{$X$}}

\newcommand{\mF}{m_\rmii{F}}
\newcommand{\mDX}{m_{\rmii{D},\rmii{$X$}}}

\newcommand{\mX}{m_\rmii{$X$}}
\newcommand{\MX}{M_\rmii{$X$}}

\newcommand{\WX}{W_\rmii{$X$}}
\newcommand{\WG}{W_\rmii{$G$}}
\newcommand{\phib}{\varphi_b}
\newcommand{\phiF}{\varphi_{\rmii{F}}}
\newcommand{\GammaT}{\Gamma_\rmii{$T$}}

\newcommand{\tr}{{\rm Tr\,}}
\renewcommand{\det}{\mbox{det}}

\newcommand{\dd}{\rm d}

\newcommand{\dumbbell}{\mathrel{\bullet\!\!\!-\!\!\!\bullet}}
\usepackage{pifont}
\newcommand{\daisy}{\selectfont \ding{95}}

\def\Lwidth{1}

\newcommand{\pic}[1]{\;\parbox[c]{30pt}{\begin{picture}(30,30)(0,0)
\SetWidth{1.0}\SetScale{1.0} #1 \end{picture}}\;}

\newcommand{\picb}[1]{\;\parbox[c]{45pt}{\begin{picture}(45,30)(0,0)
\SetWidth{1.0}\SetScale{1.0} #1 \end{picture}}\;}

\def\Textt(#1,#2,#3){\Text(#1,#2)[t]{{$\scriptstyle #3$}}}
\def\Textb(#1,#2,#3){\Text(#1,#2)[b]{{$\scriptstyle #3$}}}
\def\Textl(#1,#2,#3){\Text(#1,#2)[l]{{$\scriptstyle #3$}}}
\def\Textr(#1,#2,#3){\Text(#1,#2)[r]{{$\scriptstyle #3$}}}
\def\Texttl(#1,#2,#3){\Text(#1,#2)[tl]{{$\scriptstyle #3$}}}
\def\Textbl(#1,#2,#3){\Text(#1,#2)[bl]{{$\scriptstyle #3$}}}
\def\Texttr(#1,#2,#3){\Text(#1,#2)[tr]{{$\scriptstyle #3$}}}
\def\Textbr(#1,#2,#3){\Text(#1,#2)[br]{{$\scriptstyle #3$}}}

\def\fextB(#1,#2){%
  \def\Lgex{#2}%
  \def\Lgez{\Lge3}%
  \def\Lgey{\Lge2}%
  #1(0,15)(7.5,15)
  \ifx\Lgex\Lgez
    #2(45,15)(37.5,15)
  \else
    \ifx\Lgex\Lgey
      #2(45,15)(37.5,15)
    \else
      #2(37.5,15)(45,15)
    \fi
  \fi}

\def\fextT(#1,#2){
  \def\Lgex{#2}%
  \def\Lgez{\Lge3}%
  \def\Lgey{\Lge2}%
  #1(0,0)(22.5,0)
  \ifx\Lgex\Lgez
    #2(45,0)(22.5,0)
  \else
    \ifx\Lgex\Lgey
      #2(45,0)(22.5,0)
    \else
      #2(22.5,0)(45,0)
    \fi
  \fi}

\def\Aglx(#1,#2)(#3,#4,#5){\PhotonArc(#1,#2)(#3,#4,#5){\Lwidth}
  {6.283 #3 mul 360 div #4 #5 sub #4 #5 sub mul sqrt mul Ldensity mul}}
\def\Lglx(#1,#2)(#3,#4){\Photon(#1,#2)(#3,#4){\Lwidth}
  {#1 #3 sub #1 #3 sub mul #2 #4 sub #2 #4 sub mul add sqrt Ldensity mul}}
\def\Axx(#1,#2)(#3,#4,#5){\DashCArc(#1,#2)(#3,#4,#5){3}}
\def\Lxx(#1,#2)(#3,#4){\DashLine(#1,#2)(#3,#4){3}}
\def\Asai(#1,#2)(#3,#4,#5){\CArc(#1,#2)(#3,#4,#5)}
\def\Lsai(#1,#2)(#3,#4){\Line(#1,#2)(#3,#4)}

\def\Vtxttxt(#1,#2,#3,#4){\picb{
  #1(0,15)(22.5,15)%
  #2(45,15)(22.5,15)%
  \Vertex(22.5,15){3}%
  \Textbr(5,20,#3)%
  \Textbl(35,20,#4)%
}}

\def\TopoVR(#1){\pic{%
  #1(15,15)(15,0,180)%
  #1(15,15)(15,180,360)%
}}

\def\TopoSBo(#1,#2,#3){
  \fextB(#1,#1)
  #2(22.5,15)(15,0,180)%
  #3(22.5,15)(15,180,360)}
\def\TopoSB(#1,#2,#3){\picb{
  \TopoSBo(#1,#2,#3)}}

\def\TopoST(#1,#2){\picb{
  \fextT(#1,#1)
  #2(22.5,15)(15,-90,270)}}

\def\TopoSTtxt(#1,#2,#3,#4,#5){\picb{
  \fextT(#1,#1)
  #2(22.5,15)(15,-90,270)
  \Textr(0,0,#3)%
  \Textl(45,0,#4)%
  \Textb(22.5,32,#5)%
  }}

\def\ToptVS(#1,#2,#3){\pic{
  #1(15,15)(15,0,180)%
  #2(15,15)(15,180,360)%
  #3(30,15)(0,15)%
}}

\def\SPropCircProp(#1,#2){\picb{
  #1(0,15)(15.5,15)%
  #1(30.5,15)(45,15)%
  \GCirc(22.5,15){8}{0.75}
  \Text(22.5,15)[c]{$\scriptstyle #2$}}}
\def\TPropCircProp(#1,#2,#3,#4){\picb{
  #1(20 60 cos 20 mul add,15 60 sin 20 mul add)%
    (20 60 cos 5 mul add,15 60 sin 5 mul add)%
  #2(20 60 cos 20 mul add,15 60 sin 20 mul sub)%
    (20 60 cos 5 mul add,15 60 sin 5 mul sub)%
  #3(0,15)(15,15)%
  \GCirc(20,15){6}{0.75}%
  \Text(20,15)[c]{$\scriptstyle #4$}}}
\def\VPropCircProp(#1,#2,#3,#4,#5){\picb{
  #1(22.5 45 cos 22.5 mul add,15 45 sin 22.5 mul add)(22.5,15)%
  #2(22.5 135 cos 22.5 mul add,15 135 sin 22.5 mul add)(22.5,15)%
  #3(22.5 225 cos 22.5 mul add,15 225 sin 22.5 mul add)(22.5,15)%
  #4(22.5 315 cos 22.5 mul add,15 315 sin 22.5 mul add)(22.5,15)%
  \GCirc(22.5,15){6}{0.75}
  \Text(22.5,15)[c]{$\scriptstyle #5$}}}
\def\TopSi(#1,#2){\SPropCircProp(#1,#2)}
\def\TopSr(#1,#2){\SPropBoxProp(#1,#2)}
\def\TopTi(#1,#2,#3,#4){\TPropCircProp(#1,#2,#3,#4)}
\def\TopVi(#1,#2,#3,#4,#5){\VPropCircProp(#1,#2,#3,#4,#5)}

\def\TopoTBl(#1,#2,#3){\picb{
  #1(0,15)(10,15)
  #1(22.5 30 cos 22.5 mul add,15 30 sin 22.5 mul add)(35,15)%
  #1(22.5 30 cos 22.5 mul add,15 30 sin 22.5 mul sub)(35,15)%
  #2(22.5,15)(12.5,0,180)
  #3(22.5,15)(12.5,180,0)}}

\def\TopoTC(fex(#1),#2,#3,#4){\picb{
  \def\fex(##1,##2,##3){
    ##1(22.5 60 cos 22.5 mul add,15 60 sin 22.5 mul add)%
       (22.5 60 cos 12.5 mul add,15 60 sin 12.5 mul add)%
    ##2(22.5 60 cos 22.5 mul add,15 60 sin 22.5 mul sub)%
       (22.5 60 cos 12.5 mul add,15 60 sin 12.5 mul sub)%
    ##3(0,15)(10,15)
  }\fex(#1)
  #2(22.5,15)(12.5,-60,60)%
  #3(22.5,15)(12.5,180,300)%
  #4(22.5,15)(12.5,60,180)%
 }}

\def\TopoVBlr(fex(#1),#2,#3){\picb{
  \def\fex(##1,##2,##3,##4){
    ##1(22.5 30 cos 22.5 mul sub,15 30 sin 22.5 mul add)(10,15)%
    ##2(22.5 30 cos 22.5 mul sub,15 30 sin 22.5 mul sub)(10,15)%
    ##3(22.5 30 cos 22.5 mul add,15 30 sin 22.5 mul add)(35,15)%
    ##4(22.5 30 cos 22.5 mul add,15 30 sin 22.5 mul sub)(35,15)%
  }%
  \fex(#1)
  #2(22.5,15)(12.5,0,180)
  #3(22.5,15)(12.5,180,0)}}

\def\TopoVCrl(fex(#1),#2,#3,#4){\picb{
  \def\fex(##1,##2,##3,##4){
    ##1%
      (22.5 180 30 add cos 22.5 mul add,15 180 30 add sin 22.5 mul add)%
      (10,15)%
    ##2%
      (22.5 180 30 sub cos 22.5 mul add,15 180 30 sub sin 22.5 mul add)%
      (10,15)%
    ##4%
      (22.5 315 cos 22.5 mul add,15 315 sin 22.5 mul add)%
      (22.5 315 cos 12.5 mul add,15 315 sin 12.5 mul add)%
    ##3%
      (22.5 45 cos 22.5 mul add,15 45 sin 22.5 mul add)%
      (22.5 45 cos 12.5 mul add,15 45 sin 12.5 mul add)%
  }\fex(#1)
  #4(22.5,15)(12.5,180,315)%
  #3(22.5,15)(12.5,-45,45)%
  #2(22.5,15)(12.5,45,180)%
  }}

\def\TopoVD(fex(#1),#2,#3,#4,#5){\picb{
  \def\fex(##1,##2,##3,##4){
    ##1%
      (22.5 45 cos 22.5 mul add,15 45 sin 22.5 mul add)%
      (22.5 45 cos 12.5 mul add,15 45 sin 12.5 mul add)%
    ##4%
      (22.5 135 cos 22.5 mul add,15 134 sin 22.5 mul add)%
      (22.5 135 cos 12.5 mul add,15 135 sin 12.5 mul add)%
    ##3%
      (22.5 225 cos 22.5 mul add,15 225 sin 22.5 mul add)%
      (22.5 225 cos 12.5 mul add,15 225 sin 12.5 mul add)%
    ##2%
      (22.5 315 cos 22.5 mul add,15 315 sin 22.5 mul add)%
      (22.5 315 cos 12.5 mul add,15 315 sin 12.5 mul add)%
  }\fex(#1)%
  #5(22.5,15)(12.5,45,135)%
  #4(22.5,15)(12.5,135,225)%
  #3(22.5,15)(12.5,225,315)%
  #2(22.5,15)(12.5,-45,45)%
  }}

\def\scfc{0.7}  
\def\phgt{21}   
\def\pwc{21}    
\def\pwcb{31.5} 
 
\newcommand{\PIC}[4]{\;\parbox[c]{#2 pt}{\begin{picture}(#2,#3)(0,0)
\SetWidth{1.0}\SetScale{#4} #1 \end{picture}}\;}
\renewcommand{\pic}[1]{\PIC{#1}{\pwc}{\phgt}{\scfc}}
\renewcommand{\picb}[1]{\PIC{#1}{\pwcb}{\phgt}{\scfc}}

\makeatletter \@addtoreset{equation}{section} \makeatother
\renewcommand{\theequation}{\arabic{section}.\arabic{equation}}
\makeatletter
\renewcommand\section{\@startsection{section}{1}{\z@}%
  {-5.5ex \@plus -1ex \@minus -.2ex}
  {2.3ex \@plus.2ex}%
  {\normalfont\large\bfseries}}
\renewcommand\subsection{\@startsection{subsection}{2}{\z@}%
  {-3.25ex\@plus -1ex \@minus -.2ex}%
  {1.5ex \@plus .2ex}%
  {\normalfont\normalsize\bfseries}}
\renewcommand\thesection{\@arabic\c@section}
\renewcommand\thesubsection{\thesection.\@arabic\c@subsection}
\renewcommand{\@seccntformat}[1]{%
  \csname the#1\endcsname.\hspace{1.0em}}
\makeatother

\begin{document}

\flushbottom

\begin{titlepage}

\begin{flushright}
HIP-2024-27/TH\\
CERN-TH-2025-046
\end{flushright}
\begin{centering}

\vfill

{\Large{\bf
Finite-temperature bubble nucleation
\\
with shifting scale hierarchies
}}

\vspace{0.8cm}

\renewcommand{\thefootnote}{\fnsymbol{footnote}}
Maciej Kierkla%
\orcidlink{0000-0002-2785-5370}%
,$^{\rm a,}$%
\footnotemark[1]
Philipp Schicho%
\orcidlink{0000-0001-5869-7611}%
,$^{\rm b,}$%
\footnotemark[2]
Bogumi{\l}a {\'S}wie{\.z}ewska%
\orcidlink{0000-0003-0169-211X}%
,$^{\rm a,}$%
\footnotemark[3]
\\
Tuomas V.~I.~Tenkanen%
\orcidlink{0000-0002-3087-8450}%
,$^{\rm c,}$%
\footnotemark[4]
and Jorinde van de Vis%
\orcidlink{0000-0002-8110-1983}
$^{\rm d,}$%
\footnotemark[5]

\vspace{0.8cm}

$^\rmi{a}$%
{\em
Faculty of Physics, University of Warsaw, \\
Pasteura 5, 02-093 Warsaw, Poland\\}
\vspace{0.3cm}

$^\rmi{b}$%
{\em
  D\'epartement de Physique Th\'eorique, Universit\'e de Gen\`eve,\\
  24 quai Ernest Ansermet, CH-1211 Gen\`eve 4, Switzerland\\}
\vspace{0.3cm}

$^\rmi{c}$%
{\em
  Department of Physics and Helsinki Institute of Physics,\\
  P.O.~Box 64, FI-00014 University of Helsinki,
  Finland\\}
\vspace{0.3cm}

$^\rmi{d}$%
{\em
  Theoretical Physics Department, CERN,\\
  1 Esplanade des Particules, CH-1211 Geneva 23, Switzerland\\}
\vspace{0.3cm}

\vspace*{0.8cm}

\mbox{\bf Abstract}

\end{centering}

\vspace*{0.3cm}

\noindent
Focusing on supercooled phase transitions in models with classical scale symmetry, we formulate a state-of-the art
framework for computing the bubble nucleation rate,
accounting for the presence of varying energy scales.
In particular, we examine the limitations of derivative expansions in constructing
a thermal effective field theory for bubble nucleation.
We show that for gauge-field fluctuations,
derivative expansions diverge after the two leading orders due to
the strong variation in gauge-field masses
between the high- and low-temperature phases.
By directly computing these contributions using the fluctuation determinant,
we capture these effects while also accounting for the dominant contribution at two-loop.
Finally, we demonstrate how this approach significantly
improves nucleation rate calculations compared to leading-order results,
providing a more robust framework for predicting
gravitational-wave signals from supercooled phase transitions in models such as the SU(2)cSM.

\vfill
\end{titlepage}

{\hypersetup{hidelinks}
\tableofcontents
}

\renewcommand{\thefootnote}{\fnsymbol{footnote}}
\footnotetext[1]{maciej.kierkla@fuw.edu.pl}
\footnotetext[2]{philipp.schicho@unige.ch}
\footnotetext[3]{bogumila.swiezewska@fuw.edu.pl}
\footnotetext[4]{tuomas.tenkanen@helsinki.fi}
\footnotetext[5]{jorinde.van.de.vis@cern.ch}
\clearpage

\renewcommand{\thefootnote}{\arabic{footnote}}
\setcounter{footnote}{0}

\section{Introduction}
\label{sec:intro}
Gravitational waves (GWs), first observed from astrophysical sources by the LIGO-Virgo-Kagra collaboration~\cite{LIGOScientific:2016aoc, LIGOScientific:2016sjg, LIGOScientific:2017bnn,LIGOScientific:2017vox,LIGOScientific:2017ycc, LIGOScientific:2018mvr, LIGOScientific:2020ibl, LIGOScientific:2021usb, KAGRA:2021vkt}, constitute a tool to  directly test our understanding of the early universe.
In particular, with the advent of space-borne gravitational-wave detectors such as Laser Interferometer Space Antenna (LISA)~\cite{LISA:2017pwj, LISACosmologyWorkingGroup:2022jok}, predictions of early-universe cosmology around the electroweak epoch
will become testable. 

A phase transition associated with symmetry breaking is a phenomenon well-suited to be probed by LISA since its frequency band corresponds to energies in the early universe of the order of the electroweak scale.
Electroweak symmetry breaking proceeds via a crossover without physics
beyond the Standard Model~\cite{Kajantie:1996mn,Gurtler:1997hr,Csikor:1998eu}.
Therefore, within the Standard Model (SM) no GW aftermath of electroweak symmetry breaking is expected.

At the same time, many beyond-the-Standard-Model (BSM) models
are believed to accommodate a first-order phase transition sufficiently strong to source GWs,
often driven by
a modified scalar potential or
an extended scalar sector;
see e.g.~\cite{Caprini:2019egz} and references therein.
Thus, observing a stochastic GW background compatible with the predicted signal from such a phase transition
would be a striking signature of new physics.

The prospects of reconstructing the properties of the phase transition or of the underlying fundamental-physics model have been studied in the literature~\cite{Gowling:2021gcy, Giese:2021dnw, Boileau:2022ter, Gowling:2022pzb, Caprini:2024hue,Gonstal:2025qky}. In general, the stronger the transition, the better the prospects for observation. In particular, supercooled phase transitions in which a large amount of latent heat is released source very strong signals, likely to be observed~\cite{Randall:2006, Konstandin:2011,Hambye:2013, Jaeckel:2016,  Hashino:2016, Jinno:2016, Marzola:2017,Hashino:2018, Baldes:2018, Prokopec:2018,Marzo:2018,Mohamadnejad:2019vzg, Kang:2020jeg, Mohamadnejad:2021tke, Dasgupta:2022isg, Kierkla:2022odc, Kierkla:2023von}.
For the same reason, the predicted accuracy of reconstruction of the spectral or thermodynamic parameters is much better than for typical phase transitions~\cite{Caprini:2024hue, Gonstal:2025qky}. Recently, in~\cite{Gonstal:2025qky} it was shown that the accuracy of the reconstruction of the inverse time scale of the transition, $\beta_*/H_*$, when marginalised over the remaining parameters, should be below 10\% in a vast range of the interesting parameter space for strongly supercooled phase transitions.
This prospect calls for an improved precision in theoretical predictions. 

Theoretical predictions for GW signals from first-order phase transitions are subject to several sources of uncertainty, including, as a prime example, slow convergence of perturbation theory at high temperatures~\cite{Croon:2020cgk,Gould:2021}, signalled by a strong dependence on the renormalisation scale at low perturbative orders. Many of these issues can be mitigated by employing a systematic scheme for resummations of higher-order thermal corrections. This can be achieved by constructing a high-temperature (HT) effective field theory (EFT) using the method of HT dimensional reduction (DR)~\cite{Ginsparg:1980ef,Appelquist:1981vg,Kajantie:1995dw,Braaten:1995cm}. This method was used to demonstrate that the SM electroweak phase transition is not of first order~\cite{Kajantie:1996mn,Gurtler:1997hr,Csikor:1998eu}.
In recent years,
it has received renewed interest~\cite{Gorda:2018hvi, Niemi:2018asa, Croon:2020cgk,Niemi:2021qvp, Gould:2021,Friedrich:2022cak,Biondini:2022ggt,Gould:2023jbz, Kierkla:2023von, Lewicki:2024xan,Ekstedt:2024etx,Ramsey-Musolf:2024ykk,Gould:2024jjt,Bertenstam:2025jvd} enhanced by the availability of the numerical package {\tt DRalgo}~\cite{Ekstedt:2022bff} facilitating the use of thermal EFTs.
Dimensional reduction not only provides a systematic method for including higher-order contributions to the effective potential, it also allows for
a gauge-independent determination of
the phase transition parameters~\cite{Laine:1994zq,Croon:2020cgk,Hirvonen:2021zej,Schicho:2022wty,Gould:2023ovu}.

Driven by the exciting prospect of detecting gravitational waves
from supercooled phase transitions ---
and recognizing the limitations of the commonly used
leading-order ``daisy resummation''~\cite{Arnold:1992, Parwani:1992} ---
ref.~\cite{Kierkla:2023von} initiated a study of such transitions using HT EFT beyond leading order.
A key insight from this work was that, despite the extreme supercooling,
HT dimensional reduction can still be used to construct an effective theory for nucleation,
incorporating higher-order corrections.
The reason that the HT expansion remains applicable, is that the ``escape point'' of the transitioning field lies within the regime where the HT expansion is valid.
Comparing with earlier results of~\cite{Kierkla:2022odc} obtained using daisy resummation,
it was found that the new results for the relevant quantities describing the phase transition differ significantly from the daisy-resummed results.
For instance, the percolation temperature differs relatively by approximately 50--90\%, while
the correction to the mean bubble separation at the moment of collision is close to 50\% across the parameter space.

A particularly notable result from~\cite{Kierkla:2023von} is that including the effective operator
modifying the kinetic term in the action (the three-dimensional, thermal $Z$-factor) was found to affect the results significantly. This could indicate that the derivative expansion of the effective action, which is employed to approximate the contribution from fluctuation determinants, is not performing well. Moreover, another, related issue was encountered; the effective HT description breaks down at the tail of the bounce solution.
There, the gauge bosons ---
in the construction of the EFT assumed to be heavy compared to the scalar that undergoes the phase transitions ---
become very light.
Adopting the terminology of~\cite{Gould:2021ccf},
gauge fields are {\em scale-shifters}
across the bubble profile interpolating between the low- and high-temperature phases. Such a situation may signal problems within the EFT and especially with the derivative expansion.

The goal of the present paper is to compute the nucleation rate
for supercooled phase transitions,
consistently including the contributions from varying energy scales and
from the higher-order thermal corrections.
This will allow us to reliably predict the parameters of the phase transition and
 assess
the validity of the derivative expansion within a thermal EFT framework.
To this end,
we compute the relevant phase transition parameters,
including exact (numerical) contributions to the effective action from
the fluctuation determinants,
and compare them to results obtained via a derivative expansion.
For consistency with previous results, we use the SU(2)cSM model with classical scale symmetry,
similar to~\cite{Kierkla:2022odc, Kierkla:2023von}.

The article is organised as follows.
Section~\ref{sec:nucl-rate} reviews methods for evaluating the nucleation rate during
a first-order phase transition.
We discuss how to incorporate higher-order effects by linking them to corresponding terms in the effective action.
Additionally, we clarify the role of the derivative expansion and
its relation to fluctuations around the background field.
Section~\ref{sec:overview-expansions} summarises various expansions that are needed to perturbatively compute the thermal bubble nucleation rate in a realistic model.
In section~\ref{sec:supercool}, we formulate the method to compute the nucleation rate for a supercooled phase transition working with the SU(2)cSM model.
We review the construction of the EFT and explain higher-order corrections to the nucleation rate.
Section~\ref{sec:results} contains the numerical results of our analysis.
We demonstrate the effect of higher-order corrections to the nucleation rate on the phase-transition parameters and discuss the accuracy of various approximations.
In section~\ref{sec:discussion}, we discuss and summarise our findings.
The main text is supplemented with appendices summarising technical details of the analysis.
In appendix~\ref{sec:commutation-relations}, we provide proofs of commutation relations used in section~\ref{sec:nucl-rate}.
Appendix~\ref{sec:derivative:expansion:higher} details the computation of higher-order terms in the derivative expansion.
Appendix~\ref{sec:app-NLO-rate} explains how to apply {\tt BubbleDet} to our setting, while
appendix~\ref{sec:GY} details the application of the Gel'fand-Yaglom theorem to computation of the functional determinants.

\section{Higher-order corrections to the nucleation rate}
\label{sec:nucl-rate}

In this section,
we review the computation of the bubble nucleation rate during a first-order phase transition beyond leading order.
We clarify the connection between higher-order corrections to the nucleation rate and those to the effective action.
Additionally, we discuss the expansion of the action in small fluctuations around a background field,
as well as the momentum expansion (i.e., derivative expansion of the scalar field).
We highlight the limitations of the latter and explain the impact of these expansions on nucleation rate calculations.

\subsection{Nucleation rate and the effective action}
\label{sec:rate-NLO}

We begin by outlying the computation of the nucleation rate following~\cite{Gould:2021ccf, Ekstedt:2023sqc}.
According to Langer's nucleation theory, the thermal nucleation rate $\GammaT$ factorises into the dynamical and statistical parts~\cite{Langer:1967ax, Langer:1969bc, Langer:1974cpa, Gould:2021ccf, Ekstedt:2023sqc},
\begin{equation}
\GammaT = A_{\rm dyn} A_{\rm stat}
\,.
\end{equation}
The statistical factor captures the equilibrium process of passing the barrier due to thermal fluctuations in the system. The dynamical prefactor encapsulates the out-of-equilibrium effects related to the evolution of the system after the field emerges at the other side of the barrier;
for more details see~\cite{Ekstedt:2023sqc, Hirvonen:2024rfg,Eriksson:2024ovi}.
For our application, we are interested in the nucleation rate per unit volume.
In that case,
the rate and its corresponding factors have the following units:
\begin{align}
[\GammaT] &= T^4
  \,,&
  [A_{\rm dyn}] &= T
  \,,&
  [A_{\rm stat}] &= T^3
  \,.
\end{align}

In the present work,
we will focus on the dominant statistical part which can be computed in terms of the effective action. 
For the dynamical part, we will use the approximation of~\cite{Langer:1969bc, Hanggi:1990zz, Berera:2019uyp} implemented in {\tt BubbleDet}~\cite{Ekstedt:2023sqc}, or approximate it dimensionally as $[A_{\rm dyn}] \approx T$.
The statistical part of the nucleation rate at one-loop order
is given by~\cite{Callan:1977pt, Vainshtein:1981wh, Andreassen:2016cvx, Ekstedt:2023sqc, Ekstedt:2022tqk}
\begin{equation}
\label{eq:rateBD}
	A_{\rm stat} =
      \prod_a
      \mathcal I_a
      \mathcal V_a
      \sqrt{ \frac{\det\,{\mathcal O_a(\phiF)}}{\det'\mathcal O_a(\phib)}}
      \mathcal I_\phi
      \sqrt{ \Big|\frac{\det\,{\mathcal O_\phi(\phiF)}}{\det'\mathcal O_\phi(\phib)}\Big|}
      \, e^{-(S[\phib]-S[\phiF])}
      \,,
\end{equation}
where in the exponent the leading order (LO) action is used, which is a functional of a background field $\varphi$ corresponding to the scalar field $\phi$, undergoing the phase transition.
Throughout this work we work in Euclidean spacetime.

The product is over all fields (indexed by $a$) coupled to $\phi$.
The various operators, $\mathcal O_a$, act on the fields at quadratic order in the Lagrangian,
{\em viz.}
\begin{align}
  \label{eq:OExamples}
  \mathcal O_\phi(\f) &= -\partial^2 + V^{''}(\f)
  \,,&
  \mathcal O_a(\f) &= - \partial^2 + m_a ^2(\varphi)
  \,,
\end{align}
for the nucleating scalar field and a non-mixing field, respectively.
The derivative on the potential $V$ is with respect to the field $\phi$.
Here,
$\phib$ denotes the bounce solution,
$\phiF$ denotes the static false vacuum solution,
$\mathcal I_a$ are Jacobian factors, and
$\mathcal V_a$ are volume factors.
Both 
$\mathcal I_a$ and
$\mathcal V_a$
are only non-trivial (not equal to 1) for fields with zero modes (associated to translations).
The prime on the determinant denotes that the zero modes are omitted from the determinant.
This implies that all the square roots of the determinants with a prime have mass (or temperature) dimension $d$, but these are
cancelled by the corresponding factor $\mathcal V_a$, except for the contribution of $\phi$, which does not have a volume factor.
For the derivation of this expression, and in particular for an explanation of how to deal with the zero modes and negative eigenvalues,
we refer to~\cite{Ekstedt:2021kyx, Ekstedt:2022tqk, Andreassen:2016cvx, Bhattacharya:2024chz}.
Below, we review how the functional determinants present in the prefactor in eq.~\eqref{eq:rateBD} arise as the effective action is computed at one-loop level.

The effective action, $\Seff$,%
\footnote{%
 The effective action is commonly denoted as $\Gamma$,
but to avoid confusion with the nucleation rate, we call it $\Seff$ in this work.
} 
can be expressed using the background field method~\cite{Jackiw:1974cv, schwartz:2014book} as follows,
\begin{equation}
e^{-\Seff[\f]}=\int \mathcal{D}\chi\mathcal{D}\tilde \f e^{-S[\chi,\f+\tilde \f]}
  \,,
\end{equation}
where the integration is restricted to field configurations leading to 1PI diagrams~\cite{schwartz:2014book}, $\f$ denotes an arbitrary background field, $\tilde\f$ corresponds to fluctuations around this background and $\chi$ denotes collectively other fields present in the theory.
We can now expand the action $S$ in fluctuations about the background,
which leads to the following relation
\begin{equation}
\label{eq:eff-action-2nd-order}
e^{-\Seff[\f]}=\int\mathcal{D}\chi e^{-S[\chi, \f]}\int\mathcal{D}\tilde\f e^{-\left(\frac{\delta S}{\delta \phi}[0,\f] \tilde\f+ \frac{1}{2}\tilde\f \frac{\delta^2 S}{ \delta \phi^2}[0,\f] \tilde\f + \ldots \right)}
\,.
\end{equation}
In the first integral only the terms quadratic in $\chi$ contribute as only these can be used to construct 1PI diagrams.
In the second integral,
the first term vanishes if $\f$ is a solution of the equations of motion which extremises the action.%
\footnote{%
  For a general background, which is not a solution of the equations of motion,
  one can remove the terms linear in $\tilde\f$ as it is not possible to construct
  1PI diagrams with external $\f$ legs that would involve such vertices.
}
Since we are interested in evaluating the nucleation rate of the true vacuum bubbles,
we will indeed evaluate the action on the bounce solution which extremises  the action.
It is important to note that higher orders in the expansion in fluctuations,
beyond terms quadratic in
the fluctuation field, only contribute beyond one-loop level.
Therefore, as long as we work at one-loop order, the expression in
eq.~\eqref{eq:eff-action-2nd-order} truncated to quadratic order is exact.
Both integrals are Gaussian and thus eq.~\eqref{eq:eff-action-2nd-order} yields 
\begin{equation}
\label{eq:exp-eff-action}
e^{-\Seff[\f]}=\prod_a\frac{1}{\sqrt{\det \mathcal{O}_a(\f)}} e^{-S[0,\f]}
\,,
\end{equation}
where the $\mathcal{O}_a$ operators correspond to the
terms quadratic in the respective fields (the $\chi$ fields and the scalar field) following from the action $S$. These are the same operators that were present in the prefactor of eq.~\eqref{eq:rateBD}.
Note, that if there is no kinetic mixing,
the $\mathcal{O}_a$ operators factorise completely, whereas for the mixed sectors (e.g.\ the gauge-Goldstone sector in a general Fermi gauge) the contributions from individual fields cannot, in principle, be disentangled in a non-constant background.
We discuss this case extensively in appendix~\ref{sec:GY}.
To obtain the effective action to one-loop order, we take the logarithm of both sides of eq.~\eqref{eq:exp-eff-action} (to simplify the notation we suppress the explicit zero value of the $\chi$ fields in the classical action)
\begin{equation}
\label{eq:Seff-general}
\Seff[\f]= S[\f] + \sum_a \ln\sqrt{\det \mathcal{O}_a(\f)}= S[\f]+ \frac{1}{2} \sum_a \tr \ln \mathcal{O}_a(\f)
\,.
\end{equation}

If we assume a constant background field $\fo$, the $\mathcal{O}_a$ operators reduce to inverse background-field-dependent propagators.
For example, the contribution from a scalar field (assuming it is a mass eigenstate) reads
\begin{equation}
\Seff[\fo]=S[\fo]+\Seff^{(1)}[\fo]
\,,
\end{equation}
where
\begin{equation}
\Seff^{(1)}[\fo]=-\ln \frac{1}{\sqrt{\det (-\partial^2 + V''(\fo))}}=\frac{1}{2}\tr\ln\left(-\partial^2+V''(\fo)\right)
\,,
\end{equation}
which is the textbook expression for the one-loop contribution to the effective potential~\cite{Coleman:1973}.

\subsection{Expanding the effective action}
\label{sec:expansion}

As the evaluation of the functional determinant from eq.~\eqref{eq:Seff-general} on
a spatially varying,
inhomogeneous background is technically challenging,
many studies resort to a derivative (gradient/momentum) expansion of the one-loop effective action.
In this approximation,
contributions from fields $\chi$
which are heavier than the mass scale of the transitioning scalar,
can be expanded in momentum space with respect to $k^2/m^2_\chi(\f)$, where
$k$ is a characteristic momentum of scalar modes, and
$m_\chi$ is the mass of the heavy mode.
In position space, this leads to an effective action with higher-order derivative operators,
suppressed by powers of the mass scale $m_\chi$. 
For the transitioning scalar field itself, we cannot apply the derivative expansion
but have to compute the fluctuation determinant exactly.
In practice, however, the contribution from the scalar field is often approximated using
dimensional analysis (e.g.\ as $T^3$ or $T^3 (S_3/(2\pi)^{3/2})$ for thermal phase transitions).

The effective action in the derivative expansion can be written as
\begin{equation}
\label{eq:effective-action-momentum}
  S_{\rm eff}[\f] = \int_{\vec{x}} \Bigl[
    V _{\rm eff}(\f)
    + \frac{1}{2} Z_2(\f) \partial_\mu \f\partial^\mu \f
    + \frac{1}{2} Z_4(\f) \partial^2 \f \partial^2 \f
    + \cdots
  \Bigr]
  \,,
\end{equation}
including potential higher-derivative contributions such as
$Z_4$ and where
the dotted terms denote other higher-derivative contributions.
Here, $\int_{\vec{x}} = \int {\rm d}^d \vec{x}$
for a general spatial dimension $d$,
and
$\f$, as before, denotes a general, possibly non-constant, background field.
To describe the free energy, the expression is truncated at the zeroth-order term corresponding to a constant background ---
the effective potential.
For non-constant background configurations, such as the critical bubble, often only the tree-level kinetic term is kept,
corresponding to $Z_2(\f) \rightarrow 1$ while other higher derivative-factors are omitted.
However, in certain cases, the higher-order terms become important.
For example, in~\cite{Kierkla:2023von} we have found that the terms including
the $Z_2$ factor are of the same order in power counting as the two-loop terms needed for
preserving renormalisation-scale invariance of the nucleation rate and that they contribute significantly to the action.
On the other hand,
we have found that the momentum expansion may not be fully reliable and thus including
the full functional determinant may be necessary.

Below, we approach the construction of the effective action from two different angles, either by matching the Green's functions with external light momenta to determine the Wilson coefficients of the expansion above (this corresponds to the commonly used diagrammatic approach to construct the effective action),
or by directly expanding the expression for the determinant, eq.~\eqref{eq:Seff-general}.%
\footnote{%
  A related, third approach to derive the coefficients of
  the higher-order derivative operators is described
  in~\cite{Chakrabortty:2024wto,KorthalsAltes:2017jzj,Chapman:1994vk}.
  Therein, the effective operators are found by expanding the covariant derivative.
  At finite $T$ this is often called the heat kernel and used for constructing the QCD thermal effective theory.
}
We will show that the two approaches are equivalent.
This also means that the results obtained
in~\cite{Kierkla:2023von} can be improved by directly computing
the functional determinants, without resorting to the momentum expansion.

As the following discussion is not restricted to thermal EFTs,
we will discuss the contributions to the one-loop effective action from some general heavy scalar field $X$.
In the context of the thermal phase transition,
which we consider in
section~\ref{sec:supercool}, this field would correspond to a temporal gauge mode at the soft scale.

\subsubsection*{Derivative expansion via diagrammatic matching}

In the diagrammatic approach to computing the effective action,
we first focus on determining the $Z_2$ factor.
To this end,
we calculate the corresponding two-point 
Green's functions with light external momenta,
integrating only over the loop-momenta of the $X$-field~\cite{Hirvonen:2022jba}.
Before performing the integration,
we can Taylor-expand the propagators with respect to the light momenta,
simplifying the computation~\cite{Ekstedt:2024etx,Hirvonen:2022jba}.
Since we restrict our analysis to one-loop order,
only heavy fields contribute within the loops,
while the light scalar fields appear exclusively on the external legs.
Indeed, for the matching the light field masses can be set to zero, and hence their one-loop contributions vanish in dimensional regularisation \cite{Braaten:1995cm}.

For simplicity, consider a generic theory involving
a scalar field $\phi$,
coupled to another scalar field $X$,%
\footnote{%
  This is the Abelian analogue of the temporal gauge mode $X_0$ in the model considered in section~\ref{sec:model}.
}
through a coupling constant $g^2$,
with the classical action given by
\begin{equation}
\label{eq:classical}
S = \int_{\vec{x}} \Bigl[
    \frac{1}{2} \partial_\mu \phi \partial^\mu \phi
  + \frac{1}{2} \partial_\mu X \partial^\mu X
  + \frac{1}{2} \mX^2 X^2
  + \frac 1 4 g^2 \phi^2 X^2
  \Bigr]
  \,.
\end{equation}
In this simplified exercise,
$\phi$ is the light scalar undergoing the phase transition, while
$X$ is the heavy field being integrated out.

The effective action, defined in momentum-expansion in eq.~\eqref{eq:effective-action-momentum},
is constructed such that its tree-level vertices reproduce quantum-corrected vertices of
the classical action at one-loop order.
The effective potential $V_{\rm eff}$ and
the kinetic prefactor $Z_2$
can be determined by matching the diagrams of the theories in
eqs.~\eqref{eq:classical} and~\eqref{eq:effective-action-momentum}.
To achieve this,
we expand the IR-modes,
 $\varphi$, of the field $\phi$, around
the non-dynamical, constant background
$\fo$,%
\footnote{%
  The expansion of the fields can also be understood as separating
  IR and UV modes,
  $\phi = \f_\rmii{IR} + \f_\rmii{UV}$,
  where now
  $\f_\rmii{IR} \equiv \f$ and
  $\f_\rmii{UV} \equiv \tilde\f$
  in eq.~\eqref{eq:fieldExpand}.
}
\begin{align}
\label{eq:fieldExpand}
  \phi &= \f + \tilde\f
  \,,&
  \f &= \fo + \fluc(x)
  \,,
\end{align}
and compute the diagrams with $\fo$ appearing only in the mass of $X$ and
$\fluc$ on the external lines.
In this context, we define
$M^2 =\mX^2 +  \frac{g^2 \fo^2}{2}$
as the background-field dependent mass of the $X$-field.
By expanding the effective action~\eqref{eq:effective-action-momentum}
around the constant IR background $\fo$,
we obtain
\begin{align}
\label{eq:Seff:derivative:expand}
  \Seff[\fo + \fluc] &=
  \int_{\vec{x}}
  \sum_{n=0}^{\infty} \frac{\fluc^n}{n!}\!
  \Bigl[
      \Veff^{\rmii{$(n)$}}(\fo)
    + \frac{1}{2} (\partial_\mu \fluc)^2 Z_{2}^{\rmii{$(n)$}}(\fo)
    + \mathcal{O}(\partial^4)
  \Bigr]
  \,,
\end{align}
which is again local in $\fluc$.
Individual derivatives are given by
$f^{\rmii{$(n)$}}(\fo) = \frac{\partial^n}{\partial\f^n} f^{ }(\f)\big|_{\fo}$ and
terms linear in $\fluc$ vanish
since $\int_{\vec{x}}\fluc = 0$
given that $\fluc$ by definition only
has modes with non-zero momentum $p\neq 0$.

To compute the value of $Z_2(\f)$,
we need to match the self-energy of
the fluctuations $\fluc$ computed using the classical action~\eqref{eq:classical} to
the terms proportional to $\fluc^2$ (including derivative terms) in
the effective action~\eqref{eq:Seff:derivative:expand}.
The corresponding two-point function up to one-loop level
depends on
the tree-level and
two further diagrammatic contributions
\begin{align}
\label{eq:Z2:2pt:function}
  \hspace{5mm}
   \Vtxttxt(\Lxx,\Lxx,{\vec{q}_1,\delta\phi}\,,\,\,{\vec{q}_2,\delta\phi})
  \qquad &+
  \TopoSTtxt(\Lxx,\Asai,,,X)
  +
  \TopoSB(\Lxx,\Asai,\Asai)
  \nn[2mm]
  &=
  \Pi_{\delta\phi\delta\phi}(\vec{q}_1,\vec{q}_2)
  =
      \frac{\partial^2}{\partial \f^2} V_{\rm eff}(\fo)
    - \vec{q}_1 \cdot \vec{q}_2\, Z_2^{ }(\fo)
    + \mathcal{O}(q^4)
  \,,
\end{align}
where we only focus on contributions up to $\mathcal{O}(\partial^2)$
in comparison with eq.~\eqref{eq:effective-action-momentum}.
Here, $\vec{q}_1$ and $\vec{q}_2$ are incoming momenta and
$\Veff$ is the momentum-independent part of the self-energy
which yields a contribution in the two-point function above.
Dashed lines indicate $\fluc$ fluctuations, and
solid lines indicate heavy $X$ fields.
The momentum-dependent part of the self-energy is entirely encoded in the second diagram
which, expanded to second order in external momentum, $q = \vec{q}_{1} = -\vec{q}_{2}$, yields
\begin{eqnarray}
  \TopoSB(\Lxx,\Asai,\Asai) &=&
	\mathcal{O}(q^0)
    - \frac{g^4\fo^2}{2}
    \int_{\vec{p}} \Bigl[
      - \frac{1}{(p^2 + M^2)^3}
      + \frac{4}{d}\frac{p^2}{(p^2 + M^2)^4}
    \Bigr]
    \,q^2
  + \mathcal{O}(q^4)
  \nn &\stackrel{\mathcal{O}(q^2)}{=}&
      \frac{g^4\fo^2}{2}
    \Bigl[
        \frac{d-4}{d} I^{d}_{3}(M^2)
      + \frac{4}{d} M^2 I^{d}_{4}(M^2)
    \Bigr]q^2
  \nn &=&
   \frac{g^4\fo^2}{2}
    \frac{(d-4)(d-2)}{24 M^4}
    I^{d}_{1}(M^2)
    \,q^2
  =
    \frac{1}{192\pi} \frac{g^4\fo^2}{M^3}
    \,q^2
  \,,
\end{eqnarray}
where
the $\mathcal{O}(q^0)$ term contributes to the effective potential and where
we use the integration-by-parts (IBP) relation
$I_{n+1}^d(m^2) = (2n -d)/(2n m^2) I_{n}^d(m^2)$
to reduce the result to the one-loop master integral
with expression
\begin{align}
\label{eq:1-loop-master}
I^d_{n}(m^2) &\equiv
\int_{\vec{p}} \frac{1}{(p+m^2)^n} =
  \Big( \frac{e^\gammaE \bmu^2}{4\pi} \Big)^{\epsilon}
  \frac{(m^2)^{\frac{d}{2}-n}}{(4\pi)^{\frac{d}{2}}}
  \frac{\Gamma(n-\frac{d}{2})}{\Gamma(n)}
  \;,
\end{align}
where
$\int_{\vec{p}} = \bigl( \frac{e^\gammaE \bmu^2}{4\pi} \bigr)^{\epsilon} \int \frac{{\rm d}^d \vec{p}}{(2\pi)^d}$,
$d\to d-2\epsilon$,
and
$\gammaE$ is the Euler-Mascheroni constant.

One can now identify the functional form of $Z_2$
by comparing the effective actions~\eqref{eq:effective-action-momentum} and \eqref{eq:Seff:derivative:expand}
\begin{align}
\label{eq:Z2:factor:comparison}
    \frac{1}{2} Z_{2}(\f) (\partial_\mu \f)^2 &=
    \frac{1}{2} Z_{2}(\fo) (\partial_\mu \fluc)^2
  + \frac{1}{2} Z_{2}'(\fo) \fluc (\partial_\mu \fluc)^2
  + \dots
  \;,
\end{align}
which indicates that one could have also retrieved its derivatives from
the higher $n$-point functions in the $\fluc$-fluctuations.
In practice, this means that
by reading off the momentum-dependent contributions,
we can extract $Z_2(\f) = Z_2(\fo)\bigr|_{\fo\to\f}$
\begin{align}
\label{eq:Z2:X:diagrammatic}
  Z_{2}(\f) &=
  1 +
    \frac{1}{192\pi} \frac{g^4\f^2}{M^3}
  \,,
\end{align}
where the tree-level contribution can be directly read off from the action and
the self-energy~\eqref{eq:Z2:2pt:function}.
The corresponding higher-derivative terms can be matched analogously.
Section~\ref{sec:higherorderderiv} and
appendix~\ref{sec:derivative:expansion:higher} provide a more detailed discussion
of this expansion for the model of central interest to this paper,
with the Lagrangian defined in eq.~\eqref{eq:model:V0}.

\subsubsection*{Derivative expansion of the determinant}
Alternatively,
the coefficients in the derivative expansion can be directly computed by expanding the fluctuation determinant in
derivatives~\cite{Fraser:1984zb, Baacke:1994qp}.
We demonstrate this here for the simple model of eq.~\eqref{eq:classical} using the method presented
in~\cite{Fraser:1984zb}.

The full determinant splits into separate contributions from $\phi$ and $X$. The contribution from the scalar field $X$ to the fluctuation determinant is given by
\begin{equation}
\label{eq:Xdet}
	S^{(1)}_{\rmii{eff},\rmii{$X$}} [\f]= \frac 1 2 \tr \, \ln\Bigl(p^2 +\mX^2 + \frac 1 2 g^2 \f(x)^2 \Bigr)
  \,,
\end{equation}
where $\mX$ is field independent.
As before, we work in Euclidean momentum space and expand the field about a constant background as
$\f=\fo + \fluc$, where $\fluc$ is an arbitrary fluctuation about a constant background;
cf.\ eq.~\eqref{eq:fieldExpand}.
To recover the $Z_2$ factor computed in the previous section,
we expand the momentum-expanded expression
for the effective action~\eqref{eq:effective-action-momentum} about a constant background,
such that
$S^{(1)}_{\rmii{eff},\rmii{$X$}}[\f] = S^{(1)}_{\rmii{eff},\rmii{$X$}}[\fo + \fluc]$,
as in eq.~\eqref{eq:Seff:derivative:expand}.
One can then
extract the functional form of the $Z_2$ factor from the terms quadratic in $\fluc$ and its derivatives,
disregarding the terms corresponding to derivatives of the effective potential.
Eventually, as in eq.~\eqref{eq:Z2:factor:comparison},
the factor $Z_{2}$ is obtained by replacing the constant background $\fo$ with
the general background $\f$.

Expanding about a constant background, the trace in eq.~\eqref{eq:Xdet} becomes:
\begin{align}
  S^{(1)}_{\rmii{eff},\rmii{$X$}}[\fo + \fluc] &=
    \frac 1 2 \tr \, {\rm ln} \Bigl(
      p^2 + \mX^2 +\frac 1 2 g^2 (\fo^2 + 2\fo\fluc + \fluc^2)
    \Bigr)
  \\[1mm] &=
  \frac 1 2 \tr \biggl[
      \ln\Bigl(p^2 + \mX^2 + \frac 1 2 g^2 \fo^2 \Bigr)
    + \ln\Bigl(1  + \frac {\frac 1 2 g^2}{p^2 +\mX^2 + \frac 1 2 g^2 \fo^2} ( 2\fo\fluc + \fluc^2) \Bigr)
  \biggr]
  \nn &\equiv
  \frac 1 2 \tr \biggl[
        \ln\bigl(p^2 + M^2 \bigr)
      + \ln\Bigl(1  + \frac{\frac 1 2 g^2}{p^2 +M^2}  ( 2\fo\fluc + \fluc^2)\Bigr)
    \biggr]
  \equiv
        V_{\rmii{eff},\rmii{$X$}}^{(1)}
      + S_{\rmii{eff},\rmii{$X$} \fluc}^{(1)}
    \,,
    \nonumber
\end{align}
where we used the additivity of the trace and as before defined $M$ as
the background-field-dependent mass of the $X$-field.
The first logarithm contains the $X$-induced contribution to the one-loop effective potential.
The second term $S_{\rmii{eff},\rmii{$X$}  \fluc}^{(1)}$
contains all the $\fluc$-dependent terms,
and after an expansion in $\fluc$,
we can extract its quadratic $\mathcal{O}(\fluc^2)$ term,
\begin{align}
  S_{\rmii{eff},\rmii{$X$} \fluc}^{(1)}
  &\stackrel{\mathcal{O}(\fluc^2)}{=}
  S_{\rmii{eff},\rmii{$X$} \fluc}^{(1,2)}
  =
    \frac{1}{2} \tr\, \biggl[
        \frac{ g^2}{ 2} \frac{ 1}{p^2 +M^2}\fluc^2
      - 2\fo^2 \Bigl(  \frac{ g^2}{ 2} \Bigr)^2 \frac{1 }{p^2 +M^2}  \fluc\frac{1}{p^2 +M^2}\fluc
    \biggr]
    \,.
\end{align}
Using the commutation relations
from appendix~\ref{sec:commutation-relations},
we can rewrite this as
\begin{align}
  S_{\rmii{eff},\rmii{$X$} \fluc}^{(1,2)} &= \frac{1}{2} \tr\, \biggl[
       \frac{ g^2}{ 2} \frac{\fluc^2}{p^2 +M^2}
    - 2\fo^2 \Bigl( \frac{ g^2}{ 2}  \Bigr)^2 \biggl\{
        \frac{1}{(p^2 +M^2)^2} \fluc^2
      + \frac{1}{(p^2 +M^2)^3}[p^2,\fluc]\fluc
      \\ &
      \hphantom{{}=\frac{1}{2} \tr\, \biggl[\frac{g^2}{4} \frac{\fluc^2}{p^2 +M^2}
    -  2\fo^2 \Bigl(\frac{g^2}{4} \Bigr)^2 \biggl\{
        \frac{1}{(p^2 +M^2)^2} \fluc^2}
      + \frac{1}{(p^2+M^2)^4}[p^2,[p^2, \fluc]]\fluc
    \biggr\}
  \biggr]
  \,.
  \nonumber
\end{align}
The first two  terms
combine to the second derivative of the effective potential;
see eq.~\eqref{eq:Seff:derivative:expand}.
Hence, we are only interested in the last two terms which are of the form
\begin{align}
  \tr\biggl[
    \frac{1}{(p^2 +M^2)^3}[p^2,\fluc]\fluc
  &+ \frac{1}{(p^2+M^2)^4}[p^2,[p^2, \fluc]]\fluc
  \biggr]
  \nn[2mm] &
  =
  \tr\biggl[
  \frac{M^2 +p^2(1-\frac{4}{d})}{(p^2 +M^2)^4}\fluc \Box \fluc
    + \mathcal{O}(\partial^4)
  \biggr]
  \,,
\end{align}
where we used eqs.~\eqref{eq:commutator2} and
\eqref{eq:commutator3}, the fact that
terms odd in $p^\mu$ vanish by the trace,
and $p^\mu p^\nu =  \frac{\eta^{\mu\nu}}{d} p^2$. The $\mathcal{O}(\partial^4)$
denotes higher derivative contributions.
Using eq.~\eqref{eq:1-loop-master},
and the relations above it,
we can reduce the result to
\begin{eqnarray}
  -\frac{1}{2}\tr\biggl[
	\frac{\fo^2 g^4}{2} \frac{M^2+p^2(1-\frac 4 d)}{(p^2+M^2)^4} \fluc\Box\fluc
  \biggr]
  &=&
  -\frac{\fo^2g^4}{12} \frac{(d-4)(d-2)}{8 M^4}\int_{\vec{x}}I_1^d(M^2)\fluc\Box\fluc
  \nn &\stackrel{d=3}{=}&
   \frac{\fo^2g^4}{12}\frac{1}{32\pi}\frac{1}{M^3}\partial_\mu \fluc \partial^\mu \fluc
\,.
\end{eqnarray}
In the last step,
we assumed $d=3$ and integrated by parts in position space.
We can now read off the $Z_2$ factor for a general $\f$ as
$Z_2(\f) = Z_2(\fo)\big|_{\fo\to\f}$
\begin{equation}
\label{eq:Z2:X:expand}
  Z_2(\varphi)=
  1+
    \frac{1}{192\pi}\frac{g^4 \f^2}{M^3}
  \,,
\end{equation}
where the tree-level contribution can be directly read off from the action.
The result above agrees with the diagrammatic derivation in eq.~\eqref{eq:Z2:X:diagrammatic}.

This extended example serves to clarify the meaning of the derivative expansion of the action. First of all, it shows that if the first term with non-vanishing momenta casts doubt on the validity of the expansion, one can consider higher-order terms to check convergence.
Moreover, by computing the full functional determinant one can become independent of the momentum expansions and include the full one-loop correction to the effective action.
We will explore both of these routes in the computation of the nucleation rate for the scale-invariant model of interest in section~\ref{sec:supercool}.
Before that, in the following section, we review various scale hierarchies and expansions that need to be carefully considered in this computation.

\section{Overview of perturbative expansions}
\label{sec:overview-expansions}

Before proceeding to the computation of the nucleation rate, we summarise here the various expansions that appear in the perturbative treatment of the thermal system we consider, which exhibits multiple different scales.
These scale hierarchies form the backbone of the effective description we utilise.
The relevant expansions are:
\begin{itemize}

\item[ \daisy]
  {\em High-temperature expansion.}
  \\
  This expansion
  allows us to integrate out all but the zero Matsubara modes
  and is the cornerstone of dimensional reduction.
  It allows us to construct a 3D EFT related to the full theory by matching. In models with classical scale invariance, the EFT description for the nucleation is constructed at relatively small field values up to an escape point (of the semi-classical approximation of the bounce). At such small field values, the HT expansion is reliable~\cite{Kierkla:2023von}.

\item[ \daisy]
  {\em Coupling expansion for the fluctuations of the gauge fields.}
  \\
  These include both thermally (Debye) screened temporal gauge field components, as well as purely spatial gauge modes that are massless in the high-temperature phase and massive in the low-temperature phase, inside the bubble, due to the Higgs mechanism. Inside the bubble, the masses for these gauge field fluctuations are significantly larger than the effective thermal mass of the transitioning scalar field. Occasionally, we adopt the terminology of~\cite{Gould:2023ovu} and refer to these modes as \textit{soft}. It is the treatment of these soft modes that requires a lot of care, since their masses vary across the bubble in-between other scales. In particular, in the following sections we shall pay attention to the question whether the one-loop contribution to the nucleation rate from these modes admits a reliable derivative expansion or not. We also emphasise, that the renormalisation group running of the 3D EFT is dominantly governed by the soft modes, and since such running arises due to ultraviolet divergences first appearing at two-loop level, their accurate treatment necessitates a two-loop computation, to obtain physically meaningful, RG-scale-invariant predictions. 

\item[ \daisy]  
  {\em Coupling expansion for the fluctuations of the transitioning scalar field.}
  \\
  The mass scale of these fluctuations is the deepest perturbative scale in the infrared, and describes the lightest (still perturbative) modes above the ultrasoft, magnetostatic scale where gauge-field modes are confining.
  As we show, a perturbative treatment for this scale is relatively straightforward, since already the one-loop fluctuation determinant for these modes is fairly well suppressed, and the RG running related to this scale appears formally at the same order as the fully non-perturbative, ultrasoft effects~\cite{Ekstedt:2024etx}. 

\item[ \daisy] 
  {\em Strict expansion of the effective action around the leading order bounce solution.}
  \\
  In this expansion, a starting point is the semi-classical picture within the EFT: the leading order nucleation rate is the classical rate (for the highly occupied bosonic field modes) described by the partition function.
  Perturbative quantum corrections to the classical rate are then computed in a saddle point approximation of the 3D EFT path integral leading to the effective action, around classical field configuration for the critical bubble, the bounce solution. These corrections are computed in accord with the aforementioned coupling expansion for the gauge field modes.

\end{itemize}

We already emphasise, that it proves useful to keep these expansions separate at many occasions instead of blending them together.
However, extra care needs to be taken when
the separation of different mass scales becomes compromised.
In particular, the derivative expansion for the contribution of gauge-field modes, which is an expansion in the  momentum scale of the light transitioning scalar versus gauge field masses, becomes unreliable when gauge fields become light enough far away from the bubble centre.

\section{Nucleation rate for supercooled phase transitions}
\label{sec:supercool}

The main aim of this work is to establish a reliable procedure for computing (the statistical part of)
the bubble nucleation rate for supercooled phase transitions.
We will build upon the results of~\cite{Kierkla:2023von}, where
the nucleation EFT for a model featuring supercooling was constructed.
We will evaluate the functional determinants
contributing to
the nucleation rate
 and compare with the results obtained previously using the derivative expansion.

\subsection{The model}
\label{sec:model}

We will study the nucleation rate in the so-called
SU(2)cSM model~\cite{Hambye:2013, Carone:2013} which is a relatively minimal extension of
the SM featuring classical scale invariance.
It comprises of the scaleless SM and a new
SU(2)$_{\rmii{$X$}}$-symmetric dark sector featuring
a scalar doublet of the new gauged SU(2)$_{\rmii{$X$}}$.
The two sectors communicate via a Higgs portal interaction which also transmits the information about
radiative symmetry breaking occurring in the new sector, to the SM.
Through the Higgs portal coupling an effective mass term for the Higgs field is generated and the correct pattern of electroweak symmetry breaking is achieved. The model has been extensively treated in the literature~\cite{Hambye:2013, Carone:2013, Khoze:2014, Pelaggi:2014wba, Karam:2015, Plascencia:2016, Chataignier:2018, Hambye:2018, Baldes:2018, Prokopec:2018, Marfatia:2020, Kierkla:2022odc, Kierkla:2023von}. 
The present work builds upon the analysis of the allowed parameter space and properties of
the model presented
in~\cite{Kierkla:2022odc, Kierkla:2023von}.

The tree-level potential for the two background scalar fields is given by
\begin{equation}
\label{eq:model:V0}
	V^{(0)}(h,\varphi) =
    \frac{1}{4} \Bigl(
        \lambda_h h^4
      + \lambda_{h\varphi} h^2 \varphi^2
      + \lambda_{\varphi} \varphi^4
  \Bigr)
  \,,
\end{equation}
where $h$ denotes the background for the SM Higgs field, and $\varphi$ the new doublet.
We have already exploited the global
${\rm SU}(2) \times {\rm SU}(2)_{\rmii{$X$}}$ invariance
to align the background fields along the radial components of the full fields.

To reliably study the phase transition we need to take into account quantum and thermal corrections to the classical potential.
We will do this by constructing an HT EFT following~\cite{Kierkla:2023von},
which is reviewed in the following subsection.

\subsection{Nucleation EFT for supercooled phase transition in the derivative expansion}
\label{sec:recap}

In~\cite{Kierkla:2023von} we have constructed the three-dimensional nucleation EFT for
the SU(2)cSM. 
It has been shown that in the SU(2)cSM model the nucleation takes place along the direction of the new field~\cite{Prokopec:2018, Kierkla:2022odc}, and after the field exits the barrier close to the origin of the field space, the scalar fields roll towards the global minimum of the potential. Therefore, here we focus on the nucleation of the new scalar field.

We construct the effective theory by integrating out the hard modes of the bosonic and fermionic fields. The resulting theory is a 3D theory of the scalar and gauge Matsubara zero modes.
According to the power counting discussed in~\cite{Kierkla:2023von}, one-loop corrections of the soft gauge modes contribute at the same order to the effective potential as the tree-level scalar terms.
Therefore, they should be included in the tree-level effective potential of the nucleation EFT.
The leading order effective potential in terms of the dark scalar background field $v_3$ is given by
\begin{align}
\label{eq:Veff-EFT-LO}
V_3^{\rmii{EFT,\,LO}} &=
    \frac 1 2 m_3^2 v_3^2
    + \frac 1 4 \lambda_3^{ } v_3^4
    - \frac{1}{12\pi} \Bigl(
      6 (m^2_{\rmii{$X$},3})^{\frac{3}{2}}
    + 3 (m^2_{\rmii{$X_{0}$},3})^{\frac{3}{2}}
  \Bigr) 
  + \frac{1}{12\pi} \Big( 3 (\mDX^2)^{\frac{3}{2}} \Big)
  \,.
\end{align}
The mass term $m_3^2$ as well as the Debye mass for the temporal gauge component $\mDX^2$ are thermal masses generated by the hard thermal scale. 
The terms cubic in the fields arise by integrating out soft gauge-field modes, with
\begin{align}
  \label{eq:massGauge}
  m_{\rmii{$X$},3}^2 &= \frac 1 4 g_{\rmii{$X$},3}^2 v_3^2
  \,,&
  m_{\rmii{$X_0$},3}^2 &= \mDX^2 + \frac 1 2 h_3^{ } v_3^2
  \,.
\end{align}
The last term in eq.~\eqref{eq:Veff-EFT-LO} ensures that the potential vanishes for zero field value.
We do not include higher-order, marginal operators  in the leading order potential.
The validity of the high-temperature expansion requires that these higher-order operators are small, and we confirmed this explicitly in~\cite{Kierkla:2023von}.
The parameters of the EFT, as functions of temperature and parameters of the full theory, are given in
appendix B.2 of~\cite{Kierkla:2023von}.

We use the leading order action
\begin{equation}\label{eq:ActionLO}
  S_{\rm 3}^{\rmii{EFT,LO}}[v_3] = 4\pi\int {\dd} {\boldmath r}\ {\boldmath r}^2\left[
      \frac{1}{2}(\partial_i v_{3})^2
    + V_3^{\rmii{EFT,LO}}(v_{3})
  \right]
  \,, 
\end{equation}
to find the spherically symmetric bounce solution $v_{3,b}(r)$ with radial coordinate $r$,
i.e.\
\begin{align}
\frac{\delta}{\delta v_3} S_{\rm 3}^{\rmii{EFT,LO}}[v_{3,b}] = 0
\,.
\end{align}

In the derivative expansion,
the NLO contribution to the action from soft gauge-field modes reads
\begin{align}
\label{eq:SbounceNLO}
  S_{\rm 3}^{\rmii{EFT,NLO}}[v_{3,b}] &=
    4\pi\int {\dd} {\boldmath r}\ {\boldmath r}^2\left[
      \frac{1}{2} Z^{\rmii{NLO}}_3(v_{3,b})(\partial_i v_{3,b})^2
    + V_3^{\rmii{EFT,NLO}}(v_{3,b})
  \right]
  \,, 
\end{align}
where the integration is performed over the LO bounce configuration $v_{3,b}(r)$.%
\footnote{%
  We apply the strict expansion around the leading order bounce solution,
  as described in section~\ref{sec:overview-expansions}.
}
Diagrammatically, $V_3^{\rmii{EFT,NLO}}$ composes of two-loop diagrams with soft fields.
We point out, that this NLO expression is suppressed compared to the LO action by $\sim \gX/\pi$,
i.e.\ by the effective expansion parameter of the soft scale,
and the derivative expansion at NLO generates an effective
kinetic operator proportional to $Z^{\rmii{NLO}}_3$.

Concretely, the expressions for
$V_3^{\rmii{EFT,NLO}}(v_{3})$ and
$Z^{\rmii{NLO}}_3(v_{3})$ read%
\footnote{%
  Eq.~\eqref{eq:SbounceNLO} corrects a typo contained in
  eq.~(4.14) of~\cite{Kierkla:2023von} in
  the term $\propto g_{\rmii{$X$},3}^2 v_3^2$ in the first line.
}
\begin{align}
\label{eq:VeftNLO}
V_3^{\rmii{EFT,NLO}} &=
    \frac{1}{(4\pi)^2}\frac{3}{64} g^2_{\rmii{$X$},3} \Bigl\{
          8 m^2_{\rmii{$X$},3} \bigl(5-21\ln3 \bigr) 
        - 3 g^2_{\rmii{$X$},3} v^2_{3} \bigl(1-14\ln2 \bigr)
      \nn &
      \hphantom{{}\frac{1}{(4\pi)^2}\frac{3}{64} g^2_{\rmii{$X$},3} \Bigl(}
      + 2 \bigl(80 m^2_{\rmii{$X$},3} - 3 g^2_{\rmii{$X$},3} v^2_{3}\bigr)
        \ln\frac{\Lamd}{2 m_{\rmii{$X$},3}}
      \Bigr\}
      \nn[1mm] &
  + \frac{1}{(4\pi)^2} \Bigl\{
    \frac{3}{4} g^2_{\rmii{$X$},3} \bigl(
        6 m^2_{\rmii{$X_0$},3}
      + 4 m_{\rmii{$X$},3} m_{\rmii{$X_0$},3}
    \bigr) 
    + \frac{15}{4}  \kappa_3^{ } m^2_{\rmii{$X_0$},3}
    - \frac{3}{8} h^2_{3} v^2_3 \Bigl( 1 + 2 \ln\frac{\Lamd}{2 m_{\rmii{$X_0$},3}}\Bigr)
    \nn[1mm] &
    \hphantom{{}\frac{1}{(4\pi)^2}\biggl(}
    - \frac{3}{2} g^2_{\rmii{$X$},3} \bigl(m^2_{\rmii{$X$},3} - 4 m^2_{\rmii{$X_0$},3}\bigr)
      \ln\frac{\Lamd}{2 m_{\rmii{$X_0$},3} + m_{\rmii{$X$},3} }
  \Bigr\} 
  \nn &
  - \frac{1}{(4\pi)^2} \Bigl\{
      \frac{15}{4} \mDX^2 \kappa_3^{ }
      + \frac{3}{2} g_{\rmii{$X$},3}^2 \mDX^2\Bigl(3 + 4 \ln\frac{\Lamd}{2 \mDX}\Bigr)
  \Bigr\}
  \,,
  \\[2mm]
\label{eq:Z3}
  Z^{\rmii{NLO}}_3(v_{3}) &=
    - \frac{11}{16\pi}\frac{g_{\rmii{$X$},3}}{v_3}
    + \frac{1}{64\pi}\frac{h_3^2 v_3^2}{m_{\rmii{$X_0$},3}^3}
    \,,
\end{align}
where
$\Lamd$ is the renormalisation scale of the soft EFT,
and the spatial and temporal gauge mode masses $m_{\rmii{$X$},3}$ and $m_{\rmii{$X_0$},3}$ are given
in eq.~(\ref{eq:massGauge}).
These expressions were computed in~\cite{Kierkla:2023von} following~\cite{Gould:2023ovu} and
can readily be obtained with
{\tt DRalgo}~\cite{Ekstedt:2022bff, Fonseca:2020vke}
version {\tt v1.1} and higher.%
\footnote{%
  Starting from {\tt DRalgo} version {\tt v1.1},
  one can employ Higgs Effective Field Theory (HEFT) functionalities of
  {\tt HEFT.m} such as
  {\tt PrepareHET[]},
  {\tt CalculatePotentialHET[]}, and
  {\tt PrintActionHET[]}
  to integrate out fields that get
  large field-dependent masses in the Higgs phase.
}
In this work we have used {\tt v1.2}.

Soft-scale corrections beyond NLO arise from
terms incorporating soft corrections to the bounce configuration $v_{3,b}$,%
\footnote{%
  \label{ft:dumbbell}%
  Diagrammatically, corrections to the bounce appear as
  one-particle reducible
  {\em dumbbell} diagrams ($\dumbbell$) where the reducible line is
  the light scalar propagator and blobs are two-loop gauge field sunset diagrams~\cite{Ekstedt:2022tqk}.
}
higher-order derivative operators as well as the three-loop effective potential~\cite{Ekstedt:2024etx}.
All these contributions are suppressed as $\mathcal{O}(\gX^2/\pi^2)$ compared to LO, so they are NNLO in terms of the soft expansion, and hence we do not aim to fully incorporate them in this article.
However, to evaluate the accuracy of the derivative expansion,
in section~\ref{sec:higherorderderiv},
we inspect higher-order derivative operators.
The NLO EFT action of eqs.~\eqref{eq:ActionLO} and~\eqref{eq:SbounceNLO} was used in~\cite{Kierkla:2023von} to compute the thermodynamic parameters of the phase transition. The effect of the NLO contributions was found to be substantial, in particular, the contribution to the action from the kinetic term at NLO was large. The study of higher orders in the derivative expansion will allow us to evaluate the reliability of this result.

\subsection{Higher orders in the derivative expansion}
\label{sec:higherorderderiv}

We examine the derivative operators that emerge at
the next order in the derivative expansion at the soft scale.
The action expanded up to fourth order in gradients,
reads~\cite{Fraser:1984zb,Andreassen:2016cvx}:
\begin{align}
\label{eq:S3:higher}
S_{\rm 3}^{\rmii{EFT}}[v_{3,b}]  \supset
 \int_{\vec{x}}
 \Bigl[
      V^\rmii{EFT}(v_3)&
     + \frac{1}{2} Z_{2,3} (\partial_i v_3)^2
     + \frac{1}{2} Z_{4,3} (\partial^2 v_3)^2
     \nn* &
   + \frac{1}{2} Y_{3,3} (\partial_i v_3)^2 \partial^2 v_3
   + \frac{1}{8} Y_{4,3} (\partial_i v_3)^2(\partial_j v_3)^2
   + \mathcal{O}(\partial^6)
 \Bigr]
 \,,
\end{align}
where
$\int_{\vec{x}}\stackrel{d=3-2\epsilon}{=} 4\pi\int {\rm d}r\,r^2$,
$V^\rmii{EFT}(v_3)$ is the zero-momentum part of the effective action
 and
we omitted one-particle reducible diagrams (cf.\ footnote~\ref{ft:dumbbell}).
Having used spatial rotation and translation symmetry and integration by parts,
the operator basis in eq.~\eqref{eq:S3:higher}
is unique, and
depends only on the five independent functions,
$V^\rmii{EFT}(v_3)$,
$Z_{2,3}$,
$Z_{4,3}$,
$Y_{3,3}$, and
$Y_{4,3}$.
The concrete vector- and temporal-scalar-induced contributions to
the corresponding coefficients are
collected in appendix~\ref{sec:derivative:expansion:higher}.
Here, we report them in $d=3$ dimensions, and note that all $Z$- and $Y$-functions are UV finite.

At $\mathcal{O}(\partial^2)$,
the first unique higher-order contribution arises
from the one- and two-loop correction to the scalar two-point function and amounts to
\begin{align}
\label{eq:Zfactor}
Z_{2,3}(v_3) &=
  1
    - \frac{11}{16\pi}\frac{g_{\rmii{$X$},3}}{v_{3}}
    + \frac{1}{64\pi}\frac{h_3^2 v_{3}^{2}}{m_{\rmii{$X_0$},3}^3}
 \nn &
  + \frac{1}{(4\pi)^2}\biggl[
        \frac{731 - 108\ln2 - 216\ln3}{72}\frac{g_{\rmii{$X$},3}^2}{v_{3}^2}
      + \frac{1}{3} \frac{2g_{\rmii{$X$},3}^{2} v_{3}^{2} + 21 g_{\rmii{$X$},3}^{ } v_{3}^{ }m_{\rmii{$X_0$},3}^{ } + 44m_{\rmii{$X_0$},3}^{2}} {(g_{\rmii{$X$},3}^{ } v_{3}^{ }+ 4 m_{\rmii{$X_0$},3}^{ })^3}
      \frac{g_{\rmii{$X$},3}^{3}}{v_{3}^{ }}
 \nn[1mm] &
 \hphantom{{}+\frac{1}{(4\pi)^2}\biggl[}
    + h_{3}^{ }\biggl(
        24\frac{m_{\rmii{$X_0$},3}^{ }}{g_{\rmii{$X$},3}^{ } v_{3}^{3}}
      - \frac{1}{4}\frac{g_{\rmii{$X$},3}^{3} v_{3}^{ }}{m_{\rmii{$X_0$},3}^{3}}
      + \frac{1}{3}\frac{g_{\rmii{$X$},3}^{3} v_{3}^{ }}{m_{\rmii{$X_0$},3}^{ }}
      \frac{5 g_{\rmii{$X$},3}^{ } v_{3}^{ }+ 28 m_{\rmii{$X_0$},3}^{ }}{(g_{\rmii{$X$},3}^{ } v_{3}^{ }+ 4 m_{\rmii{$X_0$},3}^{ })^3}
      \biggr)
 \nn[1mm] &
 \hphantom{{}+\frac{1}{(4\pi)^2}\biggl[}
    + h_{3}^{2}\biggl(
        \frac{5}{64}\frac{\kappa_{3}^{ } v_{3}^{2}}{m_{\rmii{$X_0$},3}^4}
        + \frac{49}{16}\frac{1}{m_{\rmii{$X_0$},3}^{2}}
        + \frac{1}{3}\frac{g_{\rmii{$X$},3}^{ } v_{3}^{ }}{m_{\rmii{$X_0$},3}^{3}}
        + \frac{3}{16}\frac{g_{\rmii{$X$},3}^{3} v_{3}^{3}}{m_{\rmii{$X_0$},3}^{5}}
  \nn &
  \hphantom{{}+\frac{1}{(4\pi)^2}\biggl[+h_{3}^{2}\biggl(}
        + \frac{128}{3}\frac{m_{\rmii{$X_0$},3}^{ }}{(g_{\rmii{$X$},3}^{ } v_{3}^{ }+ 4 m_{\rmii{$X_0$},3}^{ })^3}
        - \frac{16}{3}\frac{1}{m_{\rmii{$X_0$},3}^{ }}
          \frac{5 g_{\rmii{$X$},3}^{ } v_{3}^{ }+ 11 m_{\rmii{$X_0$},3}^{ }}{(g_{\rmii{$X$},3}^{ } v_{3}^{ }+ 4 m_{\rmii{$X_0$},3}^{ })^2}
      \biggr)
 \nn[1mm] &
  \hphantom{{}+\frac{1}{(4\pi)^2}\biggl[}
    - \frac{11}{48}\frac{h_{3}^{3} v_{3}^{2}}{m_{\rmii{$X_0$},3}^4}
    + \frac{23}{192}\frac{h_{3}^{4} v_{3}^{4}}{m_{\rmii{$X_0$},3}^6}
    \biggr]
    + \mathcal{O}(\gX^3)
  \,,
\end{align}
where
the $\mathcal{O}(\gX^{ })$ terms are the NLO and
the $\mathcal{O}(\gX^{2})$ terms the NNLO contribution.
The one- and two-loop $d$-dimensional vector-induced result is given
in eq.~\eqref{eq:Z23:d:2loop}.
At $\mathcal{O}(\partial^4)$,
the contributions to the action~\eqref{eq:S3:higher}
are given by the one-loop corrections to
the scalar two-, three- and four-point function, respectively.
Up to $\mathcal{O}(\gX^{-1})$ they read
\begin{align}
\label{eq:higher-Z}
  Z_{4,3}(v_3) &=
    - \frac{23}{80\pi}\frac{1}{g_{\rmii{$X$},3} v_{3}^3}
    - \frac{3}{1280\pi}\frac{h_3^2 v_{3}^{2}}{m_{\rmii{$X_0$},3}^5}
  \,,
  \\
  Y_{3,3}(v_3) &=
    + \frac{359}{80\pi}\frac{1}{g_{\rmii{$X$},3} v_{3}^4}
    - \frac{3}{640\pi}\frac{h_3^2 v_{3}^{ }}{m_{\rmii{$X_0$},3}^5}
    + \frac{1}{256\pi}\frac{h_3^3 v_{3}^{3}}{m_{\rmii{$X_0$},3}^7}
  \,,
  \\
  Y_{4,3}(v_3) &=
    - \frac{127}{80\pi}\frac{1}{g_{\rmii{$X$},3} v_{3}^5}
    - \frac{3}{320\pi}\frac{h_3^2}{m_{\rmii{$X_0$},3}^5}
    + \frac{1}{64\pi}\frac{h_3^3 v_{3}^{2}}{m_{\rmii{$X_0$},3}^7}
    - \frac{7}{1024\pi}\frac{h_3^4 v_{3}^{4}}{m_{\rmii{$X_0$},3}^9}
  \,.
\end{align}
Their one-loop $d$-dimensional vector- and temporal-scalar-induced results are given in
eqs.~\eqref{eq:Z43:d:1loop}--\eqref{eq:Y43:d:1loop}.

To investigate the behaviour of the derivative expansion at the large-$r$ tail of the bounce,
where scale-shifting masses become small,
akin to~\cite{Lofgren:2021ogg, Hirvonen:2021zej},
we proceed by using the asymptotic behaviour of the bounce solution
\begin{align}\label{eq:v3b_tail}
  \Box v_{3,b} &\sim \mu^2 v_{3,b}
  \,,&
  v_{3,b}(\infty) &= 0
  \,,&
  \implies
     &&
  v_{3,b}(r) &\sim c \frac{e^{-\mu r}}{r}
  \quad
  \text{as}\;
  r \to \infty
  \,,
\end{align}
where $c$ is an undetermined constant, and
$\mu$ the characteristic mass scale of nucleation which is set by the scalar mass $m_3$.
To study the possibly problematic contributions above
we divide the region of radial integration into two domains
(i) $r\leq R$ and
(ii) $r > R$, with $R$ being larger than the characteristic size of the bounce,
i.e.\ $R > \mu^{-1}$. 
The corresponding terms for the pure vector contributions read
\begin{align}
  \label{eq:Z2:int}
  \int_{\vec{x}} \frac{(\partial_i v_{3,b})^2}{v_{3,b}^{ }}
  &\approx
    [\text{contribution from}\,r\leq R]
    - 4\pi c \mu^2 \int_{r\geq R}\!{\rm d}r\, e^{-\mu r}
      \Bigl[r + \dots\Bigr]
  \,,\\
  \int_{\vec{x}} \frac{(\partial^2 v_{3,b})^2}{v_{3,b}^{3}}
  &\approx
    [\text{contribution from}\,r\leq R]
    - \frac{4\pi}{c} \mu^4 \int_{r\geq R}\!{\rm d}r\, e^{\mu r}
      \Bigl[r^3 + \dots\Bigr]
  \,,\\
  \int_{\vec{x}} \frac{(\partial_i \phib)^2 \partial^2 v_{3,b}}{v_{3,b}^{4}}
  &\approx
    [\text{contribution from}\,r\leq R]
    - \frac{4\pi}{c} \mu^4 \int_{r\geq R}\!{\rm d}r\, e^{\mu r}
      \Bigl[r^3 + \dots\Bigr]
  \,,\\
  \int_{\vec{x}} \frac{(\partial_i \phib)^2(\partial_j v_{3,b})^2}{v_{3,b}^{5}}
  &\approx
    [\text{contribution from}\,r\leq R]
    - \frac{4\pi}{c} \mu^4 \int_{r\geq R}\!{\rm d}r\, e^{\mu r}
      \Bigl[r^3 + \dots\Bigr]
  \,,
\end{align}
where only the ``most IR divergent'' term of the integration is kept.
When integrated over the volume using the LO bounce to obtain their contribution to the action,
all operators except the first,
diverge for $r\geq R$.
In turn, all operators of $\mathcal{O}(\partial^4)$ contribute to the bounce action by the same order in $r$
which suggests that these contributions can be resummed by abandoning the derivative expansion.
Additionally,
the two-loop contributions to $Z_{2,3}$ enter the effective action through integrals of
\begin{align}
  \label{eq:Z2:2loop:int}
  \int_{\vec{x}} \frac{(\partial_i v_{3,b})^2}{v_{3,b}^{2}}
  &\approx
    [\text{contribution from}\,r\leq R]
    - 4\pi \mu^2 \int_{r\geq R}\!{\rm d}r\,
      \Bigl[r^2 + \dots\Bigr]
  \,,
\end{align}
which also diverge at the tail but now already at $\mathcal{O}(\partial^2)$.
The derivative expansion therefore breaks down earlier at higher-loop levels.
Conversely, for contributions involving temporal scalars,
the corresponding scalar mass, $m_{\rmii{$X_0$},3}$, regulates the large-$r$ integration.

For pure vector contributions,
the derivative expansion
already becomes unreliable at NLO where
$S_{3}^\rmii{EFT,NLO} \sim \gX^{-1/2}$~\cite{Kierkla:2023von}.
While the NLO contribution remains finite at the tail of the bounce,
the breakdown of the derivative expansion manifests clearly at
N$^3$LO which is the next vector-induced order where
$S_{3}^\rmii{EFT,N$^3$LO} \sim \gX^{1/2}$.
This order corresponds to
two-loop corrections to the kinetic term and
one-loop corrections to the fourth-derivative terms in the effective action.
Since the derivative expansion is an expansion in $k^2/m^2(\f)$,
its failure as $m$ approaches zero on the bounce tail
was anticipated.
By explicitly demonstrating that terms with higher powers of momenta diverge at
the tail of the bounce,
we confirm that the derivative expansion is completely unreliable beyond NLO.
Instead of attempting to salvage the derivative expansion through technical modifications,
we resum all seemingly divergent contributions by directly computing
the exact fluctuation determinants.

The vector fluctuation determinant
is the leading vector-induced contribution to the action.
Going to the next order in vector contributions requires also
going beyond
the two-loop potential
$V_{3}^{\rmii{EFT,NLO}}$
derived from the derivative expansion~\cite{Ekstedt:2021kyx}
which introduces a finite non-negligible error in the action.
We refrain from including
higher-order loop corrections without derivative expansion or
one-particle reducible, dumbbell-like (cf.~footnote~\ref{ft:dumbbell}),
contributions and
relegate their study to future work.

\subsection{Nucleation rate using functional determinants}
\label{sec:NLO-with-dets}

\begin{figure}[t]
	\centering 
	\includegraphics[width=0.7\textwidth]{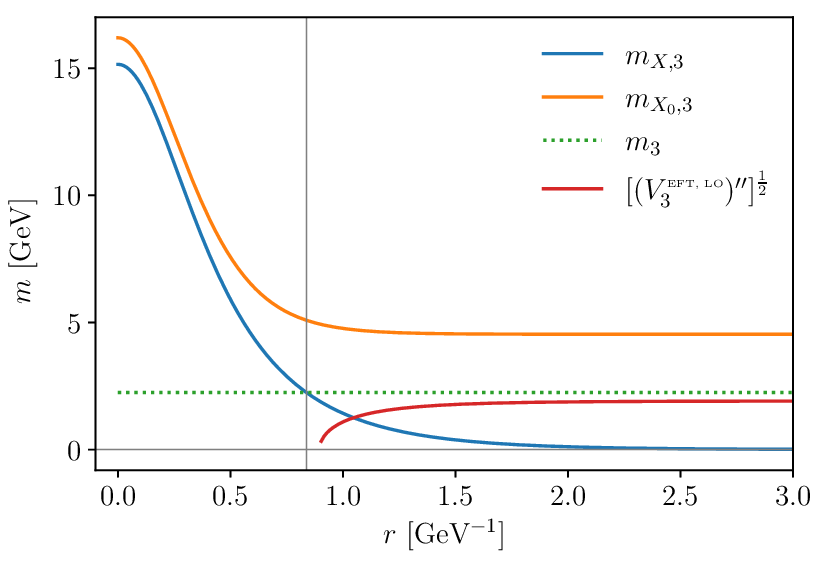}
	\caption{%
  Field-dependent masses evaluated on the bounce solution for
  a benchmark point with $\gX = 0.8$ and $\MX = 10^4$~GeV.
  Here,
  $m_{\rmii{$X$},3}$,
  $m_{\rmii{$X_0$},3}$
  corresponds to the spatial and temporal gauge modes, respectively, $m_3$ is the scalar field bare mass. The prime denotes derivatives with respect to the field $v_3$.  The vertical grey line denotes the tail of bounce solution where the gauge modes become lighter than nucleating field. 
	}
	\label{fig:scale-shifters}
\end{figure}

In principle, to construct the nucleation EFT we should integrate out all the degrees of freedom but the nucleating field~\cite{Gould:2021ccf}. We have presented the nucleation EFT for the SU(2)cSM model in section~\ref{sec:recap}. In such a construction, only the nucleating field could contribute to the prefactor of the nucleation rate (see eq.~\eqref{eq:rateBD})
containing the functional determinants, as the only dynamical field in the theory. However, in theories with gauge bosons, it is not possible to integrate out the soft gauge fields such that the resulting EFT would be valid along the entire bounce solution since at the tail of the bounce (where the scalar field goes to zero) the masses of the transverse (spatial) gauge modes tend to zero, invalidating the mass hierarchy at the foundation of the EFT. Such behaviour was referred to as scale-shifting in~\cite{Gould:2021ccf} and it is illustrated for the model considered in this work in
figure~\ref{fig:scale-shifters}.

Often, it has been claimed that scale-shifting masses do not source a large error as the contribution from the tail of the bounce to the action is suppressed~\cite{Ekstedt:2021kyx,Lofgren:2021ogg,Hirvonen:2021zej}. 
This is true for the contributions from the effective potential, as we further discuss in section~\ref{sec:RG}, but not necessarily for the finite-momentum part of the effective action.%
\footnote{%
  Compare with the results of~\cite{Moore:2000jw},
  where the corrections from the $Z$-factor were found to be well-behaved.
}
This issue is more pronounced in models with classical scale symmetry as there the scalar field has no negative mass term that would take it to the so-called supersoft scale,
below the scale of soft gauge modes~\cite{Kierkla:2023von}.

A remedy for this issue, suggested in~\cite{Gould:2021ccf}, is to compute the bounce in the leading-order approximation with the gauge modes integrated out and then evaluate the NLO nucleation rate on the LO bounce solution, including fluctuation determinants from the gauge modes. In this way, the gauge bosons contribute both to the barrier (as they should due to their large mass in the bulk of the bounce), and to the functional determinants (due to their lightness at the tail of the bounce). With this procedure, we capture the volatile nature of the spatial gauge fields. At the same time, by computing their full one-loop contributions via functional determinants we avoid using the derivative expansion, which we have already demonstrated to break down at the tail of the bounce.

We will now define the full one-loop nucleation rate which incorporates the treatment of scale-shifters explained above that we will use in section~\ref{sec:results} and compare to the derivative expansion.
Formally, we compute the (statistical part of the) nucleation rate as
\begin{align}
\label{eq:rateFactors}
A_{\rm stat} &=
\det_{\rmii{$S$}} \times
\det_{\rmii{$V$}} \times
e^{-S_{\rm 3}^{\rmii{EFT,LO}}[v_{3,b}]}  \times
e^{-\int_{\vec{x}} V_3^{\rmii{EFT,NLO}}[v_{3,b}]} \times
e^{C[v_{3,b}]}.
\end{align} 
The term
$\det_{\rmii{$S$}} \times \det_{\rmii{$V$}}$
contains the functional determinants accounting for one-loop corrections to the action from the scalars and the vectors (to be precisely defined below), as described in section~\ref{sec:rate-NLO}. The factor $e^{-S_{\rm 3}^{\rmii{EFT,LO}}[v_{3,b}]}$ constitutes the LO contribution to the nucleation rate, containing the LO action evaluated on the LO bounce solution. As described above,
in constructing our LO nucleation theory, eq.~(\ref{eq:Veff-EFT-LO}), we integrated out the gauge fields to obtain their contribution to the barrier. From this theory we obtain the leading order bounce. When computing the contribution to the fluctuation determinant for the gauge modes, we are thus double-counting contributions that were already captured in the leading order nucleation action. We remedy this by subtracting the cubic term, evaluated on the bounce with $e^{C[v_{3,b}]}$, where
\begin{align}
  C[v_{3,b}] =
  - \frac{1}{12\pi}
  \int_{\vec{x}} \Bigl[
      6 (m^2_{\rmii{$X$},3})^{\frac{3}{2}}
    + 3 (m^2_{\rmii{$X_{0}$},3})^{\frac{3}{2}}
    - 3 (\mDX^2)^{\frac{3}{2}}
  \Bigr]
  \,.
\end{align}
Finally, $e^{-\int_{\vec{x}} V_3^{\rmii{EFT,NLO}[v_{3,b}]}}$ contains the NLO soft corrections to the potential. Note, that the NLO corrections to the finite-momentum part of the action are included in the functional determinants, without resorting to the derivative expansion. See also the discussion in section~\ref{sec:RG}.

For the determinants we have introduced the following simplified notation
\begin{align}\label{eq:detS_softEFT}
  \det_{\rmii{$S$}} &=
	\mathcal I_\phi \sqrt{ \Big|\frac{\det{\mathcal O_\phi(v_{3,\rmii{F}})}}{\det'\mathcal O_\phi(v_{3,b})}\Big |}
  \,,&
  \mathcal I_\phi &= \left(\frac{ S_{\rm 3}^{\rmii{EFT,LO}}[v_{3,b}] }{2\pi}\right)^{3/2}
  \,,&
  \mathcal O_\phi &= - \partial^2 + (V_3 ^{\rmii{EFT,LO}})^{\prime\prime}
  \,,
\end{align}
for the contribution of the scalar field.
Here, $v_{3,\rmii{F}}$ denotes the false vacuum (corresponding to $v_3 = 0$).
Note, that the factor $\det_{\rmii{$S$}}$ has units of $T^3$ while all
the other terms in eq.~\eqref{eq:rateFactors} are dimensionless.
 
The contributions from the vectors, Goldstone modes (of the new scalar field $\phi$) and ghosts
are denoted as
\begin{equation}\label{eq:vecdet}
\det_{\rmii{$V$}} =
    \det_{\rmii{$X_0$}}
    \det_{\rmi{g}}
    \det_{\rmii{$X_T$}} 
    \det_{\rmii{$XG$}}
  \,.
\end{equation} 
The first factor denotes the contribution from the temporal gauge modes $X_0$, which neatly factorises from the other contributions
\begin{equation}\label{eq:detX0}
	\det_{\rmii{$X_0$}}=\biggl|\frac{
    	\det\mathcal{O}_\rmii{$X_0$}(v_{3,\rmii{F}})
    }{
    	\det\mathcal{O}_\rmii{$X_0$}(v_{3,b})
    }
  \biggr|^{3/2}
  \,.
\end{equation}
The second term denotes the contribution from the ghosts, and the third from the transverse polarisation of the spatial gauge mode, neither of which mix with the other components.
The last term denotes the contribution from the Goldstones and spatial gauge fields, whose contributions mix (see section~\ref{sec:compDet}).
For a non-constant background field, this mixing cannot be removed by a gauge choice~\cite{Ai:2020sru}  and we have checked that the contribution from the mixed determinant is significant (compared to the approximation used in~\cite{Ekstedt:2023sqc}, where the off-diagonal terms are dropped). We compute its contribution using the Gel'fand--Yaglom method, following~\cite{Ekstedt:2021kyx} and we summarise the computation in great detail in appendix~\ref{sec:compDet}. The contributions from the scalar field itself and the temporal gauge modes are computed with {\tt BubbleDet}~\cite{Ekstedt:2023sqc}. 
The contributions from ghosts and the transverse polarisation of the spatial gauge mode are also computed using our Gel'fand--Yaglom implementation, however, using certain gauge choices one could compute them using {\tt BubbleDet} as well.

\subsection{Theoretical consistency of the method}
\label{sec:RG}

\subsubsection*{Gauge-independence}
Since we aim to compute physical quantities, like the percolation temperature, from the nucleation rate, it has to be gauge-invariant.
The gauge-invariance of the nucleation rate including the contribution from the fluctuation determinant was demonstrated in~\cite{Baacke:1999sc} for the SU(2) Higgs model (this is the same model 
as the one we consider, but with positive self-coupling). The fluctuation determinant splits into contributions from physical modes, and from modes that get cancelled by the Faddeev-Popov ghosts, leaving no dependence on the gauge fixing parameter $\xi$.
Other demonstrations of the gauge-independence of the nucleation rate can be found in~\cite{Endo:2017gal}, and in derivative expansion in~\cite{Lofgren:2021ogg, Hirvonen:2021zej}.
The latter references concretely demonstrate that the factor
$e^{-\int_{\vec{x}} V_3^{\rmii{EFT,NLO}}[v_{3,b}]}$ is gauge invariant,
in accord with Nielsen-Fukuda-Kugo identities~\cite{Nielsen:1975,Fukuda:1975}.
Following~\cite{Ekstedt:2021kyx}, in our evaluation of the vector determinant we exploit the gauge-independence to choose the gauge with $\xi=1$ in which
the Goldstone-vector mixing becomes more manageable to obtain our numerical results;
see appendix~\ref{sec:compDet}.

\subsubsection*{Renormalisation scale invariance}
Since the running of the 3D EFT starts at two-loop order,
our one-loop result would still include uncancelled running (in the LO action),
if we do not include large logarithms at two loops~\cite{Gould:2021}.
Hence, to render the results consistent and reliable, we need to include the two-loop soft corrections,
which are included in the term
$ e^{- \int_{\vec{x}} V_3^{\rmii{EFT,NLO}}[v_{3,b}]}$ in eq.~\eqref{eq:rateFactors}.
These two-loop corrections were computed in eq.~\eqref{eq:VeftNLO} within the derivative expansion, whose validity we aim to study here.
Without resorting to the derivative expansion, the inclusion of the two-loop corrections is a formidable task due to the inhomogeneous scalar background.%
\footnote{%
  Such computation involves two-loop diagrams with numerically determined
  propagators using Green's functions akin
  to~\cite{Bezuglov:2018qpq,Ekstedt:2022tqk,Dashko:2024spj}.
} 
Regardless of such complications, we can still extract and include the two-loop logarithm of the renormalisation scale by
utilising the super-renormalisability of the EFT.

For this, we consider a formal expansion
\begin{align}
  -\ln \frac{A_{\rm stat}}{T^3} &=
        \mathcal{C}_0
      + \varepsilon \; \mathcal{C}_1
      + \delta \mathcal{C}_{\frac{3}{2}}
      + \mathcal{O}(\varepsilon^2)
      \,,
\end{align}
where
$\mathcal{C}_0 \equiv S_{\rm 3}^{\rmii{EFT,LO}}[v_{3,b}]$
is the leading order action,
$\mathcal{C}_1 \equiv  \int_{\vec{x}} \mathcal{X}[v_{3,b}] - \ln \det_{\rmii{$V$}} - C[v_{3,b}]  $ includes the unknown two-loop contribution $\mathcal{X}[v_{3,b}]$
(without utilising the derivative expansion), and
$\mathcal{C}_{\frac{3}{2}} \equiv - \ln \det_{\rmii{$S$}}$ is the scalar determinant.
The two-loop contribution $\mathcal{X}[v_{3,b}]$ is suppressed compared to LO by the soft expansion parameter
$\varepsilon \sim \gX/\pi$, and within the derivative expansion
$\mathcal{X}[v_{3,b}] \approx V_3^{\rmii{EFT,NLO}}[v_{3,b}]$. 

Similarly, within the derivative expansion
$-\ln \det_{\rmii{$V$}} - C[v_{3,b}] \approx \frac{1}{2} Z^{\rmii{NLO}}_3(v_{3,b})(\partial_i v_{3,b})^2$.
However, here we \textit{assume} that the sum $-\ln \det_{\rmii{$V$}} - C[v_{3,b}]$ is suppressed by $\varepsilon \sim \gX/\pi$ also \textit{without} the derivative expansion: this justifies our treatment, where we have included $C[v_{3,b}]$ in the LO action where it contributes to the bounce, and when we subtract it in the sum $-\ln \det_{\rmii{$V$}} - C[v_{3,b}]$ this sum is suppressed compared to LO. 
For the scalar fluctuations, we have used an independent expansion parameter $\delta$,
but since we know how the scalar contributions scale compared to the soft expansion,
i.e.\ $\delta \sim \varepsilon^{\frac{3}{2}}$,
we can bundle together these two expansions~\cite{Gould:2023ovu}.

The renormalisation scale enters these perturbative expressions in dimensional regularisation when
renormalising the UV divergence of the two-loop sunset diagram through mass renormalisation.
As a consequence,
the mass parameter runs with the renormalisation scale
$\Lamd$, i.e.\ $m^2_3 = m^2_3(\Lamd)$ with scaling
$\Lamd \frac{dm^2_3}{d\Lamd} = \varepsilon \beta_{m^2_3} + \mathcal{O}(\varepsilon^2)$.
Concretely, the $\beta$-function is given by%
\footnote{%
  We point out, that in eq.~(4.21) in~\cite{Kierkla:2023von}
  the beta function is missing an overall factor 1/16 due to a misprint.
}
\begin{align}
  \beta_{m^2_3} = \frac{1}{(4\pi)^2}\left(
    \frac{39}{16} g^4_{\rmii{$X$},3}
  - 6 g^2_{\rmii{$X$},3} h_3^{ }
  +  \frac{3}{2} h^2_3 \right)
  \,.
\end{align}
Explicit dependence on $\ln \Lamd$ appears in $\mathcal{C}_1$ through
$\mathcal{X}[v_{3,b}]$.
The contributions
$\mathcal{C}_0$ and
$\mathcal{C}_{\frac{3}{2}}$ do not depend explicitly on $\Lamd$,
but only implicitly through the mass, as well as the bounce which depends implicitly on the mass. 
Physical quantities are independent of the renormalisation scale, i.e.%
\footnote{%
  Eq.~\eqref{eq:RG-invariance} holds when expanded order by order,
  but has an implicit, leftover RG scale-dependence otherwise.
  This leftover dependence reflects the size of higher order corrections.
  Alternatively to an order-by-order expansion, one could use optimised perturbation theory as
  $\frac{{\rm d}}{{\rm d}\ln\Lamd} \left( -\ln \frac{A_{\rm stat}}{T^3} \right)|_{\mu_\text{opt}} = 0$.
  Thus, the scale is fixed so that the rate has minimal sensitivity to the RG scale.
  This effectively resums different orders together and results in the RG-improved nucleation rate.
}
\begin{align}
\label{eq:RG-invariance}
 \frac{{\rm d}}{{\rm d}\ln\Lamd} \Bigl( -\ln \frac{A_{\rm stat}}{T^3} \Bigr) = 0
  \,.
\end{align}
Since this holds order by order in $\varepsilon$,
we get an identity at $\mathcal{O}(\varepsilon)$
\begin{align}
\Lamd \frac{\partial}{\partial \Lamd} \mathcal{C}_1
= 
\Lamd \frac{\partial}{\partial \Lamd} \mathcal{X}[v_{3,b}]
=
  -\beta_{m^2_3} \Big(
      \frac{\partial}{\partial m^2_3} \mathcal{C}_0
    + \frac{\partial v_{3,b}}{\partial m^2_3} \underbrace{\frac{\partial}{\partial v_3} \mathcal{C}_0}_{=0} \Big) =  -\beta_{m^2_3}  \frac{1}{2} v_{3,b}^2
  \,,
\end{align}
where
the derivative of $\mathcal{C}_0$ vanishes since the bounce extremises the LO action (and we have left integration over space implicit).
Hence, we conclude that the two-loop result has the following formal structure
\begin{align}
\label{eq:ActionRGOnly} 
\mathcal{X}[v_{3,b}]
&\simeq
      \mathcal{A}
      -\beta_{m^2_3} \Big(
          \frac{1}{2} v^2_{3,b} \Big)
        \ln \frac{\Lamd}{M_i(v_{3,b})}
\,,
\end{align}
where $\mathcal{A}$ is a yet undetermined contribution, which does not involve $\ln \Lamd$,
and $M_i$ formally correspond to soft gauge field masses (see below).
The point of this discussion is the realisation that in our perturbative setup,
the UV structure,
i.e.\ the logarithms of the renormalisation scale, are not affected by whether or not we apply the derivative expansion, but are completely dictated by the running of the mass within the EFT.

Thereby,
all logarithmic terms of $\mathcal{X}[v_{3,b}]$ have to align with those in $V_3^{\rmii{EFT,NLO}}[v_{3,b}]$, and we have
\begin{align}
\label{eq:VeftNLO-do-derivative-expansion}
\mathcal{X}[v_{3,b}] &= \mathcal{A}
    + \frac{1}{(4\pi)^2}\frac{3}{64} g^2_{\rmii{$X$},3} \Bigl(
      2 \bigl(80 m^2_{\rmii{$X$},3} - 3 g^2_{\rmii{$X$},3} v^2_{3}\bigr) \ln\frac{\Lamd}{2 m_{\rmii{$X$},3}}\Bigr)
      \nn &
  + \frac{1}{(4\pi)^2} \biggl\{
    -\frac{3}{8} h^2_{3} v^2_3 \Bigl( 2 \ln\frac{\Lamd}{2 m_{\rmii{$X_0$},3}}\Bigr)
    - \frac{3}{2} g^2_{\rmii{$X$},3} (m^2_{\rmii{$X$},3} - 4 m^2_{\rmii{$X_0$},3}) \ln\frac{\Lamd}{2 m_{\rmii{$X_0$},3} + m_{\rmii{$X$},3} }
  \biggr\} 
  \nn &
  - \frac{1}{(4\pi)^2} \biggl\{
      \frac{3}{2} g_{\rmii{$X$},3}^2 \mDX^2\Bigl(4 \ln\frac{\Lamd}{2 \mDX}\Bigr)
  \biggr\}
  \,.
\end{align}
This is our result for the two-loop soft contribution without the derivative expansion.
However, since $\mathcal{A}$ is left undetermined, in our numerical results in the next section, we simply resort to a replacement
$\mathcal{X}[v_{3,b}] \rightarrow V_3^{\rmii{EFT,NLO}}[v_{3,b}]$,
i.e.\
we employ the derivative expansion within the two-loop result (but not at one-loop).

\subsubsection*{Using the derivative expansion at two-loop}
\label{sec:der-exp-2loop}

Using the gradient expansion at two-loop while examining the validity of gradient expansion at one-loop requires a justification.
Let us first note that, at a given loop order, the finite-momentum contributions to the effective action are suppressed by one power of the soft expansion parameter
$\varepsilon \sim \gX/\pi$ with respect to the zero-momentum contribution,
i.e.~the effective potential.
This statement holds in the setup of our model, which can be generalised to other models featuring classical scale invariance.
This power counting was the reason why we included the one-loop
$Z_{2,3}$ factor and the two-loop contributions to the effective potential to compute the nucleation rate at NLO in~\cite{Kierkla:2023von}.
This can be also explicitly seen by examining the momentum-dependent contributions to the effective action computed within the gradient expansion;
see eqs.~\eqref{eq:Zfactor}--\eqref{eq:higher-Z}.
Hence, we can expect that in the full two-loop contribution to
the effective action, the effective potential is the dominant part, belonging to NLO in the soft expansion, while the finite-momentum part is suppressed and thus belongs to NNLO.
It is, therefore, well justified to only extract the effective potential part from the two-loop contribution to the action. 

Next, we will show that the contribution from the effective potential does not lead to a large uncertainty in the evaluation of the action on the bounce solution stemming from a breakdown of the derivative expansion.
This point was raised previously in~\cite{Gould:2021ccf},
especially for LO contributions, i.e.~$C[v_{3,b}] $.
Here, we will illustrate this behaviour for the NLO contribution from spatial gauge modes. The temporal gauge modes are always heavier than the nucleating scalar along the bounce solution.
Therefore, the derivative expansion always works for them, and
we have also confirmed this numerically.

Let us consider the part of
$V_3^{\rmii{EFT,NLO}}$ that contains only the contributions from the gauge spatial modes:
\begin{align}
	C^{\rmii{NLO}}_{\rmii{$X$},3} [v_{3}] &\equiv
	\int_{\vec{x}}    \frac{1}{(4\pi)^2}\frac{3}{64} g^4_{\rmii{$X$},3} v^2_{3}
	\Bigl(
			\underbrace{
				11 +42\ln{2} -14\ln{3} 
			}_{\equiv B}
			+34\ln{\frac{\Lamd}{ g_{\rmii{$X$},3} v_{3} }}
	\Bigr)
  \nn[2mm] &=
  \underbrace{
				\frac{B}{(4\pi)^2}\frac{3}{64}     \int_{\vec{x}}     g^4_{\rmii{$X$},3} v^2_{3}
		}_{\equiv I_1}
	  +\underbrace{
	  	         \frac{34}{(4\pi)^2} \frac{3}{64} 	\int_{\vec{x}}    g^4_{\rmii{$X$},3} v^2_{3}
	  	         \ln{\frac{\Lamd}{ g_{\rmii{$X$},3} v_{3} }}
	  	}_{\equiv I_2}
  \,.
\end{align}
To estimate the size of this contribution at the tail of the bounce solution, we can use the asymptotic form of $v_{3,b}$ as defined previously in eq.~\eqref{eq:v3b_tail}.
Then, the integrals $I_1$ and $I_2$ become:
\begin{align}
  I_1[v_{3,b}] &=
	  \frac{B}{(4\pi)}\frac{3}{64} \int_{r\geq R}\!{\rm d}r\, 
	 g^4_{\rmii{$X$},3} c^2  e^{-2 m_3 r} 
	 = \left[
	 	 \frac{B c^2}{(4\pi)}\frac{3}{64} g^4_{\rmii{$X$},3} \frac{e^{-2 m_3 r}}{2 m_3}
	  \right]
	  \xrightarrow{r\rightarrow \infty} 0
    \,, \\[2mm]
	I_2[v_{3,b}] &=
	 \frac{34}{(4\pi)} \frac{3 c^2}{64}  \int_{r\geq R}\!{\rm d}r\, 
	  g^4_{\rmii{$X$},3}   e^{-2 m_3 r} 
	   \ln{ \frac{\Lamd}{ g_{\rmii{$X$},3}  \frac{c}{r} e^{-m_3 r}   }}
     \nn
	  & =  \frac{34}{(4\pi)} \frac{3 c^2}{64} g^4_{\rmii{$X$},3}
	   \biggl[
	   		\frac{e^{-2 m_3 r} }{4m_3}
	   		-\frac{ \mbox{Ei}(-2 m_3 r)  }{2m_3}
	   		+\frac{ e^{-2 m_3 r}  \ln{ \frac{\Lamd}{ g_{\rmii{$X$},3}  \frac{c}{r} e^{-m_3 r}   }}  }{2m_3 }
	  \biggr]
	  \xrightarrow{r\rightarrow \infty} 0
    \,.
\end{align}
Thus, at large values of $r$, where the derivative expansion breaks down, the zero-momentum contribution simultaneously goes to zero. Hence, it is justified to use the two-loop corrections to the effective potential in the calculation of the nucleation rate.

\section{Numerical results}
\label{sec:results}

This section presents the results of our numerical analysis of the phase transition in
the SU(2)cSM model.
One of the main points of the present work is to scrutinise
the accuracy of various approximations to the bubble nucleation rate.
To this end,
we will present comparisons of different approaches,
which we summarise below:
\begin{enumerate}
\renewcommand\labelenumi{Approach~\arabic{enumi}:}

\item
  \label{it:daisy}
  [daisy]
  \\[1mm]
  Action computed using mere daisy resummation, with the functional determinants in the exponential prefactor approximated on dimensional grounds by $T^4$.
\begin{equation}
	\GammaT^{[\rmii{daisy}]} = T^4 e^{-S_{\rm daisy}}
  \,,
\end{equation}
where the action consists of the LO kinetic term and the daisy-resummed effective potential, given in eq.~(3.6) of~\cite{Kierkla:2023von}.

\item
  \label{it:LO}
  [LO]
  \\[1mm]
  Action at LO of the HT EFT (see eq.~\eqref{eq:Veff-EFT-LO}),
  the scalar functional determinants in the exponential prefactor
  approximated on dimensional grounds by $T^4$.
  In the construction of the EFT,
  the matching to the 4D theory is performed at
  two-loop level for the masses, and
  one-loop level for the couplings.
  Here, the gauge one-loop functional determinants are computed by using a
  derivative expansion and keeping the terms up to $\mathcal{O}(\gX^4)$.
  This approach reduces to the previous one if
  the matching relations in the construction of the EFT are
  truncated to the LO (cf.~\cite{Kierkla:2023von}).
	\begin{equation}
	\GammaT^{[\rmii{LO}]} = T^4 e^{-S_{\rm 3}^{\rmii{EFT,LO}}[v_{3,b}]},
\end{equation}
with $S_{\rm 3}^{\rmii{EFT,LO}}[v_{3,b}]$ defined in eq.~\eqref{eq:ActionLO}.
\item
  \label{it:NLO-grad}
  [NLO~$\nabla$]
  \\[1mm]
  The action computed within the HT EFT, with soft NLO corrections, in the derivative expansion.
  The scalar functional determinant in the exponential prefactor
  is approximated on dimensional grounds by $T^4$.
  The effective potential contains one- and two-loop gauge field contributions
  (up to $\mathcal O(\gX^5)$) and the kinetic term receives
  a correction from a field-dependent $Z$-factor.
  This is the approach used in~\cite{Kierkla:2023von}.
  \begin{equation}
	\GammaT^{[\rmii{NLO $\nabla$}]} =
      T^4
      e^{-S_{\rm 3}^{\rmii{EFT,LO}}[v_{3,b}] -S_{\rm 3}^{\rmii{EFT,NLO}}[v_{3,b}]},
  \end{equation}
  with $S_{\rm 3}^{\rmii{EFT,NLO}}[v_{3,b}]$ defined in eq.~\eqref{eq:SbounceNLO}.

\item
  \label{it:NLO-det}
  [NLO det]
  \\[1mm]
  The effective action computed within the HT EFT,
  the one-loop contribution computed with full functional determinants,
  supplemented with soft NLO corrections to the effective potential computed in the gradient expansion.
  The momentum-dependent NLO correction is accounted for by the functional determinants,
  without resorting to the derivative expansion.
  The scalar functional determinant is included, as well as an approximation for
  the dynamical prefactor
  evaluated using {\tt BubbleDet}.
  This is the most accurate approach considered in this work, and
  that is available in the literature so far.
  \begin{equation}
    \label{eq: Gamma [NLO det]}
    \GammaT^{[\rmii{NLO det}]} =
      A_{\rm dyn} \times
      \det_{\rmii{$S$}} \times
      \det_{\rmii{$V$}} \times
      e^{-S_{\rm 3}^{\rmii{EFT,LO}}[v_{3,b}]}\times
      e^{-V_3^{\rmii{EFT,NLO}}[v_{3,b}]} \times
      e^{ C[v_{3,b}] }
      \,,
  \end{equation}
  with $\det_{\rmii{$S$}}$, $\det_{\rmii{$V$}}$ defined in
  eqs.~\eqref{eq:detS_softEFT}--\eqref{eq:detX0} and
  $V_3^{\rmii{EFT,NLO}}[v_{3,b}]$ given in eq.~\eqref{eq:VeftNLO}.
  For a justification of including
  the two-loop contributions to the effective potential, derived in the derivative expansion,
  see the end of section~\ref{sec:RG}.
\begin{enumerate}
\item
  \label{it:NLO-det-T4}
  [NLO det $T^4$]
  \\[1mm]
  Instead of including the full scalar determinant prefactor and
  the approximate dynamical prefactor,
  the approximate prefactor of $T^4$ is used to
  evaluate the importance of the scalar prefactor by comparing with the full method.
  \begin{equation}
    \GammaT^{[\rmii{NLO det\ }T^4]} =
      T^4 \times
      \det_{\rmii{$V$}} \times
      e^{-S_{\rm 3}^{\rmii{EFT,LO}}[v_{3,b}]} \times
      e^{-V_3^{\rmii{EFT,NLO}}[v_{3,b}]} \times
      e^{ C[v_{3,b}] }
      \,.
  \end{equation}
\end{enumerate}
\end{enumerate}
By exhausting these options, we hope to learn which contributions to
the rate are important to include
to obtain quantitatively reliable results.

\subsection{Detailed comparison for one benchmark point}

\begin{figure}[t]
\centering
	\includegraphics[width=0.7\textwidth]{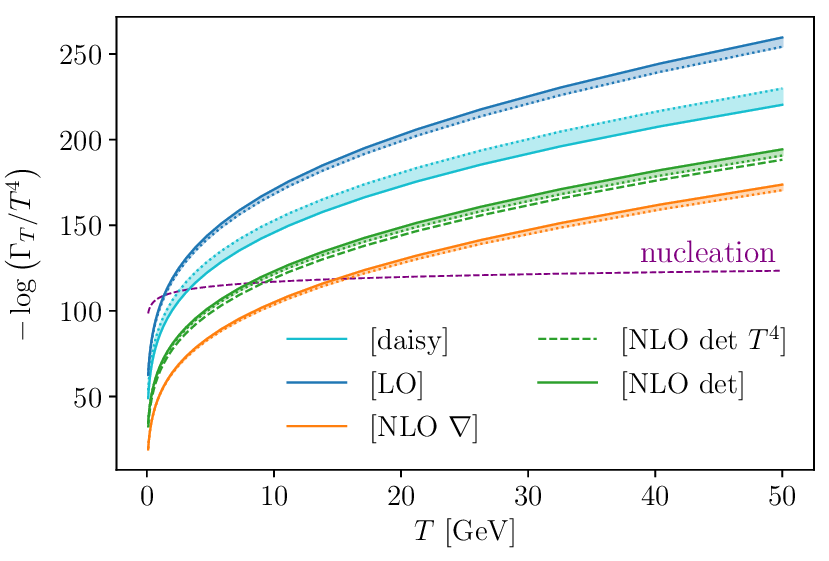}
	\caption{%
    Different approaches for computing  $\GammaT/T^4$ for a benchmark point with
    $\gX = 0.8$ and $\MX = 10^4$~GeV.
		Bands illustrate the sensitivity of different approaches to the choice of 4D RG scale at
    $\LamD = \pi T$ (solid) and
    $\LamD = 7 T$ (dotted). 
		The \hyperref[it:NLO-det-T4]{[NLO~det~$T^4$]} curve is evaluated at $\LamD = 7T$.
		The 3D scale is set to $\Lamd=T$.
   	}
	\label{fig:action-sample}
\end{figure}

In figure~\ref{fig:action-sample},
we illustrate the logarithm of the bubble nucleation rate (normalised by $T^4$) as a function of temperature,
computed using different approaches, for a fixed benchmark point with
$\gX = 0.8$ and
$\MX = 10^4$~GeV.%
\footnote{%
  The coupling $\gX$ is defined in the 4D theory at
  the energy scale $\mu=\MX$, where $\MX$
  corresponds to the tree-level approximation of the physical mass of the $X$ boson.
  These values are used to determine the other couplings of the model,
  following the procedure of~\cite{Kierkla:2022odc} and
  to compute the couplings of the HT EFT following~\cite{Kierkla:2023von}.
}
Examining the logarithm of the nucleation rate allows us to compare the effect of corrections to the action and the functional determinants in the exponential prefactors.
At the same time, one can easily evaluate the impact of various approximations on the nucleation temperature by looking at the intersection of the $-\ln( \Gamma/T^4)$ lines with the approximate nucleation-criterion line (dotted purple) for which
 $\Gamma/H^4=1$.%
\footnote{%
  This is an approximate nucleation criterion more consistent with the scenario of
  the phase transition proceeding during a phase of thermal inflation
  than the commonly adopted criterion $S_3/T\approx 140$. 
}
In the plot we present the approximations listed above with the following colour coding.
Approach~\ref{it:daisy}
\hyperref[it:daisy]{[daisy]}: turquoise,
approach~\ref{it:LO}
\hyperref[it:LO]{[LO]}: blue,
approach~\ref{it:NLO-grad}
\hyperref[it:NLO-grad]{[NLO~$\nabla$]}: orange,
approach~\ref{it:NLO-det}
\hyperref[it:NLO-det]{[NLO~det]}: green.
For approach~\ref{it:NLO-det-T4}
\hyperref[it:NLO-det-T4]{[NLO~det~$T^4$]},
we use dashed green.
For the other approaches, the solid (dotted) line corresponds to the choice of 4D RG-scale
$\LamD = \pi T \, (7 T)$.

First of all, one should note the substantial difference between results obtained at different orders in the soft expansion:
\hyperref[it:LO]{[LO]} (blue) and
\hyperref[it:NLO-det]{[NLO~det]} (green). This demonstrates that it is mandatory to include higher-order corrections to the nucleation rate to obtain precise results for thermodynamic parameters of the phase transition. 
As we can currently not include the complete next order in the soft expansion (without derivative expansion), or compare to lattice results, we have no means of confirming that our result is in fact converging towards the correct value. Some encouraging results for BSM theories can be found in~\cite{Niemi:2020hto, Gould:2023ovu, Niemi:2024axp,Ekstedt:2024etx},
where, despite large corrections at NLO, the perturbative convergence is very good, and the NLO result is close to the NNLO result. It should, however, be stressed that these comparisons were made for equilibrium quantities, and not for the nucleation rate.

Next, comparing
the green \hyperref[it:NLO-det]{[NLO~det]} and
the turquoise \hyperref[it:daisy]{[daisy]} curves one can appreciate the significant difference between the simplest  and the most advanced approaches presented here.
The nucleation temperature shifts from approximately 3.3~GeV, when computed using daisy approximation, to about 8~GeV when NLO corrections and functional determinants are included,
which indicates the insufficiency of the daisy approach.

The difference between
\hyperref[it:daisy]{[daisy]} (turquoise) and
\hyperref[it:LO]{[LO]} (blue) is caused by a different treatment of the matching relations when constructing the HT EFT. Daisy resummation corresponds to a truncation of the matching relations, as discussed in~\cite{Kierkla:2023von}.  
The \hyperref[it:daisy]{[daisy]} curve is closer to the
\hyperref[it:NLO-det]{[NLO~det]} curve than
\hyperref[it:LO]{[LO]}, which might seem counter-intuitive,
as the matching relations of
\hyperref[it:LO]{[LO]} and
\hyperref[it:NLO-det]{[NLO~det]} are identical. However, in principle there is no correlation between the higher-order corrections in the hard-scale matching relations and the NLO corrections to the potential in the soft EFT. It is, therefore, not necessary that they move the result
``in the same direction'', i.e.~matching at higher-order might increase the nucleation temperature and higher-order corrections in the EFT might decrease it, as seen here.
On the other hand, it is well known that using the HT EFT approach \emph{with} higher-order matching relations results in reduced RG-scale dependence in comparison to daisy resummation approach, and we indeed observe such behaviour in figure~\ref{fig:action-sample}, where the width of the bands indicates the sensitivity to the RG-scale.

Now we turn to the comparison of the two approaches to the NLO-corrected action:
with the full one-loop effective action, including the determinants
(\hyperref[it:NLO-det]{[NLO~det]}), and
the gradient-expansion approximation of~\cite{Kierkla:2023von}
(\hyperref[it:NLO-grad]{[NLO~$\nabla$]}).
We find that using the derivative expansion
overestimates 
the nucleation rate, resulting in an even larger difference with respect to
\hyperref[it:daisy]{[daisy]}.
Therefore, even though the inclusion of the NLO corrections to the action turns out to be crucial for a precise determination of the nucleation rate, when treated within the derivative expansion it can induce significant uncertainties.
This is one of the main findings of the present work.

It is commonly assumed that the scalar prefactor constitutes a subleading contribution to the nucleation rate,
cf.\ e.g.~\cite{Lofgren:2021ogg}.
In ref.~\cite{Ekstedt:2021kyx, Ekstedt:2023sqc}
it was shown that in certain scenarios,
however, the prefactor can be important.
In figure~\ref{fig:action-sample} we see that for our benchmark point, the scalar prefactor has a sub-dominant effect on the final result, as the solid (\hyperref[it:NLO-det]{[NLO~det]}) and dashed green lines (\hyperref[it:NLO-det-T4]{[NLO~det~$T^4$]}) are very close to each other. This is in agreement with our expectations following from power counting, but we will see in the next subsection that the difference in predictions (with and without the scalar determinant)
for percolation temperature can be as large as 40\% for other benchmark points.

\subsection{Scan of the parameter space of the SU(2)cSM model}

Figure~\ref{fig:action-sample} served as an illustration of different approaches to
compute $ \GammaT$.
Now, to generalise the above discussion, we present the results of a scan of the entire allowed parameter space.
We will show the results for the percolation temperature, $\Tp$, and
the average bubble radius at percolation normalised to the Hubble radius, $R_*H_*$, and we will compare a subset of the approaches listed at the beginning of this section.
For the calculation of $\Tp$ and $R_*H_*$ we shall follow~\cite{Kierkla:2022odc, Kierkla:2023von}.
The relative differences for $\Tp$ presented in the following figures are defined as follows
\begin{align}
	\label{eq:delta}
  \delta^{i}_{j} \Tp &=\frac{\bigl|\Tp^{i}-\Tp^{j}\bigr|}{\Tp^{j}}
  \,, &
  {\rm where}
  \quad
  \begin{aligned}
    i &\in \bigl\{
	[{\rm NLO\ det}], [{\rm NLO}\ \nabla], [{\rm NLO\ det\ }T^4]
	\bigr\}
  \,,\\
    j &\in \bigl\{[{\rm NLO \ det} ], [{\rm NLO\ det}\ T^4]\bigr\}
  \,.
  \end{aligned}
\end{align}
The relative differences for
$R_*H_*$, $\delta^i_j R_*H_*$, are defined in analogy.

The main goal of the current study is to improve the estimate of the GW signal.
In this work, we mainly focus on the validity of the derivative expansion, and therefore focus on $\Tp$ and $R_* H_*$, quantities that are directly derived from the nucleation rate.
Note that $\Tp$, while being a physical parameter describing the phase transition and setting the temperature scale for evaluating $R_*H_*$,  does not directly influence the GW spectrum.
This is true for models with a very strong phase transition, where the relevant temperature determining the peak frequency is the reheating temperature, see~\cite{Kierkla:2022odc}.
Nevertheless, we find it instructive to study the dependence of $\Tp$ on the different approximations, as this is the most direct measure of the validity of the different approximation schemes.
We refer readers that are interested in the other parameters entering in the GW signal, such as the PT strength $\alpha$  and the reheating temperature to~\cite{Kierkla:2022odc, Kierkla:2023von}.
 These quantities are not directly affected by the derivative expansion, and the results obtained before can be used.
The efficiency factor for transferring the released energy to the bubble walls and plasma, $\kappa$, depends on the value of $R_* H_*$ and would have to be recomputed to obtain the corrected spectra.
Using the results of~\cite{Lewicki:2022pdb},
this can be easily done as $\kappa \approx \frac{1}{1+R/R_{\rm eq}}$,
where $R_{\rm eq}$ denotes the radius of the bubble for which it would reach the stationary state.

\begin{figure}[t]
\centering
  	\includegraphics[width=.5\textwidth]{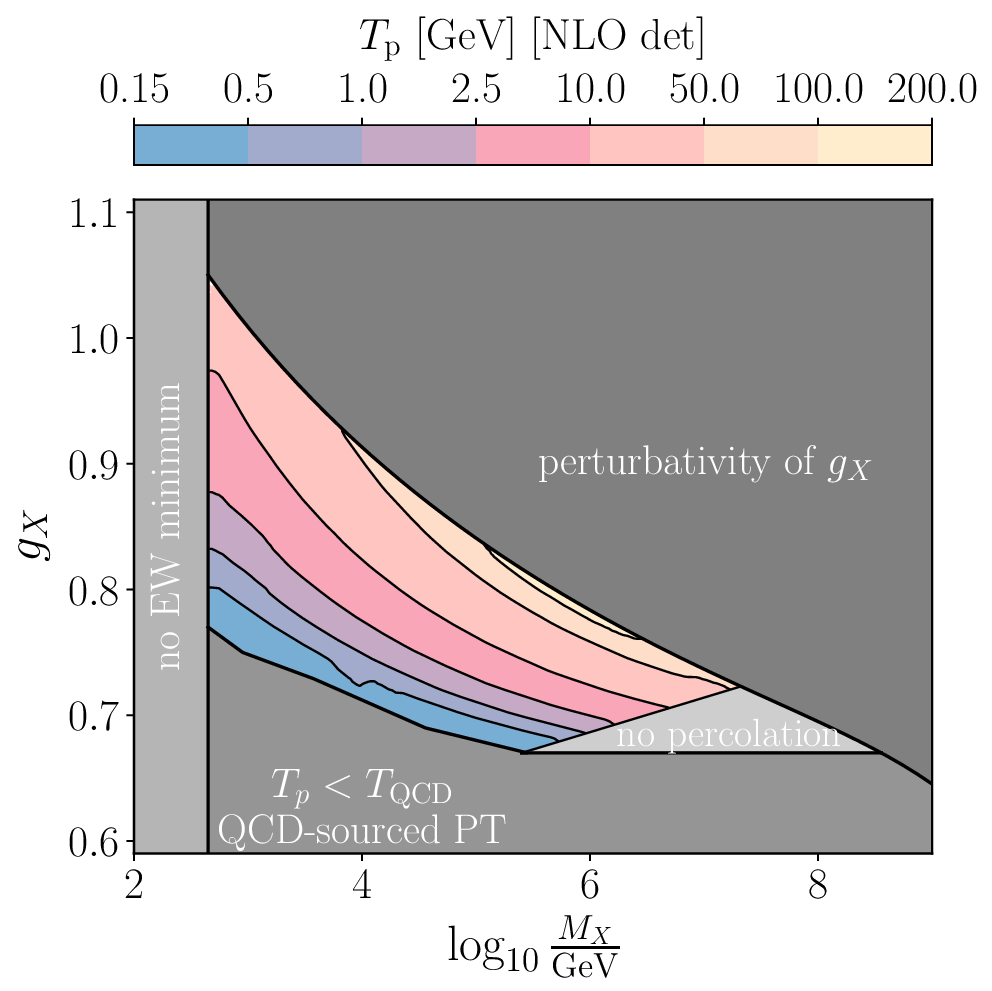}%
  	\includegraphics[width=.5\textwidth]{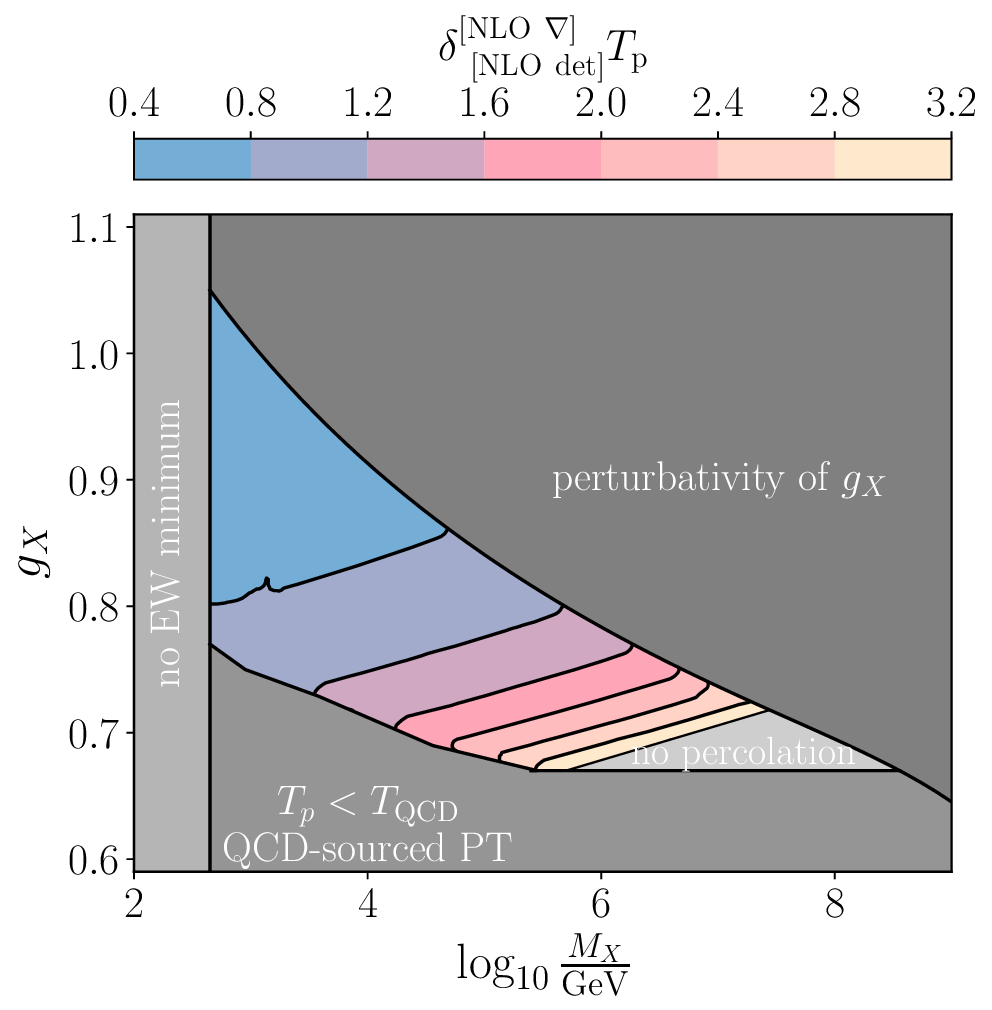}
	\caption{%
	    Predictions for the percolation temperature for different approximations to the NLO nucleation rate.
	    Left panel: \hyperref[it:NLO-det]{[NLO~det]} results for $\Tp$,
      right panel: relative difference between \hyperref[it:NLO-det]{[NLO~det]} and \hyperref[it:NLO-grad]{[NLO~$\nabla$]} results for $\Tp$.
    }
	\label{fig:Tp}
\end{figure}

\begin{figure}[t]
	\centering
  	\includegraphics[width=.5\textwidth]{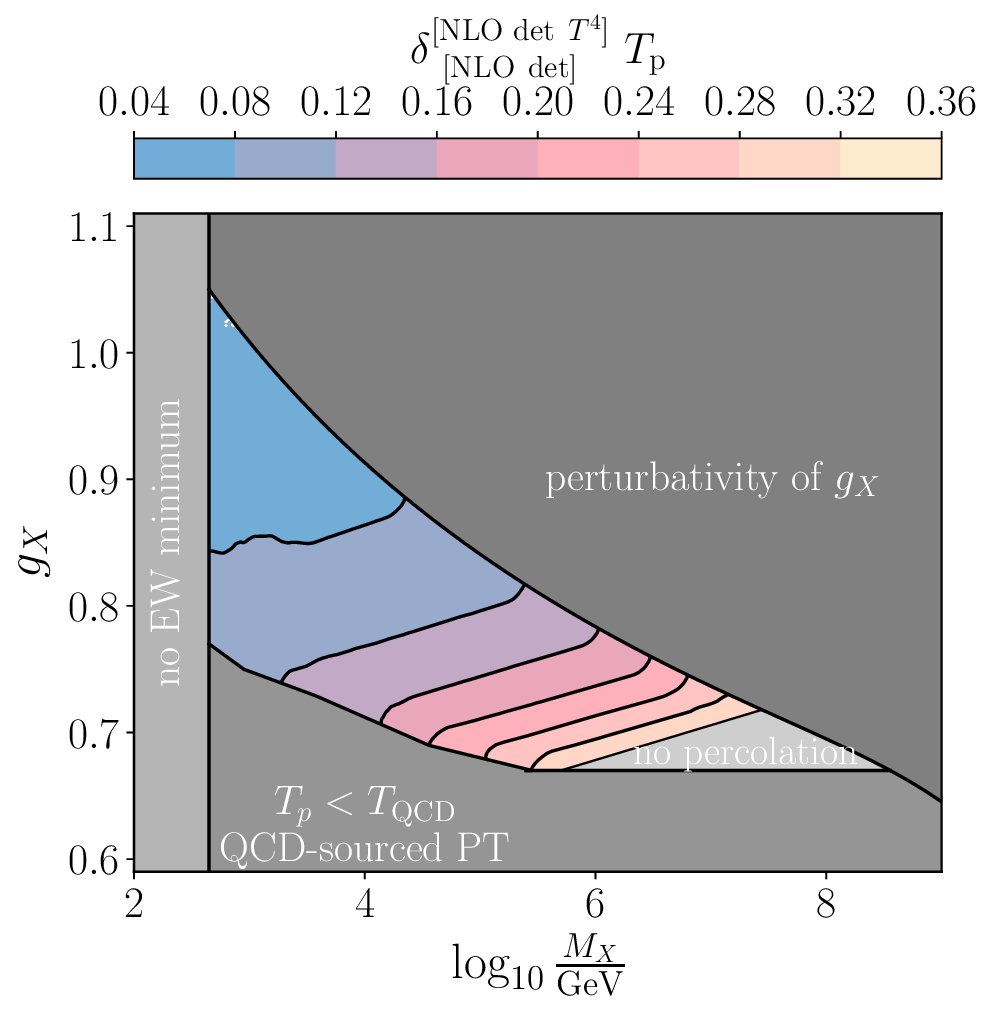}%
	\includegraphics[width=.5\textwidth]{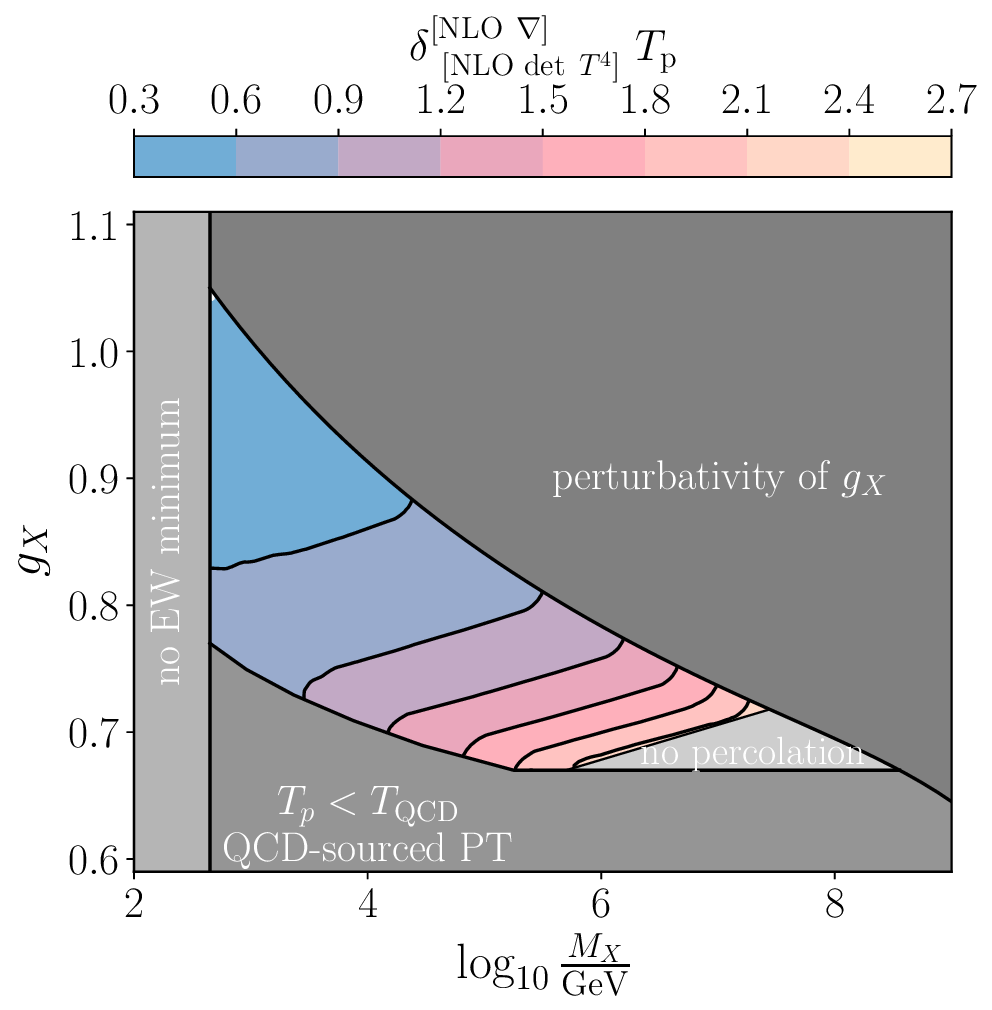}
	\caption{%
		Comparison of different approximations to the nucleation rate and the corresponding percolation temperature.
		Left panel: relative difference between \hyperref[it:NLO-det]{[NLO~det]} and \hyperref[it:NLO-det-T4]{[NLO~det~$T^4$]},
		right panel: relative difference between \hyperref[it:NLO-det-T4]{[NLO~det~$T^4$]} and \hyperref[it:NLO-grad]{[NLO~$\nabla$]}.
	}
	\label{fig:delta-Tp}
\end{figure}

\paragraph*{Comparisons of $\Tp$}
 The left panel of figure~\ref{fig:Tp} demonstrates the value of the percolation temperature in the entire allowed parameter space of our model,%
 \footnote{%
  For percolation temperatures below
  the temperature of the QCD phase transition,
  $T_{\rmii{QCD}}\approx 0.1\ \rm{GeV}$,
  it is still possible to study the phase transition in this model,
  however, we limit our analysis to the perturbative region associated with
  the electroweak phase transition.
  For details on the constraints on the parameter space see~\cite{Kierkla:2022odc}.
  Moreover, recent results~\cite{Lewicki:2024sfw} indicate that once the energy budget during the phase transition
  is treated with more care, the percolation criterion can be relaxed.
  If applied to our model, this would decrease or even remove the region excluded by non-percolation.
  However, a deepened study of this issue is beyond the scope of the present paper,
  therefore we continue using the constraints detailed in~\cite{Kierkla:2022odc}.
} computed in the most accurate approximation \hyperref[it:NLO-det]{[NLO~det]}.
 Qualitatively, the graph is similar to the results obtained with \hyperref[it:NLO-grad]{[NLO~$\nabla$]} in~\cite{Kierkla:2023von} and comparing to the results of~\cite{Kierkla:2022odc}, obtained using daisy resummation, we see that the inclusion of higher-order corrections consistently reduces the amount of supercooling, which was also indicated in figure~\ref{fig:action-sample}.
We see that a larger value of $\gX$ corresponds to a smaller percolation temperature for fixed $\MX$. For a fixed value of $\gX$, the smaller values of $\MX$ correspond to the largest amount of supercooling.

\begin{figure}[t]
\centering
 	\includegraphics[width=.5\textwidth]{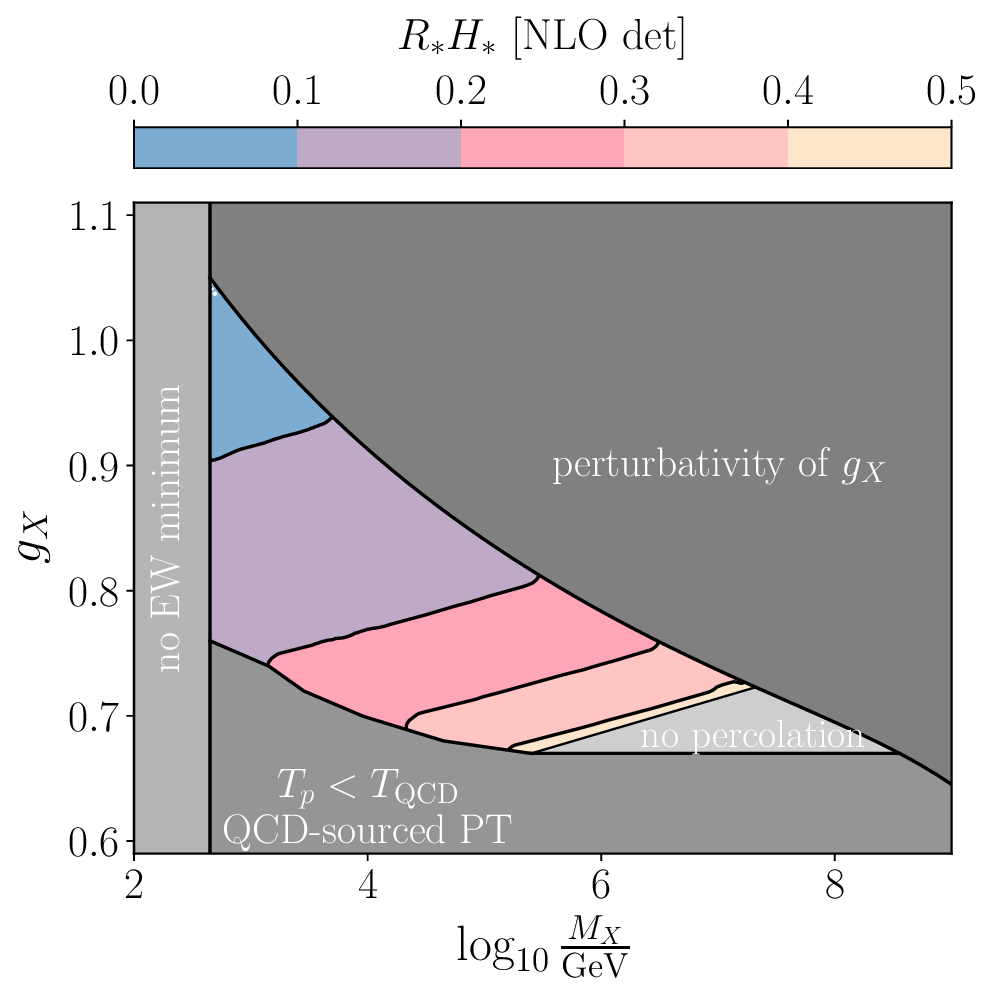}%
 	\includegraphics[width=.5\textwidth]{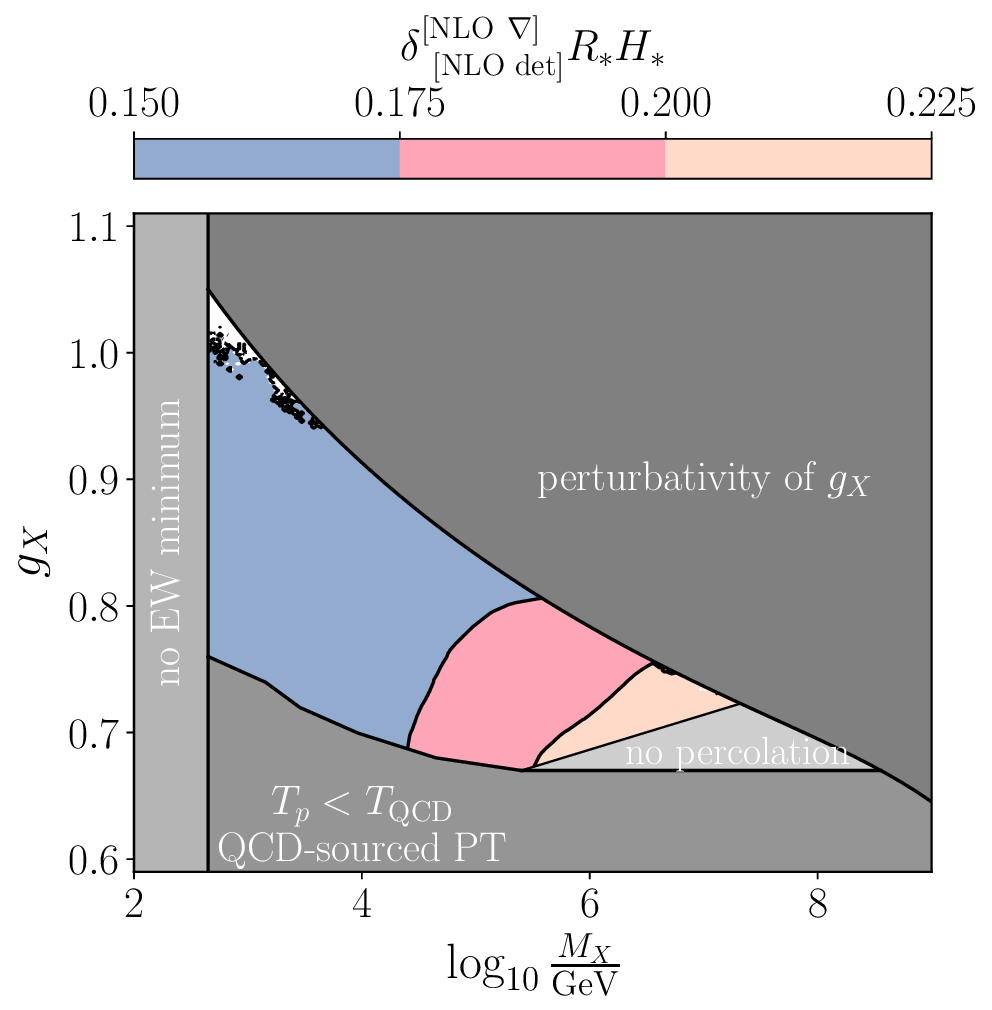}
	\caption{%
    Mean bubble radius at percolation, normalised to Hubble scale, $R_*H_*$.
    Left panel: obtained with \hyperref[it:NLO-det]{[NLO~det]},
    right panel: relative difference between
    \hyperref[it:NLO-det]{[NLO~det]} and
    \hyperref[it:NLO-grad]{[NLO~$\nabla$]}.
    The small white region in the corner of the parameter space is due to some numerical instabilities that do not influence the overall picture.
  }
	\label{fig:RH}
\end{figure}

A quantitative comparison between the
\hyperref[it:NLO-det]{[NLO~det]} and
\hyperref[it:NLO-grad]{[NLO~$\nabla$]} approaches is made in
the right panel of figure~\ref{fig:Tp}.
The inclusion of the full fluctuation determinant consistently decreases the percolation temperature compared to the gradient-expanded approach, and the significant difference between \hyperref[it:NLO-det]{[NLO~det]} and \hyperref[it:NLO-grad]{[NLO~$\nabla$]} observed for a single benchmark point in figure~\ref{fig:action-sample} persists throughout the entire parameter space. 
As in~\cite{Kierkla:2023von}, we see that the corrections are the largest in the region of large $\MX$ and small $\gX$, where the phase transition is the strongest
and the slowest (i.e.\ the size of bubbles at collision is the largest, see figure~\ref{fig:RH}).
Throughout the plot, the difference in the percolation temperatures varies from
40\% to more than 300\%, demonstrating the importance of including the determinants.

In figure~\ref{fig:delta-Tp},
we investigate the difference between the full
\hyperref[it:NLO-det]{[NLO~det]} and
\hyperref[it:NLO-grad]{[NLO~$\nabla$]} approaches in further detail.
We distinguish two sources of the relative difference observed in the right panel of figure~\ref{fig:Tp}: the approximation of the scalar determinant and contribution to $A_{\rm dyn}$ by $T^4$ as well as the gradient expansion for the gauge modes.
In the left panel of figure~\ref{fig:delta-Tp},
we evaluate the impact of the former approximation.
We see that the scalar pre-factor constitutes
a sub-dominant contribution to the observed differences.
This is in agreement with our expectations, as the one-loop contribution from the scalar is a higher-order effect than the one-loop contribution from the gauge modes.
Nevertheless, the relative differences associated to
the scalar pre-factor are non-negligible and range from 4\% to almost 40\%.

The right panel of figure~\ref{fig:delta-Tp} demonstrates the difference between
\hyperref[it:NLO-grad]{[NLO~$\nabla$]} and
\hyperref[it:NLO-det-T4]{[NLO~det~$T^4$]}.
In both approaches, the scalar pre-factor is approximated as $T^4$, and the graph thus allows us to observe directly the effect of the gradient expansion applied to the gauge, ghost and Goldstone modes.
We observe that the relative differences are much larger than in the left panel of the graph, and we can thus conclude that the gradient expansion applied to the gauge, ghost and Goldstone modes is the dominant source of difference observed in the right panel of figure~\ref{fig:Tp},
confirming our concerns about the validity of the derivative expansion at the bubble tail.
Indeed, the main cause of the breakdown of the derivative expansion are the spatial gauge modes, that do not have a Debye mass.
We have checked explicitly that applying the derivative expansion to
the temporal gauge modes only leads to  negligible deviations of order 0.01\%.

We emphasise, that in all comparison plots the relative differences between the more and less precise results are largest in the large-mass, small-coupling region.
Part of the reason for this behaviour was explained in~\cite{Kierkla:2023von}:
as the couplings are defined at the scale
$\LamD = \MX$, they have to be RG-evolved to the thermal scale.
As a result, the region of large $\MX$ corresponds to relatively large couplings, which explains why the perturbative expansion is less accurate in this region. In the region of largest supercooling, an additional effect enters.
As can be seen in figure~\ref{fig:action-sample}, the action changes more rapidly for small temperatures. Consequently, similar deviations in the action will lead to relatively larger deviations in observables at lower temperatures.

\begin{figure}[t] 
	\centering
	\includegraphics[width=.55\textwidth]{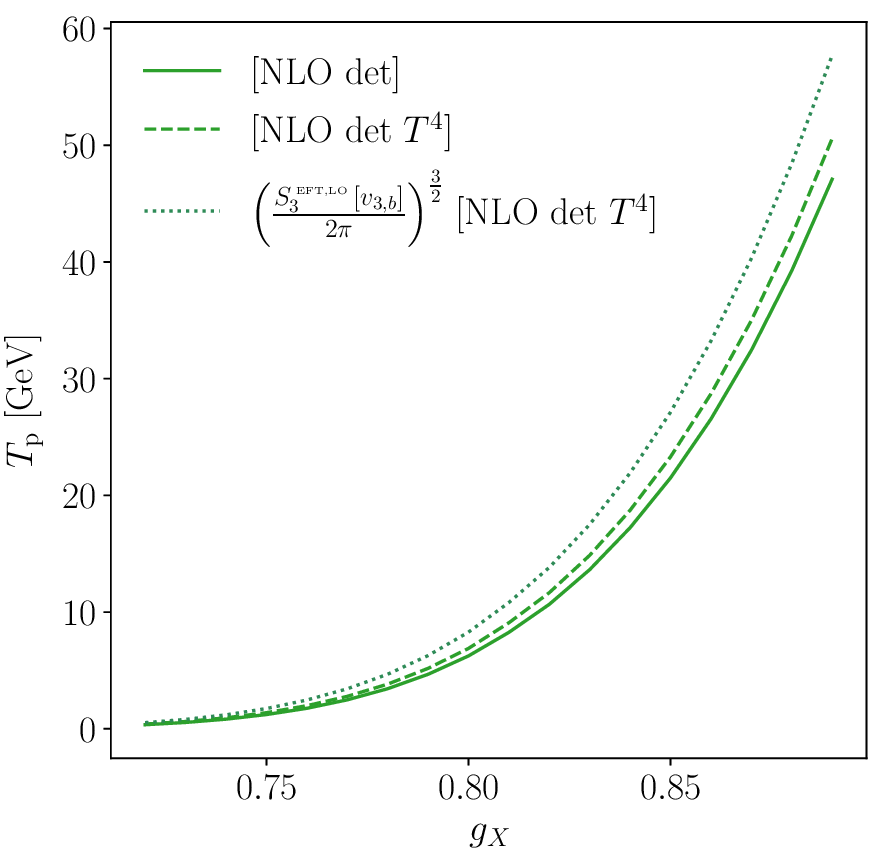}
	\caption{%
    Percolation temperature as a function of $\gX$ for fixed $\MX = 10^{4.05} $~GeV,
    obtained with different approximations to the scalar determinant
    in the prefactor of the nucleation rate.
    Here,
    $\mathcal I_\phi = \bigl(S_{\rm 3}^{\rmii{EFT,LO}}[v_{3,b}]/(2\pi)\bigr)^{3/2}$
    is the Jacobian factor obtained by
    the removal of scalar zero modes.
	}
	\label{fig:Tp_NLOdet_J}
\end{figure}
A commonly used approximation for the scalar determinant contribution also contains the
$\left( S/(2\pi) \right)^{3/2}$ factor which originates from the removal of scalar zero modes.%
\footnote{%
  Formally this factor corresponds to
  a Jacobian obtained by going to collective coordinates;
  see e.g.~\cite{Ekstedt:2023sqc}.
}
In our soft EFT this factor appears in the nucleation rate as
$\mathcal I_\phi$ in eq.~\eqref{eq:detS_softEFT}.
Then one would approximate the prefactor containing the scalar determinant and the dynamical contribution as:
	\begin{align}
		A_{\rm dyn}\det_{\rmii{$S$}}  \simeq T\cdot T^3 \left( 
		\frac{ S_{\rm 3}^{\rmii{EFT,LO}}[v_{3,b}] }{2\pi} 
		\right)^{3/2}
    \,.
\end{align}
We have found that omitting this factor leads to a better approximation of the scalar determinant in our model. 
In figure~\ref{fig:Tp_NLOdet_J},
we can see that predictions for the percolation temperature for a slice of
constant $\MX$ obtained by using solely $T^4$ match
the \hyperref[it:NLO-det]{[NLO~det]} noticeably better.
The reason for this is that the full scalar contribution is indeed an $\mathcal{O}(1)$ number times $T^3$, while the term $(S_{\rm 3}^{\rmii{EFT,LO}}[v_{3,b}]/(2\pi))^{3/2}$ alone is not $\mathcal{O}(1)$ thus leading to an overestimation of the nucleation rate.
We have confirmed that the same conclusion holds throughout the entire parameter space.

\paragraph*{Comparisons of $R_* H_*$}
In the left panel of figure~\ref{fig:RH},
we present the average radius of a bubble at percolation (normalised to the Hubble radius), $R_*H_*$, which directly influences GW spectra predicted from a phase transition.
Again, the qualitative results match those of~\cite{Kierkla:2023von}, with the large-$\gX$, small-$\MX$ corner corresponding to the smallest bubbles, and the small-$\gX$, large-$\MX$-corner corresponding to bubbles almost as large as the horizon scale.
The right panel of figure~\ref{fig:RH} demonstrates the relative differences between
\hyperref[it:NLO-det]{[NLO~det]} and
\hyperref[it:NLO-grad]{[NLO~$\nabla$]}.
As in~\cite{Kierkla:2023von},
the relative differences are much smaller than for $\Tp$, and
they range from 15\%  to about 25\%.
We note that again the main source of deviations is the gradient expansion of the gauge, ghost and Goldstone modes, and the approximation of the scalar pre-factor by $T^4$
causes at most a 6\% difference (and less in almost the entire parameter space).

The observed differences between
\hyperref[it:NLO-det]{[NLO~det]} and
\hyperref[it:NLO-grad]{[NLO~$\nabla$]} would not result in large changes in the predictions for GW spectra. However, as indicated by~\cite{Gonstal:2025qky} the accuracy of reconstructing $\beta_*/H_*$ (when marginalised over other parameters), which is directly related to $R_*H_*$, can be below 10\% throughout the interesting parameter space for supercooled phase transitions.
Then, the theoretical error associated with using the gradient expansion, not to mention the daisy resummation, would exceed the error of reconstruction. Moreover, it can drastically impact the predictions of the abundance of primordial black holes produced during supercooled phase transition, as the abundance has exponential sensitivity to the value of
$R_* H_*$~\cite{Lewicki:2023ioy, Gouttenoire:2023naa, Lewicki:2024ghw, Lewicki:2024sfw};
see also~\cite{Franciolini:2025ztf} for a recent update.

\section{Discussion}
\label{sec:discussion}

Motivated by the promising observational prospects of supercooled phase transitions,
the aim of this work has been to develop a consistent theoretical framework for providing precise predictions of
the thermodynamic parameters of such phase transitions.
We have built upon the results of our previous analysis in~\cite{Kierkla:2023von},
where the methods of high-temperature EFTs were used for supercooled phase transitions for the first time.
Within that framework, higher-order corrections (in the soft expansion)  were included,
both in the effective potential and in the kinetic part of the effective action via the field-dependent $Z_2$ factor.

We identified two challenges of applying the thermal EFT approach to supercooled transitions.
The first challenge arises from the breakdown of the mass hierarchy of different fields,
upon which the EFT approach relies.
As the symmetric phase is approached (at the tail of the bounce solution)
the hierarchy is invalidated because
both the scalar field and the transverse degrees of freedom of the gauge field become massless.
This issue is known as the problem of scale-shifting fields;
see e.g.~\cite{Gould:2021ccf}.
The second challenge stems from the $Z_2$ factor correcting the kinetic term,
which becomes divergent in the symmetric phase.
Although its contribution to the effective action is regularised by the derivative terms
vanishing as the symmetric phase is approached,
its contribution is large and not fully trustworthy.
The derivation of the $Z_2$ factor is based on the gradient expansion of the effective action,
which relies on the same mass hierarchy as the construction of the EFT.
Our aim in this paper was thus to compute the bubble nucleation rate without resorting to the problematic gradient expansion.

The breakdown of the derivative expansion is particularly severe in models with classical scale symmetry.
There, the absence of a negative mass term for the scalar causes
the mass hierarchy between the scalar and gauge modes to collapse more rapidly near
the tail of the bounce (the symmetric phase).
However, this issue is not limited to scale-invariant models;
in any gauge theory, transverse gauge modes always become massless in the symmetric phase,
rendering the derivative expansion unreliable.

To approach the issue of scale-shifters and the derivative expansion for supercooled transitions,
we have implemented the Gel'fand-Yaglom method of computing the functional determinants.
We have included both the scalar prefactor
(computed using {\tt BubbleDet}~\cite{Ekstedt:2023sqc}) and
the contributions from the gauge modes,
for which we developed an in-house code capable of handling the mixed gauge-Goldstone sector.

Our findings can be summarised in several crucial points:
\begin{itemize}
  \item[\resizebox{13pt}{!}{$\ToptVS(\Aglx,\Aglx,\Lglx)$}]
    Including higher-order terms in the soft expansion is
    mandatory for reliable predictions for supercooled phase transitions.
  \item[$\nabla$]
    The derivative expansion introduces significant errors,
    making it inadequate for precision studies.
  \item[\resizebox{13pt}{!}{$\TopoVR(\Axx)$}]
    Including the full scalar contribution to the effective action
    can source a noticeable difference in the results ---
    though still subleading compared to the uncertainties from the gradient expansion.
  \item[$\mathcal{I}$]
    As a byproduct, we also demonstrated in our model that a commonly used approximation for
    the exponential prefactor, featuring the Jacobian $(S_{\rm 3}^{\rmii{EFT,LO}}[v_{3,b}] /2\pi)^{3/2}$,
    yields results that are less accurate than the ones obtained using
    the simple $T^4$ prefactor derived from dimensional analysis.
    This specific conclusion is model-dependent; see e.g. \cite{Ekstedt:2021kyx} for examples where the approximation including the Jacobian is better.
\end{itemize}

Thus, we conclude that functional determinants must be computed exactly in
the quest for reliable predictions for supercooled phase transitions
to be probed by LISA-generation GW experiments.
While our study focuses on BSM models with classical scale invariance,
we anticipate similar conclusions to hold for other types of models
that accommodate thermal first-order phase transitions.
Our computations go beyond the accuracy achieved so far in perturbation theory and
our findings strongly motivate to push existing methods to study such phase transitions further,
and to develop new techniques to understand quantum field theories at extreme temperatures with multiple scale hierarchies.

\section*{Acknowledgements}
We thank
Oliver Gould,
Joonas Hirvonen,
Johan L{\"o}fgren,
Marco Matteini,
Miha Nemev\v{s}ek
and
Lorenzo Ubaldi
for illuminating discussions,
as well as
Nicklas Ramberg and
Daniel Schmitt for discussions on primordial black hole prospects
of supercooled phase transitions.
We are especially grateful to Andreas Ekstedt
who helped us with the implementation of the fluctuation determinant for the mixing vector and Goldstone fields. MK was supported by the Polish National Agency for Academic Exchange through Bekker NAWA program, project number BPN/BEK/2023/1/00311/U/00001.
The work of MK and B\'S was funded by the National Science Centre, Poland,
through the OPUS project number 2023/49/B/ST2/02782.
PS was supported by
the Swiss National Science Foundation (SNSF) under grant PZ00P2-21599.
JvdV was supported by the Dutch Research Council (NWO), under project number VI.Veni.212.133.
Work of TT was funded by the European Union (ERC, CoCoS, 101142449).
This research was funded in part by the National Science Centre, Poland, through the OPUS project number 2023/49/B/ST2/02782.
For the purpose of Open Access, the authors have applied a CC-BY public copyright licence to any Author Accepted Manuscript version arising from this submission.

\section*{Data availability statement}
The data used to prepare the plots presented in this article is publicly available at~\cite{SRKWL2_2025}.

\appendix

\renewcommand{\thesection}{\Alph{section}}
\renewcommand{\thesubsection}{\Alph{section}.\arabic{subsection}}
\renewcommand{\theequation}{\Alph{section}.\arabic{equation}}

\section{Commutation relations for the expansion of the determinant}
\label{sec:commutation-relations}

In this appendix we will prove some identities used in section~\ref{sec:expansion} in the derivation of the $Z_2$ factor via the expansion of the functional determinant. Note that all of the relations presented in this appendix hold both in Minkowski and Euclidean space.

Let us prove here the identity
\begin{equation}
\phi \frac{1}{p^2 +M^2} = \frac{1}{p^2 +M^2}\phi + \frac{1}{(p^2 +  M^2)^2} [p^2,\phi] + \frac{1}{(p^2 +M^2)^3}[p^2,[p^2 ,\phi]] + \cdots,\label{eq:commutator1}
\end{equation}
in the derivative expansion ($p^2/M^2 \ll 1$).
First, we expand the denominator in $p^2/M^2$
\begin{align}
	\phi \frac{1}{p^2 + M^2} = \phi \frac{1}{M^2}\Bigl(1 - \frac{p^2}{M^2} + \frac{p^4}{M^4} + \cdots \Bigr)
  \,,
\end{align}
with the dots denoting terms higher order in $p^2/M^2$.
Now we move the field to the right, and add the compensating commutators:
\begin{align}
	\phi \frac{1}{p^2 +M^2} =& \frac{1}{M^2}\Bigl( \phi - \frac{p^2}{M^2} \phi + \frac{1}{M^2}[p^2, \phi] + \frac{p^4}{M^4} \phi - \frac{1}{M^4} [p^4 ,\phi] + \cdots \Bigr)
  \\
	=& \frac{1}{M^2}\Bigl( \phi - \frac{p^2}{M^2} \phi + \frac{1}{M^2}[p^2, \phi] + \frac{p^4}{M^4} \phi - \frac{2 p^2}{M^4} [p^2 ,\phi] + \frac{1}{M^4} [p^2,[p^2,\phi]] + \cdots \Bigr)
  \nonumber
  \,.
\end{align}
Rearranging terms gives
\begin{align}
	\phi \frac{1}{p^2 +M^2} =&  \frac{1}{M^2} \Bigl(
      \Bigl(1 - \frac{p^2}{M^2} + \frac{p^4}{M^4} \Bigr) \phi
    + \frac{1}{M^2} \Bigl(1 - \frac{2p^2}{M^2} \Bigr)[p^2, \phi]
    + \frac{1}{M^4} [p^2, [p^2,\phi]] + \cdots
  \Bigr)
  \nn
	=& \frac{1}{p^2 + M^2}\phi + \frac{1}{(p^2 + M^2)^2} [p^2,\phi] + \frac{1}{(p^2+M^2)^3}[p^2,[p^2 ,\phi]] + \cdots
  \,,
\end{align}
which is the identity we wanted to prove.

Next, we also want to demonstrate that 
\begin{align}
  \label{eq:commutator2}
	&[p^2, \phi] = \Box \phi - 2i p^\mu \partial_\mu \phi
  \,,\\
  \label{eq:commutator3}
	&[p^2,[p^2,\phi]] = \Box \Box \phi + 4 i p^\mu \partial_\mu \Box \phi - 4p^\mu p^\nu \partial_\mu \partial_\nu \phi
  \,.
\end{align}
For this, we use that the momentum operator is the generator of translations
\begin{equation}
	\label{eq:pf-comutator}
	[p^\mu,\phi] = -i \partial^\mu \phi
  \,.
\end{equation}
Then we have
\begin{equation}
	[p^2 , \phi] = 2 p_{\mu} [p^\mu, \phi] - [p_{\mu},[p^\mu,\phi]] = -2ip^\mu \partial_\mu \phi + \Box \phi
  \,,
\end{equation}
and the second relation also follows easily
\begin{equation}
	[p^2,[p^2,\phi]] = \Box \Box \phi - 4 i p^\mu \partial_\mu \Box \phi -4 p^\mu p^\nu \partial_\mu\partial_\nu \phi
  \,.
\end{equation}
Note that the signs of the terms odd in $p^{\mu}$ in
the two equations above differ from the ones presented in~\cite{Fraser:1984zb}.
This difference can be traced back to the sign of the right-hand side of the commutation relation in eq.~\eqref{eq:pf-comutator}.
These sign differences, however, do not affect the final result as the terms odd in $p^\mu$ vanish once integrated over the momentum space.

\section{Details on the higher-order action in the derivative expansion}
\label{sec:derivative:expansion:higher}

In section~\ref{sec:higherorderderiv},
higher orders in the derivative expansion are reported.
In this appendix,
we detail the corresponding
vector- and temporal-scalar-induced
one-loop contributions to
$Z_{2,3}$,
$Z_{4,3}$,
$Y_{3,3}$, and
$Y_{4,3}$
in dimensional regularisation.
Due to their complexity in general dimensions,
two-loop contributions are shown only for
vector-induced terms, while results that also include temporal scalar contributions
in $d=3$ are presented in section~\ref{sec:higherorderderiv}.

\begin{figure}[t]
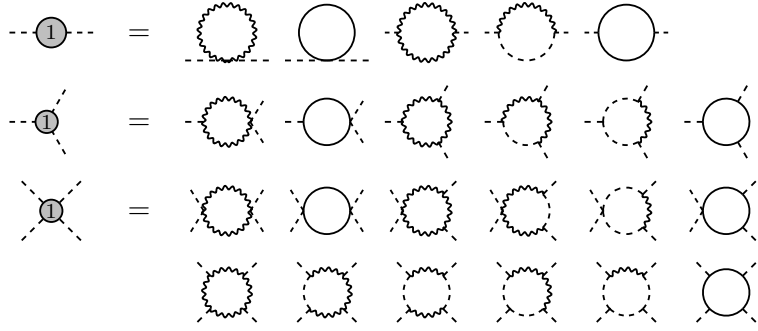

  \centering
  \begin{eqnarray*}
    \TopSi(\Lxx,1) &=&
      \TopoST(\Lxx,\Aglx)
      \TopoST(\Lxx,\Asai)
      \TopoSB(\Lxx,\Aglx,\Aglx)
      \TopoSB(\Lxx,\Aglx,\Axx)
      \TopoSB(\Lxx,\Asai,\Asai)
    \nn[3mm]
    \TopTi(\Lxx,\Lxx,\Lxx,1) &=&
      \TopoTBl(\Lxx,\Aglx,\Aglx)
      \TopoTBl(\Lxx,\Asai,\Asai)
      \TopoTC(fex(\Lxx,\Lxx,\Lxx),\Aglx,\Aglx,\Aglx)
      \TopoTC(fex(\Lxx,\Lxx,\Lxx),\Aglx,\Axx,\Aglx)
      \TopoTC(fex(\Lxx,\Lxx,\Lxx),\Aglx,\Axx,\Axx)
      \TopoTC(fex(\Lxx,\Lxx,\Lxx),\Asai,\Asai,\Asai)
    \nn[3mm]
    \TopVi(\Lxx,\Lxx,\Lxx,\Lxx,1) &=&
      \TopoVBlr(fex(\Lxx,\Lxx,\Lxx,\Lxx),\Aglx,\Aglx)
      \TopoVBlr(fex(\Lxx,\Lxx,\Lxx,\Lxx),\Asai,\Asai)
      \TopoVCrl(fex(\Lxx,\Lxx,\Lxx,\Lxx),\Aglx,\Aglx,\Aglx)
      \TopoVCrl(fex(\Lxx,\Lxx,\Lxx,\Lxx),\Aglx,\Axx,\Aglx)
      \TopoVCrl(fex(\Lxx,\Lxx,\Lxx,\Lxx),\Axx,\Aglx,\Axx)
      \TopoVCrl(fex(\Lxx,\Lxx,\Lxx,\Lxx),\Asai,\Asai,\Asai)
    \nn[2mm] &&
      \TopoVD(fex(\Lxx,\Lxx,\Lxx,\Lxx),\Aglx,\Aglx,\Aglx,\Aglx)
      \TopoVD(fex(\Lxx,\Lxx,\Lxx,\Lxx),\Aglx,\Aglx,\Axx,\Aglx)
      \TopoVD(fex(\Lxx,\Lxx,\Lxx,\Lxx),\Axx,\Aglx,\Axx,\Aglx)
      \TopoVD(fex(\Lxx,\Lxx,\Lxx,\Lxx),\Aglx,\Axx,\Axx,\Aglx)
      \TopoVD(fex(\Lxx,\Lxx,\Lxx,\Lxx),\Axx,\Axx,\Axx,\Aglx)
      \TopoVD(fex(\Lxx,\Lxx,\Lxx,\Lxx),\Asai,\Asai,\Asai,\Asai)
  \end{eqnarray*}
  \caption{%
    Non-vanishing one-loop contributions to the
    two-, three-, and four-point functions at the soft EFT
    of SU(2) with a fundamental scalar
    corresponding to the
    higher-derivative factors
      $Z_{2,3}$,
      $Z_{4,3}$,
      $Y_{3,3}$, and
      $Y_{4,3}$ from eqs.~\eqref{eq:Z23:d:2loop}--\eqref{eq:Y43:d:1loop}
    in the effective action~\eqref{eq:S3:higher}.
    Dashed lines denote soft scalars,
    wiggly lines spatial vector bosons, and
    solid lines temporal scalars.
    Diagrams are generated with {\tt Axodraw}~\cite{Collins:2016aya}.
  }
\label{fig:Z:Y:3}
\end{figure}
The one-loop contributions to the coefficients of eq.~\eqref{eq:S3:higher}
are computed
via the diagrams in figure~\ref{fig:Z:Y:3}.
In general dimensions, they read
\begin{align}
  \label{eq:Z23:d:2loop}
  Z_{2,3}(v_3) &=
      1
    + \frac{(d-1) (d^2 - 6 d + 20)}{4}
      \frac{I^{d}_{1}(m_{\rmii{$X$},3})}{v_{3}^2}
    + \frac{(d-4)(d-2)}{16}
      \frac{h_3^{2} v_{3}^{2}}{m_{\rmii{$X_0$},3}^4}
      I^{d}_{1}(m_{\rmii{$X_0$},3})
    \nn &+
      r_1(d) \frac{I^{d}_{1}(m_{\rmii{$X$},3})^2}{v_{3}^4}
    + r_2(d) \frac{g_{\rmii{$X$},3}^2}{v_{3}^2} S(m_{\rmii{$X$},3},0,0)
    + r_3(d) \frac{g_{\rmii{$X$},3}^2}{v_{3}^2} S(m_{\rmii{$X$},3},m_{\rmii{$X$},3},m_{\rmii{$X$},3})
  \,,\\[1mm]
  \label{eq:Z43:d:1loop}
  Z_{4,3}(v_3) &=
      \frac{(d-2) (d-1) (d^2 - 10d + 44)}{20}
      \frac{I^{d}_{1}(m_{\rmii{$X$},3})}{g_{\rmii{$X$},3}^2 v_{3}^4}
    \nn &+
      \frac{(d-6)(d-4)(d-2)}{320}
      \frac{h_3^{2} v_{3}^{2}}{m_{\rmii{$X_0$},3}^6}
      I^{d}_{1}(m_{\rmii{$X_0$},3})
  \,,\\[1mm]
  \label{eq:Y33:d:1loop}
  Y_{3,3}(v_3) &=
    \frac{(d-6) (d^5 - 12d^4 + 64d^3 - 18d^2 + 160d - 240)}{30d}
    \frac{I^{d}_{1}(m_{\rmii{$X$},3})}{g_{\rmii{$X$},3}^2 v_{3}^5}
    \nn &+
      \frac{(d-6)(d-4)(d-2)}{960}
      \Bigl(
          6
          + (d-8) \frac{h_{3}^{ }v_{3}^{2}}{m_{\rmii{$X_0$},3}^{2}}
      \Bigr)
      \frac{h_3^{2} v_{3}^{ }}{m_{\rmii{$X_0$},3}^{6}}
      I^{d}_{1}(m_{\rmii{$X_0$}})
  \,,\\[1mm]
  \label{eq:Y43:d:1loop}
  Y_{4,3}(v_3) &=
    \frac{(d-6) (d-4) (d-2) (d^3 - 11d^2 + 38d + 212)}{60}
    \frac{I^{d}_{1}(m_{\rmii{$X$},3})}{g_{\rmii{$X$},3}^2 v_{3}^6}
    \\ &+
      \frac{(d-6)(d-4)(d-2)}{3840}
      \Bigl[
          40
        + (d-8)\Bigl(
          16 \frac{h_{3}^{ }v_{3}^{2}}{m_{\rmii{$X_0$},3}^{2}}
        + (d-10)\frac{h_{3}^{2}v_{3}^{4}}{m_{\rmii{$X_0$},3}^{4}}
        \Bigr)
      \Bigr]
      \frac{h_3^{2}}{m_{\rmii{$X_0$},3}^{6}}
      I^{d}_{1}(m_{\rmii{$X_0$},3})
  ,
  \nonumber
\end{align}
where
$m_\rmii{$X$}$ and
$m_{\rmii{$X_0$},3}$ are the mass eigenvalues of
spatial and temporal gauge fields at the soft scale.
We also included the vector-induced two-loop contribution to the two-point function at $\mathcal{O}(k^2)$.
The rational coefficients used in the 2-loop result of $Z_{2,3}$ are given by
\begin{align}
  r_1(d) &=
  \frac{(d-1)(12d^7 - 253d^6 + 2241d^5 - 10819d^4 + 30957d^3 - 53320d^2 + 51834d -21456)}{4 (d-5) (d-3)d}
  \,, \nn
  r_2(d) &=
  \frac{9 (d-1) (2d-7)}{4}
  \,, \nn
  r_3(d) &=
      \frac{(d-1) (12d^4 - 119d^2 + 458d^2 - 800d + 546)}{6d}
  \,.
\end{align}
Here,
$I_{1}^{d}$ is given in eq.~\eqref{eq:1-loop-master}
and
the master integral
$S(m_1,m_2,m_3)$ is the 2-loop massive sunset diagram given in $d=3-2\epsilon$ via
\begin{align}
  S^{ }_3(m_1, m_2, m_3) &\equiv
  \int_{p,q} \frac{1}{
    [p^2+m^2_1]
    [q^2+m^2_2]
    [(p+q)^2+m^2_3]}
  \nn &
  = \frac{1}{(4\pi)^2} \bigg(\frac{1}{4\epsilon} + \frac{1}{2} + \ln \Big( \frac{\Lamd}{m_1 + m_2 + m_3} \Big) \bigg)
  + \mathcal{O}(\epsilon)
  \;.
\end{align}
Diagrammatically the different coefficient can be extracted by
imposing momentum conservation after computing the corresponding
$n$-point functions,
with ingoing momenta $\vec{q}_{1},\dots,\vec{q}_{4}$,
{\em viz.}
\begin{align}
  \TopSi(\Lxx, )
  &\supset
  \vec{q}_{1}^{ } \cdot \vec{q}_{2}^{ } Z_{2,3}^{ }(v_3)
    + q_1^2 q_2^2 Z_{4,3}(v_3)
    \,,\\[2mm]
  \TopTi(\Lxx,\Lxx,\Lxx, ) &\supset
    \Biggl[\sum_{i>j,k\neq i,j} \vec{q}_i^{ } \cdot \vec{q}_j^{ } q_{k}^{2}\Biggr] Y_{3,3}^{ }
    \,,\\[2mm]
  \TopVi(\Lxx,\Lxx,\Lxx,\Lxx, ) &\supset
    \Biggl[\sum_{i>j,k>l} \vec{q}_i^{ } \cdot \vec{q}_j^{ } \vec{q}_k^{ } \cdot \vec{q}_l^{ }\Biggr] Y_{4,3}^{ }
  \,,
\end{align}
where
in the sum all momenta are assumed to be different
($\vec{q}_i\neq\vec{q}_j\neq\vec{q}_k\neq\vec{q}_l$).
Here, we compute one-loop corrections to the $n$-point functions as in figure~\ref{fig:Z:Y:3}.
However, in general, at each further loop order they receive additional corrections,
as exemplified by the two-loop vector-induced correction to
$Z_{2,3}$ in eq.~\eqref{eq:Z23:d:2loop}.

\section{One-loop fluctuation determinant via {\tt BubbleDet} approach}
\label{sec:app-NLO-rate}

We can compute the parts of the rate  in eq.~\eqref{eq:rateBD}
that involve no kinetic mixing between different fields
using {\tt BubbleDet}~\cite{Ekstedt:2023sqc}.
First, we obtain the bounce solution $v_{3,b}$ and the corresponding action from the leading order potential $V_3^{\rmii {EFT, LO}}$ of the nucleation EFT.
We use the bounce solver of {\tt CosmoTransitions}~\cite{Wainwright:2011} for this as
the {\tt BubbleConfig.fromCosmoTransitions} method is then able to automatically obtain the information about the relevant bubble background. Then,
{\tt BubbleDet} computes the statistical part of the nucleation rate~\eqref{eq:rateBD}
using the Gel'fand-Yaglom method.
It returns
\begin{equation}
	{\tt findDeterminant()} = \sum_i
	\left [
		\frac{{\rm dof}(d,s_i,n_i)}{2} \ln{\left| \frac{\det'(-\partial^2 + W_i(r))}{\det (-\partial^2 + W_i(\infty))}\right|} - \ln \mathcal I_i \mathcal V_i
	\right]
	 - \ln A_{\rm dyn}
   \,,
\end{equation}
where the sum runs over the contributing fields,
$d$ denotes the dimensions,
$s_i$ the spin of the field and
$n_i$ its internal degrees of freedom.
The remaining notation has been defined in section~\ref{sec:rate-NLO}, and note that here and in the following, we use $W$ to denote the non-derivative part of the operators $\mathcal O$.
Note that the contributions from all fields are dimensionless, except for the contribution from
the scalar undergoing the phase transition, which has mass dimension~3. 
It is also important to pass {\tt Thermal = true} to {\tt BubbleDet.BubbleDeterminant} in case of nucleating scalar to get the dynamical prefactor. 

The output of {\tt findDeterminant()} can be then included in the nucleation rate in the following way:
\begin{align}
\label{eq:GammaNLO}
	\Gamma_T = T^4 e^{-S_0 - \sum_i \left [\frac{{\rm dof}(d,s_i,n_i)}{2} \ln{\left| \frac{\det'(-\nabla^2 + W_i(r))}{\det (-\nabla^2 + W_i(\infty))}\right|} - \ln \mathcal I_i \mathcal V_i \right] - \ln A_{\rm dyn} - 4\ln T}
  \,.
\end{align}

{\tt BubbleDet} also has the option to apply the derivative expansion.
This approximates the result of {\tt findDeterminant()} as
\begin{equation}
	S_1  = \int d^dx\left(V_{(1)}(\phib) - V_{(1)}(\phiF) \right) + \int d^dx \left(\frac 1 2 Z_{(1)}(\phib) \nabla_\mu \phib \nabla_\mu \phib \right)
  \,,
\end{equation}
where the first part is the LO contribution and the second part the NLO contribution. If the gradient expansion works well, this should reproduce the same result as the full fluctuation determinant. This could be illustrated for the fluctuation determinant of the temporal gauge mode where the derivative expansion works reliably because the temporal gauge mode is always heavier than the nucleating scalar.

\section{Gel'fand-Yaglom method for fluctuation determinants}
\label{sec:GY}

Let us now review how the fluctuation determinant appearing in the nucleation rate
eq.~\eqref{eq:rateFactors} can be computed. 
First, we will discuss the case where the determinant is a product of single-field fluctuation determinants.
This case was implemented in {\tt BubbleDet}. 
Then we discuss the case of multiple mixing fields, cf.\ section~\ref{sec:NLO-with-dets}.
To address mixing, we have implemented the approach discussed in~\cite{Ekstedt:2021kyx}, which we will review here.

\subsection{Fluctuation determinant of a single field}

First, we review the computation of ratios of fluctuation determinants of the form
\begin{align}
  &
	\frac{
			\mathcal {\rm det} \mathcal O_a(\phib)
		}{
			 {\rm det} \mathcal O_a (\phiF)
		}
    \,, &
    \mathcal O_a(\f) &= -\partial^2 + W(\varphi)
    \,,
\end{align}
where some examples of the function $W$ are given in eq.~\eqref{eq:OExamples}.
Since the following discussion is not unique to thermal phase transitions,
we will denote the background field as $\f$ rather than $v_3$,
and we will use $\phiF$ and $\phib$ to denote the false vacuum and bounce solution respectively.
In the following, we drop the subscript $a$ to avoid notational clutter.
Note that for the current discussion we are assuming that the fluctuation operator does not have zero modes.
This condition does not hold for the scalar field and Goldstone bosons, and
we refer the reader to
e.g.~\cite{Baacke:1993, McKane:1995vp, Dunne:2005rt, Baacke:2008zx, Falco:2017ceh, Ekstedt:2021kyx, Ekstedt:2023sqc,Bhattacharya:2024chz}
for a description on how to deal with these zero modes.

We can use the spherical symmetry of the bounce solution to expand the eigenfunctions on the basis of spherical harmonics~\cite{Baacke:1993, Baacke:2008zx, Ekstedt:2023sqc},
and write the fluctuation determinant as
\begin{align}
  \label{eq:detPartialWave}
	\frac{
			\mathcal
      {\rm det} \mathcal O(\phib(r))}{
			{\rm det} \mathcal O(\phiF(r))
		}
  &=
	 \prod_{l=0}^\infty \left( \frac{
      {\rm det} \mathcal O^{l}(\phib(r))}{
      {\rm det} \mathcal O^{l}(\phiF(r))}\right)^{{\rm deg}(l)}
  \,,&
  {\rm deg}(l) &= \frac{(d+2l-2)\Gamma(d+l-2)}{\Gamma(d-l)\Gamma(l+1)}
  \,,
\end{align}
where 
\begin{align}
  \mathcal O^{l}(\f (r) ) &= -\partial_l^2 + W(\f (r))
  \,,&
  \partial_l^2 &= \partial_r^2 + \frac{d-1}{r}\partial_r - \frac{l(l+d-2)}{r^2}
  \,.
\end{align}
We are interested in the interval $r\in [0, \infty]$.
Then, the eigenfunctions $\psi_l$ of the fluctuation operators should satisfy
the following boundary conditions
\begin{align}
  \label{eq:BCs}
  \psi_l(r)\rvert_{r=0}&=0
  \,, &
  \lim_{r \rightarrow \infty} \psi_l(r) &= 0
  \,.
\end{align} 

Rather than finding the infinite set of eigenvalues to evaluate the determinant for each value of $l$, we will make use of the Gel'fand-Yaglom theorem which greatly simplifies the computation.
We quickly review it here.%
\footnote{%
  In fact, here we use the theorem in a form that is suitable when
  $W$ corresponds to a matrix, which was first introduced by
  Forman~\cite{Forman1987},
  while the Gel'fand-Yaglom theorem corresponds to the one-dimensional case.
}

We are interested in the ratio of determinants
\begin{equation}
	\frac{{\rm det}[-L_2 + W^{(1)}(r)]}{{\rm \det}[-L_2 + W^{(2)}(r)]}
  \,,
\end{equation}
where $L_2$ denotes a second order differential operator (as $\partial_l ^2 $ in our case).
For simplicity, we write the $W^{(i)}$ as a function of $r$,
rather than functions of solutions to the equation of motion.
The eigenfunctions $\psi^{(i)}_\lambda$ (the label $i$ denotes whether the eigenvalue corresponds to the operator with $W^{(1)}$ or $W^{(2)}$)
defined by 
\begin{align}
	\bigl(-L_2 + W^{(i)}(r)\bigr) \psi^{(i)}_\lambda = \lambda \psi^{(i)}_\lambda
  \,,
  \label{eq:eigenPsi}
\end{align}
satisfy some boundary conditions at $r = r_0,r_1$, which can be written in the following form
\begin{equation}
	M \begin{pmatrix} \psi^{(i)}_\lambda(r_0) \\ \dot \psi^{(i)}_\lambda(r_0) \end{pmatrix} + N\begin{pmatrix} \psi^{(i)}_\lambda(r_1) \\ \dot \psi^{(i)}_\lambda(r_1) \end{pmatrix} = 0
  \,,
\end{equation}
where $M,N$ are $2\times 2 $ matrices, i.e.\
two rows correspond to
two boundary conditions and two columns to the pair of
$(\psi^{(i)}(r), \partial \psi^{(i)}(r)) \equiv \Psi^{(i)}$.
In principle, $M$ and $N$ should be given a label $i$, but since they are identical for $i=1,2$,
for our case of interest, we drop this here. 

Now, let $y^{(i)}_{n} (r)$ denote fundamental solutions of the homogeneous equation:
\begin{equation}
	\bigl(-L_2 + W^{(i)}(r)\bigr) y^{(i)}_n(r) = 0
  \,,
\end{equation}
where $n = 1,2$ runs over the number of fundamental solutions.
These fundamental solutions are characterised by their initial conditions:
\begin{align}
  y^{(i)}_1(0) &=1\,,& \dot y^{(i)}_1(0) &= 0\,,\\
	y^{(i)}_2(0) &=0\,,& \dot y^{(i)}_2(0) &= 1\,.
\end{align}
The number of fundamental solutions is 2 times the dimension of $W^{(i)}(r)$.
We can now define the matrix $Y$ containing all fundamental solutions and
their derivatives and also use it to express the initial conditions:
\begin{align}
  Y^{(i)}(r) &=
	\begin{pmatrix}
		y^{(i)}_1(r) && y^{(i)}_2(r) \\
		\dot y^{(i)}_1(r) && \dot y^{(i)}_2(r)
	\end{pmatrix}
  \,,&
  Y^{(i)}(0)&=1.
\end{align}
Gel'fand-Yaglom theorem now states that the ratio of determinants of fluctuation operators can be obtained as
\begin{align}
	\frac{{\rm det}[-L_2 + W^{(1)}(r)]}{{\rm \det}[-L_2 + W^{(2)}(r)]} 
	= \frac{
			{\rm det}[M + N Y^{(1)}(r_1)]
		}{
			{\rm det}[M + NY^{(2)}(r_1)]
		}
  \,.
\end{align}
Let us now apply this result to the determinant for $\mathcal O^{l}$ operators with the boundary conditions as in eq.~\eqref{eq:BCs}. First we see that the matrices $M$ and $N$ have the form
\begin{align}
    M &= \begin{pmatrix} 1 && 0 \\ 0 && 0 \end{pmatrix}
    \,, &
    N &= \begin{pmatrix} 0 && 0 \\ 1 && 0 \end{pmatrix}
    \,,
\end{align}
for $r_0=0$ and $r_1=\infty$.
Then we can obtain:
\begin{align}
	M+NY^{(i)}(\infty) =
  \begin{pmatrix} 1 && 0 \\ y^{(i)}_1(\infty) && y^{(i)}_2(\infty) \end{pmatrix}
  \,,
\end{align}
which leads to the simple expression 
\begin{equation}
	\det [ M+NY^{(i)}(\infty)] = y^{(i)}_2(\infty)
  \,.
\end{equation}
This gives the desired ratio of determinants simply by
\begin{align}
	\frac{{\rm det}[-\partial_l ^2 + W^{(1)}(r)]}{{\rm \det}[-\partial_l ^2 + W^{(2)}(r)]}
	= \frac{ y_2 ^{(1)}(\infty) }{ y_2 ^{(2)}(\infty) }
  \,.
\end{align}
This can directly be applied to the ratio of
determinants appearing in eq.~\eqref{eq:detPartialWave}.
The fundamental solution $y_2 ^{i}$ of the $i$-th operator we are interested in is defined as:
\begin{align}
  \bigl[-\partial_l ^2 + W^{(i)}(r)\bigr] y^{(i)}_2 (r) &= 0
  \,, &
  \mbox{with boundary conditions}
  \quad
  y_2 (0) &= 0
  \, \mbox{ and }
  \,
  \dot{y}_2(0) =1
  \,.
\end{align}
Thus, in a nutshell, Gel'fand--Yaglom theorem simplifies the problem of calculation of a ratio of two determinants to solving two differential equations and taking infinity limits of the solutions.
In our application, $W^{(1)}$ corresponds to $W(\f_b)$ and $W^{(2)}$ to $W(\f_{\rmii{F}})$.
 Since $W^{(2)}$ is constant, one can solve for $y_2 ^{(2)}$ analytically, which further simplifies the numerical implementation. 
 
In {\tt Bubbledet}, the differential equations for the low-$l$ modes are solved numerically using Gel'fand--Yaglom theorem, and a WKB approximation is used to determine the solutions for large $l$.%
\footnote{%
  Technically {\tt Bubbledet} also uses the WKB approximation for low-$l$
  values to speed up the convergence of computations
  (see section 4.2 of~\cite{Ekstedt:2023sqc}).
}

\subsection{Generalisation for matrices -- Forman theorem}

Before delving into the more complex case of the mixing determinant,
let us first briefly discuss the generalisation of
the theorem to higher dimensions ---
typically, cases where $\mathcal{O}$ is a matrix acting on multiple fields.
This generalisation was developed by Forman~\cite{Forman1987} and is known as
\textit{Forman theorem}, see e.g.~\cite{McKane:1995vp, Falco:2017ceh}.

Let us promote $\mathcal{O}$ to a $k \times k $ matrix (in the case relevant to us, $k = 3$).
We can also rewrite the boundary conditions of
 $\Psi^{(i)} = (
    \psi^{(i)}_1, \cdots ,
    \psi^{(i)}_k,
    \partial \psi^{(i)}_1 ,\cdots ,
    \partial \psi^{(i)}_k)$
 $=(\psi^{(i)}, \partial \psi^{(i)})_b$
 at points $r_0, r_1$ in 
terms of matrices $M, N$ of dimensions $2k \times 2k$
\begin{align}\label{eq:bc_M_N}
	M^{ab} \Psi^{(i)}_b(r_0) + N^{ab} \Psi^{(i)}_b(r_1)=0
  \,,
\end{align}
where $a,b = 1,\dots,2k$.

The $2k \times 2k$ matrix $Y^{(i)}$ is now given by
\begin{equation}
	Y^{(i)}(x) = \begin{pmatrix} \vec y_1^{(i)}(r) && \vec y_2^{(i)} (r) && \cdots && \vec y_{2r}^{(i)} (r)  \\
	\dot{\vec y}_1^{(i)}(r) && \dot{\vec y}_2^{(i)} (r) && \cdots && \dot{\vec y}_{2r}^{(i)} (r)
	 \end{pmatrix}
   \,,
\end{equation}
where the $\vec y_{1,2, \cdots k}^{(i)}(x)$ are $k$-component solutions to
$\mathcal O\, \vec{y}_{1,2, \cdots k}^{(i)} = 0$.
The initial conditions  are $Y^{(i)}(0)=1$. 
Then the ratio of determinants can be computed as:
\begin{align}
	\frac{{\rm set}[-L_2 + W^{(1)}(r)]}{{\rm set}[-L_2 + W^{(2)}(r)]} =
	\frac{
		\det \left[ M + N Y^{(1)}(\infty) \right]
	}{
		\det \left[ M + N Y^{(2)}(\infty) \right]
	}
  \,.
\end{align}

\subsection{Application to the Goldstone-vector determinant}
\label{sec:compDet}

The goal of this section is to illustrate the computation of the Goldstone-vector determinant $\det_{\rmii{$V$}}$ which (in $R_\xi$ gauge) is a product of the following terms,
cf.\ eq.~\eqref{eq:vecdet}:
\begin{align*}
  \det_{\rmii{$X_0$}}\,,
  &&& \text{the determinant of the temporal gauge mode}\,,\nn
		\det_{\rmii{$XG$}}\,,
  &&& \text{the determinant of mixing gauge and Goldstone modes}\,,\nn
		\det_{\rmii{$X_T$}}\,,
  &&& \text{the determinant of transverse gauge modes}\,,\nn
		\det_{\rmi{g}}\,,
  &&& \text{the determinant of the ghost modes}\,.
\end{align*}
The temporal gauge mode
does not kinetically mix with the other fields and the determinant can be calculated using Gel'fand-Yaglom theorem, or using {\tt{BubbleDet}} by treating the $X_0$ as a massive scalar without zero modes. As we shall show in a moment, the transverse gauge modes do not mix with Goldstones and the determinant of
$\det_\rmii{$X_T$}$ reduces to a one-dimensional problem as well.
Let us now discuss how to obtain the remaining part of the vector-Goldstone determinant.

We shall follow the approach of~\cite{Ekstedt:2021kyx}.
We work explicitly in $d=3$ and therefore denote the field as $v_3$.
First we decompose the gauge and
Goldstone fields into partial waves in the following way:
\begin{align} \label{eq:gauge_goldstone_basis}
  X_\mu(x) &=\sum_{l=0}^{\infty}\left[ \aS(r) n_\mu+\aL(r) \frac{r}{\sqrt{l(l+1)}} 
		\partial_\mu+\aT(r) \epsilon_{\mu \nu \alpha} x_v \partial_\alpha\right] Y_{l m}(\phi, \theta)
  \,,
  \\[2mm]
  \Phi_{\rmii{$G$}}(x) &=\aG(r) Y_{l m}(\phi, \theta)
  \,,\qquad 
  n_\mu=\frac{x_\mu}{r}
    \,.
\end{align}
The transverse gauge fluctuation $\aT$ is independent of the others,
i.e.\ the corresponding fluctuation operator becomes
\begin{equation}
		\mathcal{O}^l _{\rmii{$X_T$}} (v_3(r))  
	= \left(-\partial^2_l +  \WX(v_{3}(r)) \right)
  \,,
\end{equation}
and thus can be computed using one-dimensional 
Gel'fand-Yaglom theorem. 
The ghost determinant, which follows from the ghost operator
\begin{align}
	\mathcal{O}^l_{\rmi{g}}  (v_3(r))
	= \Bigl(-\partial^2_l + \frac{1}{\xi} \WX( v_3(r)) \Bigr)
  \,,
\end{align}
can be made identical to the transverse mode operator by a gauge choice $\xi=1$.

The remaining part of the determinant can be written in
the $(\aS, \aL, \aG )$ basis.
In the 
$R_\xi$-gauge,
the determinant decomposed in partial waves is thus:
\begin{equation}
	\det_{\rmii{$XG$}} = \left[ 
		\prod_{l=0}^{\infty}
    \frac{
      \det\mathcal{O}^l _{\rmii{$XG$}} (v_{3,b}(r)) }{
      \det\mathcal{O}^l _{\rmii{$XG$}} (v_{3,\rmii{F}}(r)) }
	\right]^{-\frac12}
  \,.
\end{equation}
The fluctuation operator can be written in a matrix form
\begin{align}
  \mathcal{O}^l _{\rmii{$XG$}} &=
	\begin{bmatrix}
		-\partial^2_l +\frac{2}{r^2}+\WX &
		-2 \frac{L}{r^2} &
		( \WX ^\prime)^{\frac{1}{2}} -(\WX)^{\frac{1}{2}} \partial \\
		-2 \frac{L}{r^2} & -\partial^2_l+ \WX & -(\WX)^{\frac{1}{2}} \frac{L}{r} \\
		(\WX)^{\frac{1}{2}} \frac{2}{r}+2 ( \WX ^\prime)^{\frac{1}{2}} +(\WX)^{\frac{1}{2}} \partial_r & 
		-(\WX)^{\frac{1}{2}} \frac{L}{r^2} & -\partial^2_l+\WG
	\end{bmatrix}
  \nn[2mm]
	 &+\Bigl(1-\frac{1}{\xi}\Bigr)
	\begin{bmatrix}
		\partial_r^2+\frac{2}{r} \partial_r-\frac{2}{r^2} & -\frac{L}{r}\left(\partial_r-\frac{1}{r}\right) & 0 \\
		\frac{L}{r}\left(\frac{2}{r}+\partial_r\right) & -\frac{L}{r^2} & 0 \\
		0 & 0 & 0
	\end{bmatrix}
  \nn[2mm]
	 &+\frac{1}{\xi}
	\begin{bmatrix}
		0 & 0 & (\WX ^\prime)^{\frac{1}{2}}+(\WX)^{\frac{1}{2}} \partial_r \\
		0 & 0 & (\WX)^{\frac{1}{2}} \frac{L}{r} \\
		-(\WX)^{\frac{1}{2}} \partial_r-(\WX)^{\frac{1}{2}} \frac{2}{r} & (\WX)^{\frac{1}{2}} \frac{L}{r} & 
		\WX 
	\end{bmatrix}
  \,,
\end{align}
where
we omitted the superscript on $W$,
but whenever relevant the $W$s will be distinguished by the argument
$v_{3,b}$ or $v_{3,\rmii{F}} = 0$.
Additionally, we used
\begin{align}
  \WX &= 4 m_{\rmii{$X$},3}^2
  \,,&
  \WG &= \frac{1}{v_3(r)} \bigl(V_3^{\rmii {EFT,LO}}[v_3(r)]\bigr)'
  \,,&
  L^2 &= l(l+1)
  \,.
\end{align}
As noted in~\cite{Ekstedt:2021kyx}, the matrix greatly simplifies for the choice $\xi = 1$,
\begin{align}
	\mathcal{O}^l _{\rmii{$XG$}} = 
	\begin{bmatrix}
		-\partial_l^2+\frac{2}{r^2}+\WX & -2 \frac{L}{r^2} & 2 ( \WX ^\prime)^{\frac{1}{2}} \\
		-2 \frac{L}{r^2} & -\partial_l^2+\WX & 0 \\
		2  ( \WX ^\prime)^{\frac{1}{2}}& 0 & -\partial_l^2+\WG+\WX
	\end{bmatrix}.
\end{align}
Now,
it is convenient to change the basis as
$\mathcal M^l \rightarrow \mathcal R^{-1} \mathcal M^l \mathcal R$  using the matrix:
\begin{align}
	\mathcal{R}=\frac{1}{\sqrt{2 l+1}}\left(\begin{array}{ccc}
		\sqrt{l} & -\sqrt{l+1} & 0 \\
		\sqrt{l+1} & \sqrt{l} & 0 \\
		0 & 0 & 1
	\end{array}\right).
\end{align}
The operator for vector-Goldstone fluctuations becomes:
\begin{align}
	\mathcal{O}^l _{\rmii{$XG$}} =
	\begin{bmatrix}
    -\nabla_l^{-}+\WX & 0 & \hphantom{+} 2  \sqrt{\frac{l}{2 l+1}} \WX\prime  \\
		0 & -\nabla_l^{+}+ \WX & - 2  \sqrt{\frac{l+1}{2 l+1}} \WX^\prime  \\
		2  \sqrt{\frac{l}{2 l+1}} \WX^\prime  & -2  \sqrt{\frac{l+1}{2 l+1}} \WX^\prime  & 
		-\partial_l^2+ \WG + \WX
	\end{bmatrix},
\end{align}
with 
\begin{equation}
	\nabla_l^{\pm} = \partial_r^2 + \frac{2}{r} \partial_r - \frac{(l \pm 1)(l + 1 \pm 1)}{r^2}
  \,.
\end{equation}
We will now discuss how to find the false-vacuum and bounce solutions for $l\neq 0$.
The partial waves $l=0$ correspond to zero modes.
A description of their treatment can be found in~\cite{Ekstedt:2021kyx}.

\subsubsection{Calculating the determinant of $\mathcal{O}^l _{\rmii{$XG$}}$ for the case $l\neq 0$}

As we have mentioned before, one can find the analytical expression for operators with constant $W$s. This is indeed the case for the determinant evaluated on the false vacuum solutions.
Now we shall derive these false-vacuum solutions and then show how to use them to obtain the final ratio of determinants.

\paragraph{Obtaining the false-vacuum solutions}

Consider the false-vacuum fluctuation operator, where $v_{3,\rmii{F}} = 0$:
\begin{align}
  \mathcal{O}^l _{\rmii{$XG$}} (v_{3,\rmii{F}}) =
	\begin{bmatrix}
		-\nabla_l^{-} & 0 & 0 \\
		0 & -\nabla_l^{+}  & 0 \\
		0 & 0 & -\partial_l^2+ \WG(0)
	\end{bmatrix}.
\end{align}
We see that all the terms proportional to $\WX$ are now zero, since gauge bosons do not have a mass in the false vacuum.
Now we want to find the analytical solutions in our basis for the false-vacuum case, i.e.\
$a_{\alpha;\rmii{F}} \equiv  (\aSfv, \aLfv, \aGfv)$,
and use them to calculate the desired ratio of determinants.
The solutions will be distinct for every $l$, but to avoid clutter, we drop the $l$-label.

We see that all of the operators have a similar form, and
indeed in practice,
we need to solve only the equation for $\aGfv$
\begin{align}
	\left[
		-\partial_l ^2 + \WG(0)
	\right]
	\aGfv (r) = 0
  \,,
\end{align}
where $\WG(0) = \mF ^2$, with $\mF$ the mass of Goldstone bosons in the false vacuum.
Note,
that $\aGfv$ corresponds to the homogeneous solution that was denoted by $\psi_0$
before (cf.\ eq.~\eqref{eq:eigenPsi}).
Then the other solutions can be obtained by setting $\mF = 0$ and
$l\rightarrow l\pm1$. Let us thus focus on the $\aGfv(r)$
case.
We need to solve the  equation:
\begin{align}
	-\left[
    \partial_r^2
  + \frac{2}{r} \partial_r
  - \frac{l(l+1)}{r^2}
- \mF ^2 \right]  \aGfv(r) = 0
  \,,
\end{align}
with boundary conditions such that $a_{\rmii{$G$}}(r) \rightarrow r^l$ as $r \rightarrow 0$.
Note,
that other solutions also should obey that boundary condition.
We can now use a trick and write the
$\aGfv(r) = r^l \tilde{a}_{\rmii{$G$}}(r) $.
Then the equation becomes:
\begin{align}
	  r^{l-1} \left[
	  r \partial_r^2 + 2 (l+1)\partial_r - r \mF ^2
  \right]\tilde{a}_{\rmii{$G$}}(r) = 0
  \,,
\end{align}
and its solution is
\begin{align}
  \tilde{a}_{\rmii{$G$}}(r) &= 
	  C_1 r^{\frac{1}{2} (-2 l-1)} J_{\frac{1}{2} (2 l+1)}(-i \mF r)
	+ C_2 r^{\frac{1}{2} (-2 l-1)} Y_{\frac{1}{2} (2 l+1)}(-i \mF r)
  \,,\\[2mm] 
	\Rightarrow \aGfv(r) &= \frac{
      C_1 J_{l+\frac{1}{2}}(-i \mF r)
    + C_2 Y_{l+\frac{1}{2}}(-i \mF r)}{\sqrt{r}}
  \,,
\end{align}
where
$J_\beta$,
$Y_\beta$ denote Bessel functions of the first and second kind, respectively.

Let us now determine the values of both $C_1$ and $C_2$ from the boundary conditions.
First of all, the second term in $\aGfv$ blows up at zero, so we have to set $C_2 =0$.
Then, to determine $C_1$, first we use the relation
between
the Bessel function $J_\beta$ and
the modified Bessel function $I_\beta$:
\begin{align}
	J_\beta (i z) = i^\beta I_\beta (z).
\end{align}
Setting $z=-\mF r$ and
$\beta = l +\frac{1}{2}$ (note that $\beta$ is non-integer) and using
$I_\beta (-x) = e^{i\pi \beta} I_\beta (x) $,
we further get:
\begin{align}
	  J_{l +\frac{1}{2}} (-i \mF r) &=
    i^{l +\frac{1}{2}} I_{l +\frac{1}{2}}(-\mF r)
	= i^{l +\frac{1}{2}} e^{i\pi (l +\frac{1}{2}) } I_{l +\frac{1}{2}}( \mF r)
	= i^{l+\frac{3}{2}} (-1)^l I_{l +\frac{1}{2}}( \mF r)
  \,.
\end{align}
Since the solution for $\aGfv(r)$ then takes the form
\begin{eqnarray}
  \aGfv(r) &=&
    \frac{C_1 i^{l+\frac{3}{2}} (-1)^l I_{l+\frac{1}{2}}(\mF r)}{\sqrt{r}}
  \nn &\stackrel{r\to0}{=}&
    \frac{C_1 (-1)^l i^{l+\frac{3}{2}} 2^{-l-\frac{1}{2}} (m r)^{l+\frac{1}{2}}}{\sqrt{r}
	\Gamma \left(l+\frac{3}{2}\right)} + \mathcal{O}(r^2)
  \stackrel{!}{=} r^l
   \,,
\end{eqnarray}
we could find $C_1$
by first expanding the solution in a series around $r=0$
and second using the boundary condition $\aGfv (r\rightarrow0)=r^l$,
to obtain
\begin{align}
	C_1 = (-1)^{\frac{5}{4}-l} i^{-l} (2r)^{l+\frac{1}{2}}  \Gamma
	\Bigl(l+\frac{3}{2}\Bigr) (m r)^{-l-\frac{1}{2}}
  \,.
\end{align}
Finally, by collecting all pieces together, the full solution becomes:
\begin{align}
	\aGfv(r) = \frac{2^{l+\frac{1}{2}} \mF ^{-l-\frac{1}{2}} \Gamma \left(l+\frac{3}{2}\right)
	I_{l+\frac{1}{2}}(\mF r)}{\sqrt{r}}
  \,.
\end{align}

As mentioned at the beginning, the remaining solutions can be easily obtained by expanding around $\mF=0$ and setting
$l\rightarrow l\pm1$:
\begin{align}
  \aSfv(r) &= r^{l-1}
  \,,&
  \aLfv(r) &= r^{l+1}
  \,.
\end{align}

\paragraph{Obtaining the solutions for fluctuations around the bounce}

Our goal now is to find fundamental solutions
$y$ which we can then use to compute the determinant.
First, the set of homogeneous equations for the fluctuation operator of our interest is
\begin{align}
	\mathcal{O}^l _{\rmii{$XG$}} (v_{3,b}) a^l _{\alpha} =
	\begin{bmatrix}
		  - \nabla_l^{-}+\WX & 0 & \hphantom{+} 2 \sqrt{\frac{l}{2 l+1}} \WX ^\prime
    \\
		0 & -\nabla_l^{+}+ \WX & -2 \sqrt{\frac{l+1}{2 l+1}} \WX^\prime
    \\
    2  \sqrt{\frac{l}{2 l+1}} \WX^\prime  &
    - 2\sqrt{\frac{l+1}{2 l+1}} \WX^\prime &
		- \partial^2_l + \WG + \WX
	\end{bmatrix}
	\begin{bmatrix}
		\aS ^l \\ \aL^l \\ \aG^l
	\end{bmatrix}
	= 0
  \,.
\end{align}
Note,
that again the solutions $a^l_\alpha$ correspond to the homogeneous solutions that we denoted by $\psi_0$ above (cf.\ eq.~\eqref{eq:eigenPsi}).
As we want to calculate the ratio of determinants associated with the bounce and false vacuum solutions,
it will be useful to consider the solutions normalised by the false vacuum ones
\begin{align}
	T^l _\alpha (r) \equiv
	\frac{
		 a^l _{\alpha; b} (r)
	}{
    a^l _{\alpha;\rmii{F}} (r)
	}
  \,.
\end{align}
As before, for $a_{\alpha; b} ^l$,
we impose boundary conditions such that
$a_{\alpha;b} ^l (r) \rightarrow r^l$ as $r \rightarrow 0$.
Thus, the $T_\alpha ^l$ obey the following boundary conditions:
\begin{align}
  \partial_r T^l_\alpha(0) &= 0
  \,,&
  T^l_\alpha(\infty) &= 0
  \,.
\end{align}

As in the previous one-dimensional Gel'fand-Yaglom example,
we pick $r_0=0$ and $r_1 = \infty$ and express the boundary conditions in terms of
$M$ and $N$ matrices, which in our $d=3$ case are $6\times 6$
dimensional:
\begin{align}
	\underbrace{
		\begin{bmatrix}
			0 & 0 & 0 & 0 & 0 & 0 \\
			0 & 0 & 0 & 0 & 0 & 0 \\
			0 & 0 & 0 & 0 & 0 & 0 \\
			0 & 0 & 0 & 1 & 0 & 0 \\
			0 & 0 & 0 & 0 & 1 & 0 \\
			0 & 0 & 0 & 0 & 0 & 1 \\
		\end{bmatrix}
	}_{M}
	\begin{bmatrix}
		\TS^l (0) \\ \TL^l(0) \\ \TG^l(0) \\ \partial \TS^l(0) \\ \partial \TL^l(0) \\ \partial \TG^l(0)
	\end{bmatrix}
	+ \underbrace{
		\begin{bmatrix}
			1 & 0 & 0 & 0 & 0 & 0 \\
			0 & 1 & 0 & 0 & 0 & 0 \\
			0 & 0 & 1 & 0 & 0 & 0 \\
			0 & 0 & 0 & 0 & 0 & 0 \\
			0 & 0 & 0 & 0 & 0 & 0 \\
			0 & 0 & 0 & 0 & 0 & 0 \\
		\end{bmatrix}
	}_{N}
	\begin{bmatrix}
		\TS^l(\infty) \\
    \TL^l(\infty) \\
    \TG^l(\infty) \\
    \partial \TS^l(\infty) \\
    \partial \TL^l(\infty) \\ 
		\partial \TG^l(\infty)
	\end{bmatrix} 
	=0
  \,.
\end{align}
The last necessary ingredient is the matrix $Y$ containing fundamental solutions.
For a $3\times3$ fluctuation
operator there exist $6$ fundamental solutions $y^i = (\yS, \yL, \yG)^i$ which are defined as:
\begin{align}
	Y(r) = 
	\begin{bmatrix}
		\yS^1 & \yS^2 &\dots & \yS^6 \\ 
		\yL^1 & \yL^2 & \dots & \yL^6 \\
		\yG^1 & \yG^2 & \dots & \yG^6 \\
		\partial \yS^1 & \partial \yS^2 & \dots & \partial \yS^6 \\ 
		\partial \yL^1 & \partial \yL^2 & \dots & \partial \yL^6 \\
		\partial \yG^1 & \partial \yG^2 & \dots & \partial \yG^6 
	\end{bmatrix}
	\xrightarrow{r\rightarrow0}  1
  \,.
\end{align}
Now according to Forman's theorem, we need to compute the following combination of fundamental solutions:
\begin{align}
	\label{eq:detVG_final_detY}
	\det[ M+NY(\infty)] &= 
	\begin{vmatrix}
		\yS^1 & \yS^2 & \yS^3 & 0 & 0 & 0 \\
		\yL^1& \yL^2 & \yL^3 & 0 & 0 & 0 \\
		\yG^1 & \yG^2 & \yG^3 & 0 & 0 & 0 \\
		0 & 0 & 0 & 1 & 0 & 0 \\
		0 & 0 & 0 & 0 & 1 & 0 \\
		0 & 0 & 0 & 0 & 0 & 1 \\
	\end{vmatrix} _{r\rightarrow \infty}
  \\[2mm]
	&= \Bigl[
      \yS^1 \left( \yL^2 \yG^3 - \yL^3 \yG^2 \right)
	  - \yS^2 \left( \yL^1 \yG^3 - \yL^3 \yG^1 \right)
	  + \yS^3 \left( \yL^1 \yG^2 - \yL^2 \yG^1 \right)
  \Bigr]_{r\rightarrow \infty}
  \,.
  \nonumber
\end{align}
Thus, we see that for $l$-th operator we need to
obtain 3 sets of solutions $(\yS, \yL, \yG)$, and
to find the ratio of fluctuation determinants simply evaluate
the expression~\eqref{eq:detVG_final_detY}.

\paragraph{Equations for fundamental solutions}
We want to find the fundamental solutions associated with equations for $T^l$ that we have introduced before. Using the false-vacuum solutions $a^l_{\alpha;\rmii{F}}$
we then write the equations as follows
\begin{align}
    \bigl[-\nabla_l^{-}+\WX \bigr] \TS^l
  + \biggl[2  \sqrt{\frac{l}{2 l+1}} \WX ^\prime \biggr]
  \biggl(\frac{\aGfv^l }{\aSfv^l}\biggr) \TG^l &= 0
    \,, \\[2mm]
    \bigl[-\nabla_l^{+}+ \WX \bigr] \TL^l
	+ \biggl[-2  \sqrt{\frac{l+1}{2 l+1}} \WX^\prime \biggr]
  \biggl(\frac{ \aGfv^l}{\aLfv^l}\biggr)  \TG^l &= 0
    \,, \\[2mm]
    \bigl[-\partial_l^2+ \WG + \WX \bigr]
    \biggl(\frac{\aGfv^l }{r^l} \biggr) \TG^l
	+ \biggl[2 \sqrt{\frac{l}{2 l+1}} \WX^\prime \biggr]
    \biggl(\frac{\aSfv ^l}{r^l} \biggr)
		\TS^l
    &
    \nn
  - \biggl[2 \sqrt{\frac{l+1}{2 l+1}} \WX^\prime \biggr]
    \biggl(\frac{\aLfv^l}{r^l} \biggr)
    \TL^l &= 0
    \,.
\end{align}
These equations can be solved numerically and by choosing the initial conditions,
we can obtain the three sets of
fundamental solutions that we need:
\begin{align}
  y^1(r) &=
	\begin{bmatrix}
		\yS^1(r) \\ \yL^1(r) \\ \yG^1(r)
	\end{bmatrix}=
	\begin{bmatrix}
		\TS^l(r)\\ \TL^l(r)\\ \TG^l(r)
	\end{bmatrix}^1
  \,, &
  \mbox{with }
	\begin{bmatrix}
		\TS^l(0)\\ \TL^l(0)\\ \TG^l(0)
	\end{bmatrix}^1
  &=
	\begin{bmatrix}
		1\\ 0\\ 0
	\end{bmatrix}
  &
  \mbox{ and }
	\begin{bmatrix}
		\partial \TS^l(0)\\ \partial\TL^l(0)\\ \partial\TG^l(0)
	\end{bmatrix}^1
  &=
	\begin{bmatrix}
		0\\ 0\\ 0
	\end{bmatrix}
  \,,\\[2mm]
  y^2(r) &=
	\begin{bmatrix}
		\yS^2(r) \\ \yL^2(r) \\ \yG^2(r)
	\end{bmatrix}=
	\begin{bmatrix}
		\TS^l(r)\\ \TL^l(r)\\ \TG^l(r)
	\end{bmatrix}^2
  \,, &
  \mbox{with }
	\begin{bmatrix}
		\TS^l(0)\\ \TL^l(0)\\ \TG^l(0)
	\end{bmatrix}^2
  &=
	\begin{bmatrix}
		0\\ 1\\ 0
	\end{bmatrix}
  &
	\mbox{and }
	\begin{bmatrix}
		\partial \TS^l(0)\\ \partial\TL^l(0)\\ \partial\TG^l(0)
	\end{bmatrix}^1
  &=
	\begin{bmatrix}
		0\\ 0\\ 0
	\end{bmatrix}
  \,,\\[2mm]
  y^3(r) &= \begin{bmatrix}
		\yS^3(r) \\ \yL^3(r) \\ \yG^3(r)
	\end{bmatrix}=
	\begin{bmatrix}
		\TS^l(r)\\ \TL^l(r)\\ \TG^l(r)
	\end{bmatrix}^2
  \,, &
  \mbox{with }
	\begin{bmatrix}
		\TS^l(0)\\ \TL^l(0)\\ \TG^l(0)
	\end{bmatrix}^3
  &=
	\begin{bmatrix}
		0\\ 0\\ 1
	\end{bmatrix}
  &
	\mbox{and }
	\begin{bmatrix}
		\partial \TS^l(0)\\ \partial\TL^l(0)\\ \partial\TG^l(0)
	\end{bmatrix}^1
  &=
	\begin{bmatrix}
		0\\ 0\\ 0
	\end{bmatrix}
  \,.
\end{align}
After finding these solutions numerically, all we need to do is to evaluate their
$r\rightarrow\infty$
limits and use the expression from eq.~\eqref{eq:detVG_final_detY}.
Thus, the final result for the $l$-th ratio of
vector-Goldstone mixing modes fluctuation determinants is:
\begin{align}
  \frac{ \det\mathcal{O}^l _{\rmii{$XG$}} (v_{3,b}) }{ \det\mathcal{O}^l _{\rmii{$XG$}} (v_{3,\rmii{F}}) } &=
	\frac{
		\rm{Det} \mathcal{M}^l
	}{
    \rm{Det} \mathcal{M}^l _\rmii{FV} 
	}
  \nn &
	=
	 [\yS^1 \left( \yL^2 \yG^3 - \yL^3 \yG^2 \right)
	- \yS^2 \left( \yL^1 \yG^3 - \yL^3 \yG^1 \right)
	+ \yS^3 \left( \yL^1 \yG^2 - \yL^2 \yG^1 \right)]_{r\rightarrow \infty}
  \,.
  \nonumber
\end{align}

\subsection{One-loop vector contribution to the effective action}

Knowing how to obtain the value of the gauge-Goldstone determinant,
we now want to explicitly write how it can be used to calculate the nucleation rate.
The functional determinant we have obtained is a one-loop gauge contribution to the effective action. It is a natural choice to include it in the exponential of the nucleation rate,
so the rate in eq.~\eqref{eq: Gamma [NLO det]} would contain the following term:
\begin{equation}
	\GammaT^{[\rmii{NLO det}]} \sim
	e^{
		-S_{\rm 3}^{\rmii{EFT,LO}}[v_{3,b}]
    -  \ln \rm{det}_{\rmii{$V$}}
	}
  \,.
\end{equation}
Then, taking together all the results from previous sections we arrive at the following expression for the vector-Goldstone determinant:
\begin{align}
  \ln \det_{\rmii{$V$}} =
	\underbrace{
		\ln\left( -2 \phi _\infty \varphi_b (r=0)  \right)
	}_{l=0} +
	\sum^{\infty } _{l=1} 
	(2 l +1)
	\biggl(
		\ln
			\frac{ \det\mathcal{O}^l _{\rmii{$XG$}} (v_{3,b}) }{ \det\mathcal{O}^l _{\rmii{$XG$}} (v_{3,\rmii{F}}) }
		- \ln
			\frac{ \det\mathcal{O}^l _{\rmii{$X_T$}} (v_{3,b}) }{ \det\mathcal{O}^l _{\rmii{$X_T$}}(v_{3,\rmii{F}}) }
	\biggr)
\,  ,
\end{align}
where we have used the fact that the ghost contribution is minus 2 times the contribution from the transverse gauge mode. The zero mode is contained in the $l=0$ term in the partial wave decomposition and the first term in the expression  above can be evaluated by using
{\tt BubbleDet} and the function
{\tt findLogPhiInfinity()};
for details see~\cite{Ekstedt:2023sqc}. 

In practice, the sum over $l$ is evaluated up to some finite $L_{\rm{max}}$ and
then one uses a WKB approximation for
$l > L_{\rm{max}}$~\cite{Ekstedt:2023sqc, Ekstedt:2021kyx}.%
\footnote{See~\cite{Baratella:2025dum} for a recent progress in regularising the functional determinants in $D=4$.}
Here, we write the leading order terms explicitly:
\begin{align}
	\ln \det_{\rmii{$V$}} \xrightarrow{l \gg 1}
  -3 \int {\rm d}r~r ( \Delta \WX(r))
  - \int {\rm d}r~r ( \Delta \WG(r) -\Delta \WX(r) )
	+ \mathcal{O}(l^{-3}) \, ,
\end{align}
where $\Delta W_a(r) \equiv W_a(r) - W_a(\infty)$.

{\small
\bibliographystyle{bib/utphys.bst}
\bibliography{bib/conformal-bib+GW.bib}
}
\end{document}